\RequirePackage{lineno} 

\documentclass[twocolumn,showpacs,preprintnumbers,amsmath,amssymb,superscriptaddress]{revtex4}

\usepackage{graphicx}
\usepackage{dcolumn}
\usepackage{bm}
\usepackage[chatter]{rotating}

\def\bea{\begin{eqnarray}}
\def\eea{\end{eqnarray}}
\newcommand{\deta}{\ensuremath{\eta_\Delta}}
\newcommand{\dphi}{\ensuremath{\phi_\Delta}}

\newenvironment{color}[3]{
}{
}
\newcommand{\red}[1]       {\begin{color}{1}{0}{0}{#1}\end{color}}

\newcommand{\cyan}[1]      {\begin{color}{0}{1}{1}{#1}\end{colJor}}

\begin{document}

\setpagewiselinenumbers
\modulolinenumbers[5]
\linenumbers

\preprint{Version 6.8}

\title{Anomalous centrality evolution of 
two-particle  angular correlations\\  
from Au-Au collisions at $\sqrt{s_{\rm NN}}$ = 62 and 200 GeV}

\affiliation{Argonne National Laboratory, Argonne, Illinois 60439, USA}
\affiliation{Brookhaven National Laboratory, Upton, New York 11973, USA}
\affiliation{University of California, Berkeley, California 94720, USA}
\affiliation{University of California, Davis, California 95616, USA}
\affiliation{University of California, Los Angeles, California 90095, USA}
\affiliation{Universidade Estadual de Campinas, Sao Paulo, Brazil}
\affiliation{University of Illinois at Chicago, Chicago, Illinois 60607, USA}
\affiliation{Creighton University, Omaha, Nebraska 68178, USA}
\affiliation{Czech Technical University in Prague, FNSPE, Prague, 115 19, Czech Republic}
\affiliation{Nuclear Physics Institute AS CR, 250 68 \v{R}e\v{z}/Prague, Czech Republic}
\affiliation{University of Frankfurt, Frankfurt, Germany}
\affiliation{Institute of Physics, Bhubaneswar 751005, India}
\affiliation{Indian Institute of Technology, Mumbai, India}
\affiliation{Indiana University, Bloomington, Indiana 47408, USA}
\affiliation{Alikhanov Institute for Theoretical and Experimental Physics, Moscow, Russia}
\affiliation{University of Jammu, Jammu 180001, India}
\affiliation{Joint Institute for Nuclear Research, Dubna, 141 980, Russia}
\affiliation{Kent State University, Kent, Ohio 44242, USA}
\affiliation{University of Kentucky, Lexington, Kentucky, 40506-0055, USA}
\affiliation{Institute of Modern Physics, Lanzhou, China}
\affiliation{Lawrence Berkeley National Laboratory, Berkeley, California 94720, USA}
\affiliation{Massachusetts Institute of Technology, Cambridge, MA 02139-4307, USA}
\affiliation{Max-Planck-Institut f\"ur Physik, Munich, Germany}
\affiliation{Michigan State University, East Lansing, Michigan 48824, USA}
\affiliation{Moscow Engineering Physics Institute, Moscow Russia}
\affiliation{NIKHEF and Utrecht University, Amsterdam, The Netherlands}
\affiliation{Ohio State University, Columbus, Ohio 43210, USA}
\affiliation{Old Dominion University, Norfolk, VA, 23529, USA}
\affiliation{Panjab University, Chandigarh 160014, India}
\affiliation{Pennsylvania State University, University Park, Pennsylvania 16802, USA}
\affiliation{Institute of High Energy Physics, Protvino, Russia}
\affiliation{Purdue University, West Lafayette, Indiana 47907, USA}
\affiliation{Pusan National University, Pusan, Republic of Korea}
\affiliation{University of Rajasthan, Jaipur 302004, India}
\affiliation{Rice University, Houston, Texas 77251, USA}
\affiliation{Universidade de Sao Paulo, Sao Paulo, Brazil}
\affiliation{University of Science \& Technology of China, Hefei 230026, China}
\affiliation{Shandong University, Jinan, Shandong 250100, China}
\affiliation{Shanghai Institute of Applied Physics, Shanghai 201800, China}
\affiliation{SUBATECH, Nantes, France}
\affiliation{Texas A\&M University, College Station, Texas 77843, USA}
\affiliation{University of Texas, Austin, Texas 78712, USA}
\affiliation{University of Houston, Houston, TX, 77204, USA}
\affiliation{Tsinghua University, Beijing 100084, China}
\affiliation{United States Naval Academy, Annapolis, MD 21402, USA}
\affiliation{Valparaiso University, Valparaiso, Indiana 46383, USA}
\affiliation{Variable Energy Cyclotron Centre, Kolkata 700064, India}
\affiliation{Warsaw University of Technology, Warsaw, Poland}
\affiliation{University of Washington, Seattle, Washington 98195, USA}
\affiliation{Wayne State University, Detroit, Michigan 48201, USA}
\affiliation{Institute of Particle Physics, CCNU (HZNU), Wuhan 430079, China}
\affiliation{Yale University, New Haven, Connecticut 06520, USA}
\affiliation{University of Zagreb, Zagreb, HR-10002, Croatia}

\author{G.~Agakishiev}\affiliation{Joint Institute for Nuclear Research, Dubna, 141 980, Russia}
\author{M.~M.~Aggarwal}\affiliation{Panjab University, Chandigarh 160014, India}
\author{Z.~Ahammed}\affiliation{Variable Energy Cyclotron Centre, Kolkata 700064, India}
\author{A.~V.~Alakhverdyants}\affiliation{Joint Institute for Nuclear Research, Dubna, 141 980, Russia}
\author{I.~Alekseev}\affiliation{Alikhanov Institute for Theoretical and Experimental Physics, Moscow, Russia}
\author{J.~Alford}\affiliation{Kent State University, Kent, Ohio 44242, USA}
\author{B.~D.~Anderson}\affiliation{Kent State University, Kent, Ohio 44242, USA}
\author{C.~D.~Anson}\affiliation{Ohio State University, Columbus, Ohio 43210, USA}
\author{D.~Arkhipkin}\affiliation{Brookhaven National Laboratory, Upton, New York 11973, USA}
\author{G.~S.~Averichev}\affiliation{Joint Institute for Nuclear Research, Dubna, 141 980, Russia}
% ...

\collaboration{STAR Collaboration}\noaffiliation

\date{\today}

\begin{abstract}
We present two-dimensional (2D) two-particle angular correlations on relative pseudorapidity $\eta$ and azimuth $\phi$  for charged particles from Au-Au collisions at $\sqrt{s_{\rm NN}} = 62$ and 200~GeV with transverse momentum $p_t \geq 0.15$~GeV/$c$, $|\eta| \leq 1$ and $2\pi$ azimuth. Observed correlations include a {same-side} (relative azimuth $< \pi/2$) 2D peak, a closely-related away-side azimuth dipole, and an azimuth quadrupole conventionally associated with elliptic flow.
The same-side 2D peak and away-side dipole are explained by semihard parton scattering and fragmentation (minijets) in proton-proton and peripheral nucleus-nucleus collisions. 
Those structures follow N-N binary-collision scaling in Au-Au collisions until mid-centrality, where a transition to a qualitatively different centrality trend occurs within a small centrality interval. 
Above the transition point the number of same-side and away-side correlated pairs increases rapidly {relative to} binary-collision scaling, the $\eta$ width of the same-side 2D peak also increases rapidly ($\eta$ elongation) and the $\phi$ width actually decreases significantly. 
Those centrality trends are in marked contrast with conventional
expectations for jet quenching in a dense medium.
%
%
The observed centrality trends are compared to perturbative QCD
predictions computed in {\sc hijing}, which serve as a theoretical
baseline, and to the expected trends for semihard parton scattering
and fragmentation in a thermalized opaque medium predicted by
theoretical calculations and phenomenological models.
%
We are unable to reconcile a semihard parton scattering and fragmentation origin for the observed correlation structure and centrality trends with heavy ion collision scenarios which invoke rapid parton thermalization.
If the collision system turns out to be effectively opaque to
few-GeV partons the present observations would be inconsistent 
with the minijet picture discussed here.



\end{abstract}


\maketitle

\section{Introduction}
\label{Sec:Intro}

Many conventional theory descriptions of central collisions at the Relativistic Heavy Ion Collider (RHIC) full energy invoke the basic assumption that copious parton (mainly gluon) production during initial nucleus-nucleus (A-A) contact and subsequent parton rescattering lead to a color-deconfined, locally-thermalized quark-gluon plasma~\cite{qgp1,qgp2}. Hydrodynamic models~\cite{hydro1,hydro2,hydro3,hydro4}, claims of ``perfect liquid'' formation~\cite{perfect1,perfect2,perfect3,perfect4}, and the relevance of lattice QCD predictions to RHIC data all rely on assumed formation of a rapidly-thermalized QCD medium.  However, experimental confirmation of that assumption remains an open question. Although the constituents of the system may interact strongly, thermalized matter may not emerge in the time available in relativistic collisions~\cite{kovchegov}. Experimental study of possible rapid thermalization
is one of the goals of this paper.

RHIC heavy ion collisions are studied as a function of nucleus size A, collision energy and centrality to search for evidence that an approximately linear superposition of nucleon-nucleon (N-N) interactions~\cite{powerlaw} expected for peripheral A-A collisions evolves with increasing size, energy and centrality to a collective system of dense, strongly-interacting QCD matter. In reports by the four RHIC experiments~\cite{starwp,phenixwp,phoboswp,brahmswp} it was argued that observations are consistent with a collective thermalized medium.  

High-$p_t$ jet tomography was proposed to probe the conjectured QCD medium. Hard-scattered partons produced in large-$Q$ interactions during initial \mbox{A-A} contact 
[where $Q$ is the parton (actually dijet) energy scale]
are nominally well-understood probes of collision dynamics and QCD medium properties (i.e., described by perturbative QCD or pQCD)~\cite{highpt}. The underlying assumption is that formation of a QCD medium should modify parton scattering and fragmentation to hadrons and may thereby produce deviations of corresponding hadron distributions (single-particle spectra and correlations) from binary-collision scaling~\cite{starwp,phenixwp}. Much attention has therefore been paid to high-$p_t$ systematics (e.g., reduced high-$p_t$ hadron yields~\cite{starraa}, suppression of jet-related away-side azimuth correlations~\cite{staras}) interpreted to reveal strong parton energy loss~\cite{highpt}. But those results do not distinguish thermalization scenarios from other possibilities~\cite{kovchegov}.


In this paper we utilize two-particle angular correlations among all accepted charged particles and focus on those structures associated with semihard parton scattering and fragmentation~\cite{minijet}, referred to as {\em minijet angular correlations}. Those structures provide a complementary approach to medium studies. 
Inference of jet structure (minijets) from minimum-bias (all particles in the $p_t$ acceptance) angular correlations~\cite{axialci,ptxptci,bubbles,phobosci} differs qualitatively from high-$p_t$ jet methods in that the minijet analysis does not depend on an {\em a priori} jet model. No ``trigger particle'' (parton proxy) is required and no ``associated-particle'' $p_t$ cuts are imposed. In the absence of trigger-associated $p_t$ cuts all minijet  hadrons, which strongly overlap on $p_t$ those hadrons produced by soft processes (e.g.,  participant nucleon fragmentation along the collision axis), are accepted in the analysis. 

The phrase ``minijet contribution'' refers in the present context to 
the distribution of correlated hadron fragments from a minimum-bias parton energy spectrum averaged over a given A-A (or N-N) event ensemble. 
Because the parton spectrum is rapidly falling ($\sim 1/p_t^{6}$), with an observed lower bound near 3 GeV, the apparent minimum-bias parton spectrum is nearly monoenergetic~\cite{fragevo}. The term ``minijets'' then corresponds experimentally to jets localized near the 3 GeV lower bound (equivalent to parton energy scale $Q \approx$ 6 GeV), consistent with the original usage~\cite{sarc,kll}. 
Minijets (minimum-bias jets) are further discussed in App.~\ref{minijets}. 


In this analysis we report experimental tests of the
local-thermalization hypothesis and conjectured bulk
medium properties using minijets as probes of the system.
By analogy with Brownian motion~\cite{brown} minijet probes
(small-Q gluons) are just ``large'' enough (sufficiently
energetic) to manifest as hadronic correlations (minijets)
yet ``small'' enough to provide good sensitivity to local
medium properties and dynamics (e.g., other semihard partons)
{~\cite{jethydro}.

It is essential to establish a theoretical baseline prediction for
minijet correlations. In the absence of medium effects such correlations
should correspond to a linear superposition of N-N collisions
(binary collision scaling) as described by the Glauber model of
A-A collisions (Glauber linear superposition or GLS). Minijets may be strongly modified in more-central collisions or even vanish in an opaque thermalized medium~\cite{kll,mjmuel,nayak,mjshin}.
The goal of this analysis is to determine where measured minijet correlations agree with  baseline predictions (no medium effects) obtained from perturbative QCD as represented by the {\sc hijing} Monte Carlo~\cite{hijing} and to quantify any deviations from that baseline as a function of collision energy and centrality. Our results are further discussed in terms of the expected centrality trends for semihard parton scattering and fragmentation in dense, strongly-interacting media predicted by theoretical calculations and phenomenological models.

Angular correlations among the products from nuclear collisions are revealed by two-dimensional (2D) angular {\em autocorrelations} (Sec.~\ref{Sec:Analysis}) defined on pseudorapidity and azimuth difference variables $\eta_{\Delta} \equiv \eta_1 - \eta_2$ and $\phi_{\Delta} \equiv \phi_1 - \phi_2$~\cite{auto,axialcd,ptedep}.  Correlation sources include hadronic resonances, elliptic flow, quantum statistics (HBT) and semihard parton scattering (minijets). In proton-proton ({\it p-p}) collisions the observed angular correlations, when viewed using pair-wise $p_t$ cuts~\cite{jeffpp1,aspect}, are comprised of simple geometric structures: (i) a same-side ($\phi_\Delta < \pi/2$) 2D peak at the origin on $(\eta_\Delta,\phi_\Delta)$, (ii) an away-side ridge in the form of dipole $\cos(\phi_\Delta-\pi)$, and (iii) a 1D peak on $\eta_\Delta$ centered at the origin. (i) and (ii), with hadron $p_t > 0.35$ GeV/$c$ (for \mbox{{\it p-p}} collisions), are interpreted together as minijet angular correlations, and (iii) falls mainly below hadron $p_t = 0.5$ GeV/$c$~\cite{minijet,jeffpp1,aspect,kll,mjmuel,nayak,mjshin}.


Other correlation analyses have been performed with RHIC data, but most have focused on specific features of angular correlations. Several PHENIX studies (e.g.,~\cite{phenix}) were restricted to 1D azimuth correlations. Other STAR and PHOBOS analyses have imposed so-called trigger-associated $p_t$ cuts (e.g.,~\cite{trigger}) which retain only part of the jet structure and reduce or exclude other contributions. One other analysis~\cite{phoboscorr} does consider $p_t$-integral 2D angular correlations (albeit over a restricted centrality range) and is discussed further in Sec.~\ref{unique}.

The STAR Collaboration previously reported measurements of minimum-bias 2D angular correlations for charged-particle pairs from Au-Au collisions at 130~GeV~\cite{axialci}.  Significant correlation structures from several sources were reported, including those interpreted as 
minijet contributions. 
Centrality variation of the same-side 2D peak was inconsistent with expectations from jet-quenching theory~\cite{kll,mjmuel,nayak,mjshin}. Instead of diminishing with increasing \mbox{Au-Au} centrality (as expected in jet quenching scenarios), the same-side peak amplitude increased strongly with centrality, and the azimuth width decreased instead of increasing. Most surprisingly, the width on relative pseudorapidity $\eta_\Delta$ increased more than 2-fold from peripheral to central collisions. However, the limited statistics of the 130~GeV Au-Au data did not permit detailed study of the centrality dependence of the correlation structure.


In the present analysis the method of Ref.~\cite{axialci} has been applied to charged hadron production from minimum-bias Au-Au collisions at $\sqrt{s_{\rm NN}}$ = 62 and 200~GeV~\cite{star}. A preliminary report of results was presented in~\cite{daugherity}. The much larger data volume (compared to the 130~GeV data) and two collision energies make possible a detailed study of the centrality and energy dependence of correlation systematics. 
The new results confirm our previous observation of unexpected centrality trends~\cite{axialci}, which in retrospect constitute the discovery of $\eta$ broadening of the same-side peak, but also reveal for the first time the onset of strong deviations from binary-collision scaling at a specific Au-Au centrality common to both energies. 


Taken together, our analysis results reveal that the
correlation structure of interest (minijet structure) evolves with centrality according
to a simple Glauber linear-superposition baseline, consistent
with no novelty in A-A collisions compared to p-p, up to
a specific centrality point where evolution of several parameters undergoes
a sharp transition (large slope changes within a small
centrality interval) to a qualitatively different smooth
trend. The large increase in jet-like
structure above the transition point relative to the GLS
trend contrasts with expectations of strong jet quenching
in more-central A-A collisions~\cite{sarc,mjmuel,nayak,jethydro}. The anomalous centrality evolution then consists of the sharp transition
and the unexpected increase in jet-like correlations in
more-central Au-Au collisions relative to theoretical
expectations, as discussed in Secs.~\ref{Sec:Anomalous} and \ref{anomdisc}.

Given the discovery of anomalous
centrality evolution involving correlations interpreted in
p-p collisions in terms of minijets we wish to test various
theoretical collision scenarios, especially those assuming
rapid thermalization to form a dense bulk medium nominally
opaque to jets. We hypothesize that pQCD minijet structure
should follow a GLS reference trend in A-A collisions
unless modified by interactions leading to thermalization.
We determine to what extent jet structure is modified from
p-p to central Au-Au collisions relative to the theoretical
baseline.  Perturbative QCD-based Monte Carlo model {\sc hijing}~\cite{hijing}
(without jet quenching) provides a nominal GLS theory baseline.
In thermalization scenarios we expect to see strong reduction
and other modifications (symmetric peak broadening) or even
extinction of jet-like correlations. If those expectations
are not met we may question the theoretical assumptions.
Recent correlation predictions from the transport model {\sc ampt}~\cite{ampt}
and event-wise hydrodynamic model {\sc nexspherio}~\cite{nexsph} are
discussed in that context in Sec.~\ref{other}.

This paper is organized as follows: The analysis method, data selection and measured angular correlations are described in Secs.~\ref{Sec:Analysis}-\ref{Sec:Dist} respectively. The model-fitting procedure, fit results, and systematic uncertainties are presented in Secs.~\ref{Sec:Model}-\ref{Sec:Errors} respectively.  Results and implications for heavy ion collision interpretations are discussed in Sec.~\ref{Sec:Diss}, and a summary and conclusions are presented in Sec.~\ref{Sec:Summ}. Further analysis details are presented in five Appendices.

\section{Analysis Method}
\label{Sec:Analysis}


Number correlations (reported here) on binned two-particle momentum space (as opposed to $p_t$ correlations~\cite{auto,ptedep}) are commonly reported as a ratio in each bin of the number of sibling pairs (from the same collision) to a number of reference or mixed pairs (from different but similar collisions -- see App.~\ref{ratio}). The ratio relative to unity is
\bea
\label{Eq1}
\frac{\Delta\rho}{{\rho_{\rm ref}}} &\equiv&  \frac{\rho_{\rm sib} - \rho_{\rm ref}}{{\rho_{\rm ref}}}
 =  r - 1,
\eea
where bin indices are suppressed, $\rho_{\rm sib}$ denotes the density of sibling pairs, $\rho_{\rm ref}$ is the reference density of mixed pairs, and ratio $r = \rho_{\rm sib} / \rho_{\rm ref}$. 
Expressions using binned pair counts explicitly are provided in App.~\ref{ratio}.
The {\em per-pair} measure defined in Eq.~(\ref{Eq1}) is useful for quantum correlations~\cite{hbtreview}, where e.g.~the number of correlated pairs in a bin on invariant relative momentum may be approximately proportional to the single-particle density squared. However, correlation structures associated with initial-state scattering (relative to number of participant nucleons) or hadronization (relative to final-state hadron multiplicity) are better described by a {\em per-particle} measure. Defined symbol $\Delta\rho/\sqrt{\rho_{\rm ref}}$ represents such a per-particle  measure designed specifically to test the null hypothesis that a nucleus-nucleus collision is equivalent to a Glauber linear superposition of N-N collisions. 
The statistical measure $\Delta\rho/\sqrt{\rho_{\rm ref}}$
defined by
\bea
\label{Eq2}
\frac{\Delta\rho}{\sqrt{\rho_{\rm ref}}} &\equiv& \sqrt{\rho^{\prime}_{\rm ref}} \frac{\Delta\rho}{\rho_{\rm ref}}
 = \sqrt{\rho^{\prime}_{\rm ref}} (r - 1)
\eea 
is equivalent to Pearson's normalized covariance (or correlation coefficient). The numerator $\Delta\rho$ is the covariance of fluctuating particle numbers in two single-particle histogram bins, and the denominator 
(effectively $\sqrt{\rho^\prime_{\rm ref}}$) is approximately the geometric mean of two single-particle number variances, leading (in the Poisson limit) to  {\em per-particle} normalization.
 The explicit form is given in Eq.~(\ref{Eq3}) and App.~\ref{ratio}.
The right-most expression in Eq.~(\ref{Eq2}) insures the cancelation of acceptance effects (and other experimental artifacts) in the ratio $r$. 

Prefactor $\sqrt{\rho^{\prime}_{\rm ref}}$ in Eq.~(\ref{Eq2}) is the ideal geometric-mean single-particle density absent acceptance and inefficiency effects, approximately the single-charged-particle density
$d^2\bar N_\text{ch}/d\eta d\phi$ averaged over the angular acceptance.  Both 62 and 200~GeV Au-Au
multiplicity distributions are constant on pseudorapidity 
to within 1-2\% for $|\eta| \leq 1$~\cite{phobos62,starspec200,molnarthesis}.  The prefactor can therefore be approximated by $\sqrt{\rho^{\prime}_{\rm ref}} \approx \bar{N}_{\rm ch}/\Delta\eta\Delta\phi$, where $\bar{N}_{\rm ch}$ is the {\em corrected} mean charged-particle multiplicity within the acceptance for each
centrality bin (see Tables~\ref{TableI} and \ref{TableII}), and the angular acceptance for this analysis is defined by $\Delta\eta = 2$ ($|\eta| \leq 1$) and $\Delta\phi = 2\pi$. 
The correlation measure used in this analysis is then
\bea
\label{Eq3}
&& \hspace{-.25in}\frac{\Delta\rho^{\rm (CI)}}{\sqrt{\rho_{\rm ref}}}(a,b)  =  \frac{\bar{N}_{\rm ch}}{\Delta \eta \Delta \phi } (\hat{r}_{ab} - 1)  =  \frac{\bar{N}_{\rm ch}}{2 \times 2\pi } (\hat{r}_{ab} - 1),
\eea
where CI denotes the charge-independent summation over four charge-pair combinations. $\hat{r}_{ab}$ is the sibling/mixed ratio of normalized total pair numbers  in 2D histogram bin $(a,b)$ averaged over charge-pair combinations, event-multiplicity bins (within a given centrality bin), and collision-vertex-position bins, the average being weighted by sibling-pair number as described in App.~\ref{ratio}.

Indices $(a,b)$ represent an unspecified 2D binning of 6D two-particle momentum space $(\vec{p}_1,\vec{p}_2)$. For {\it p-p} collisions it was shown that projections onto subspaces $(p_{t1},p_{t2})$ and $(\eta_1,\eta_2,\phi_1,\phi_2)$ are complementary (correlation structure is factorized with minimal information loss)~\cite{jeffpp1,aspect}.  
In this analysis, as in Ref.~\cite{axialci}, only projections onto the angular subspace are reported ($p_t$-integral correlations). 

In Ref.~\cite{axialcd} correlation structures on 2D angular subspaces $(\eta_1,\eta_2)$ and $(\phi_1,\phi_2)$ were found to be invariant on sum axes $\eta_1 + \eta_2$ and $\phi_1 + \phi_2$ within the STAR TPC acceptance. 4D angular subspace $(\eta_1,\eta_2,\phi_1,\phi_2)$ can then be simplified by projection onto difference axes $\eta_1-\eta_2$ and $\phi_1-\phi_2$ by averaging without loss of information along the sum axes within the TPC angular acceptance, thus forming a 2D angular autocorrelation~\cite{auto,ptedep}. Indices $(a,b)$ then label 2D bins on difference axes $(\eta_{\Delta},\phi_{\Delta})$. 

The autocorrelation technique in the context of nuclear collisions applies to angular correlations only, and only in the case that invariance on the sum axes is a good approximation (e.g., within restricted intervals on $\eta$). The technique does not apply to correlations on $(m_t,m_t)$ or $(y_t,y_t)$ 
for instance. We use the formal term ``autocorrelation'' initially for the purpose of definition and adopt the shorter form ``correlations''  subsequently in the text.

This analysis is unique in part because it introduces several new techniques, including (a) consideration of the full range of A-A centralities down to N-N collisions, (b) application of a statistically well-defined per-particle correlation measure, (c) definition of a Glauber linear superposition reference, (d) accurate model fits to 2D angular correlations, (e) proper control of several systematic biases including pile-up effects, distortions due to canonical suppression arising from centrality definition and distortions due to variation in position of the collision vertex and collision multiplicity. Those aspects are discussed further in Sec.~\ref{unique}.

\begin{figure*}[t]
\includegraphics[width=.24\textwidth]{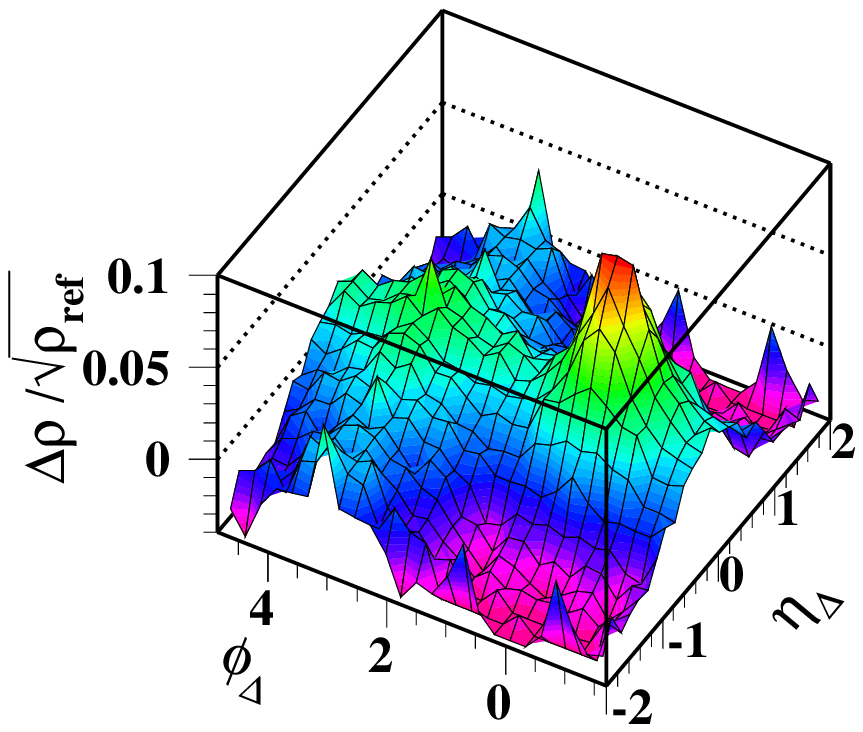} 
\put(-100,85){\bf (a)}
\includegraphics[width=.24\textwidth]{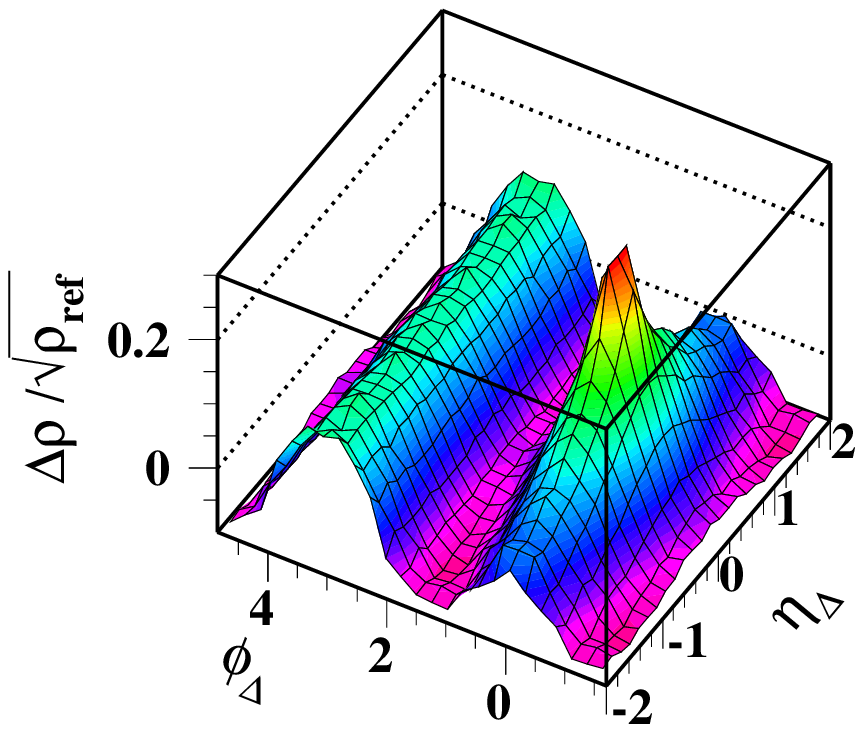}
\put(-100,85){\bf (b)}
\includegraphics[width=.24\textwidth]{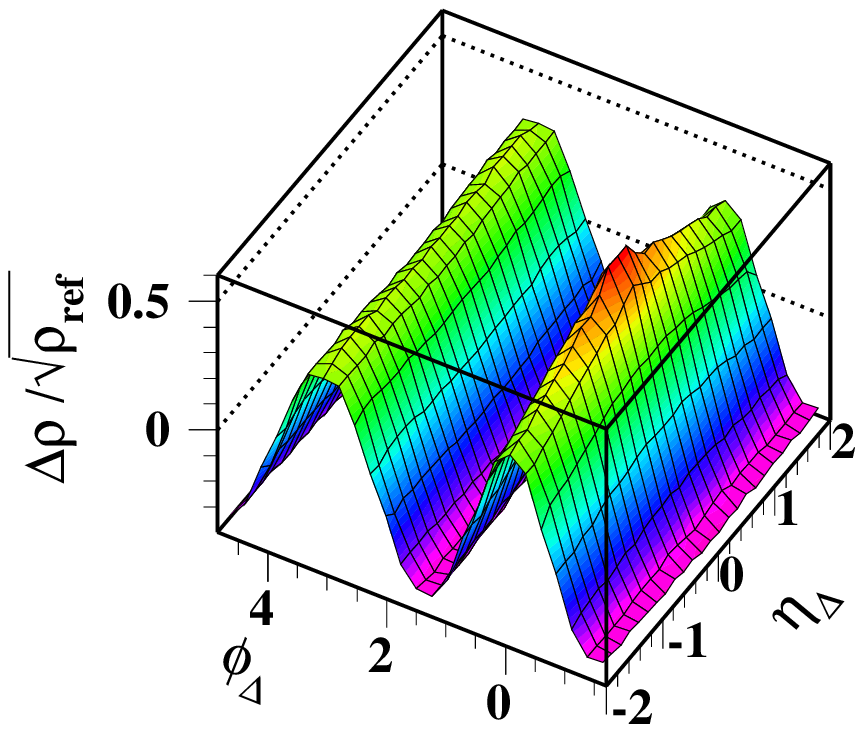}
\put(-100,85){\bf (c)}
\includegraphics[width=.24\textwidth]{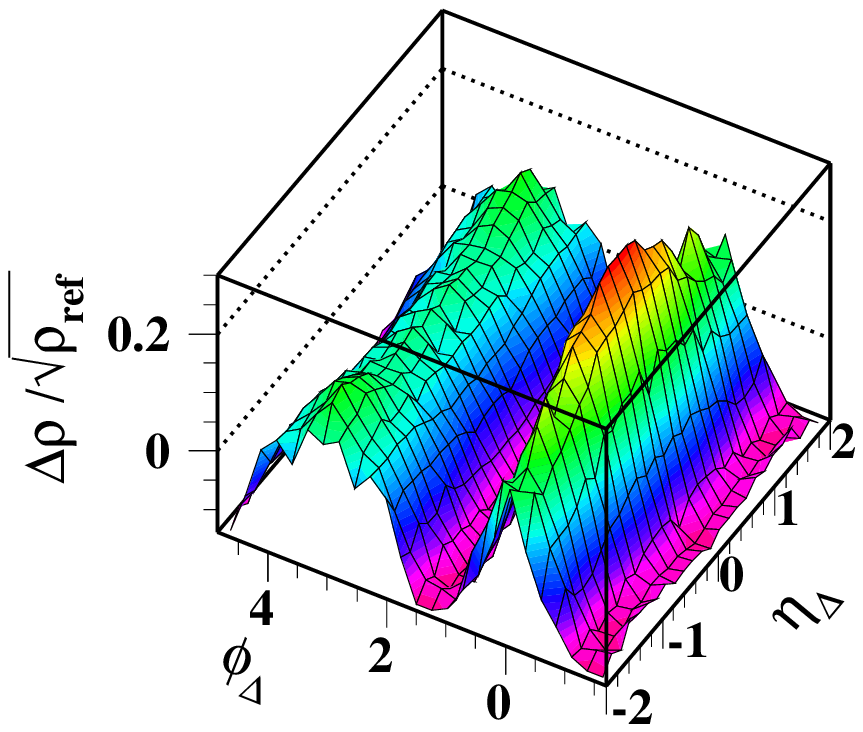}
\put(-100,85){\bf (d)}\\
\includegraphics[width=.24\textwidth]{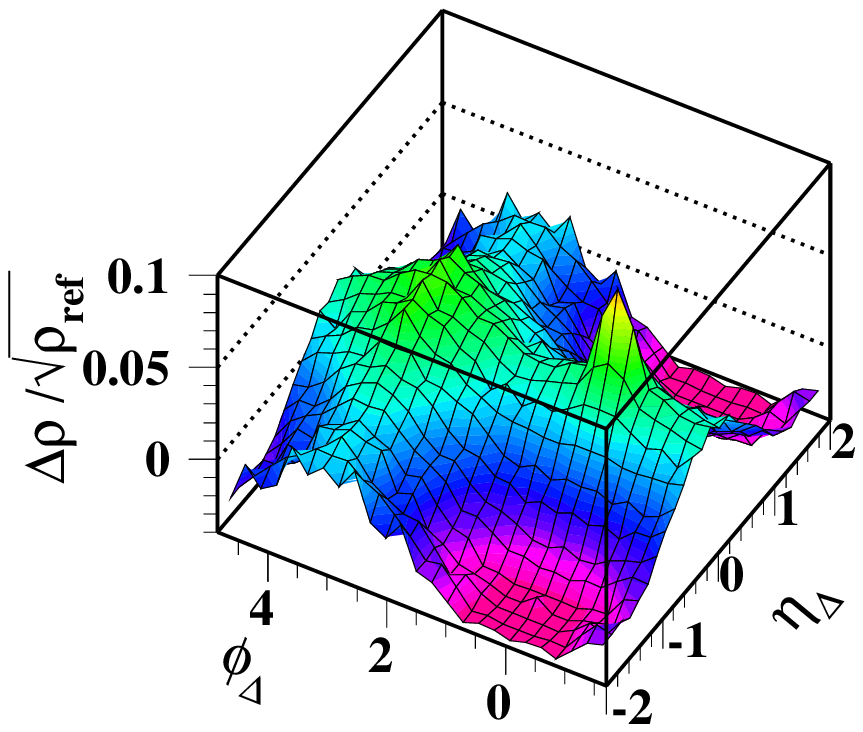}
\put(-100,85){\bf (e)}
\includegraphics[width=.24\textwidth]{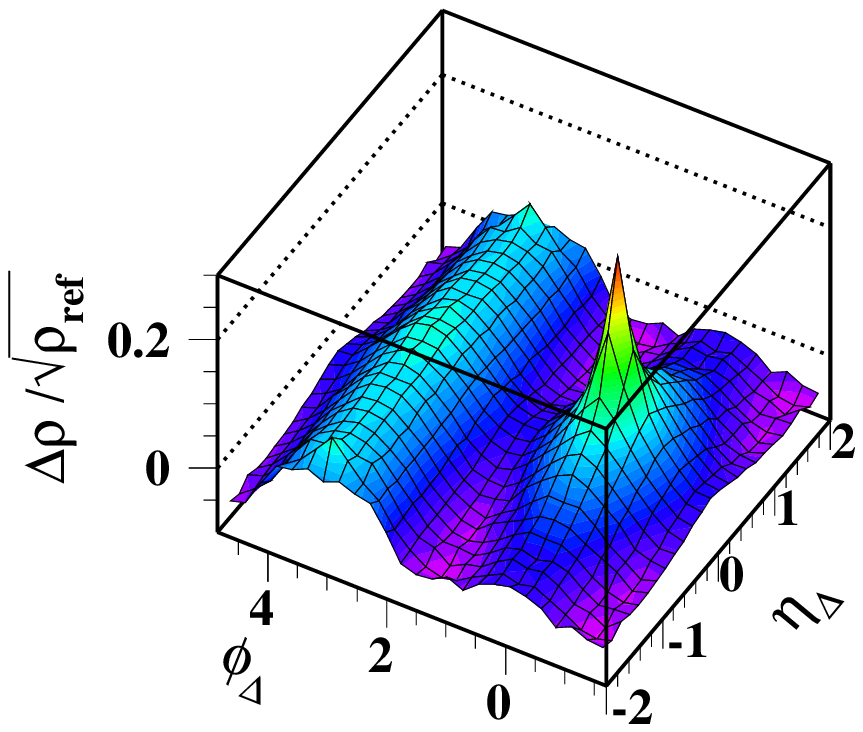}
\put(-100,85){\bf (f)}
\includegraphics[width=.24\textwidth]{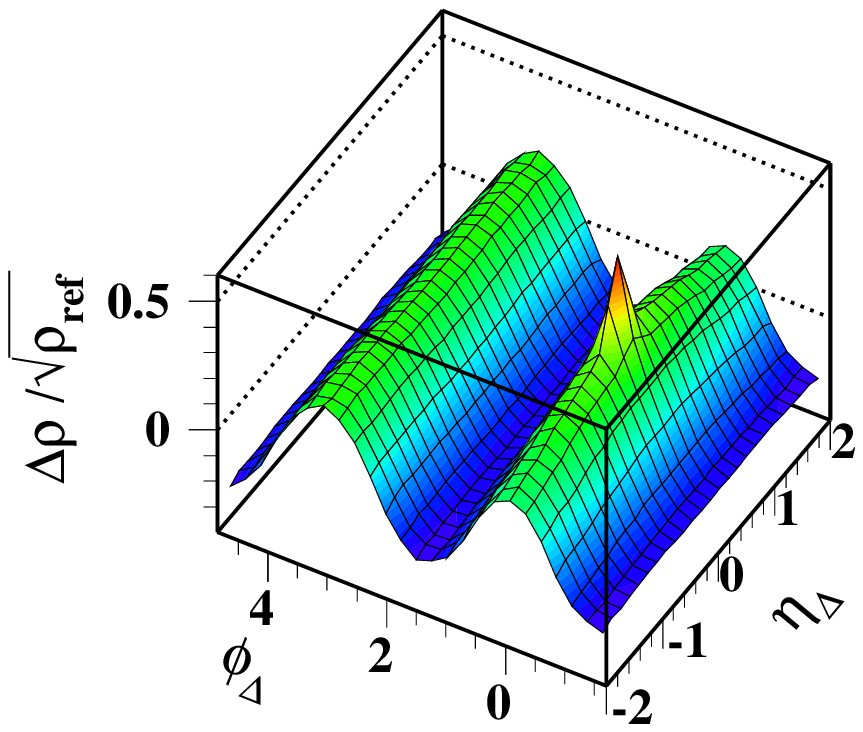}
\put(-100,85){\bf (g)}
\includegraphics[width=.24\textwidth]{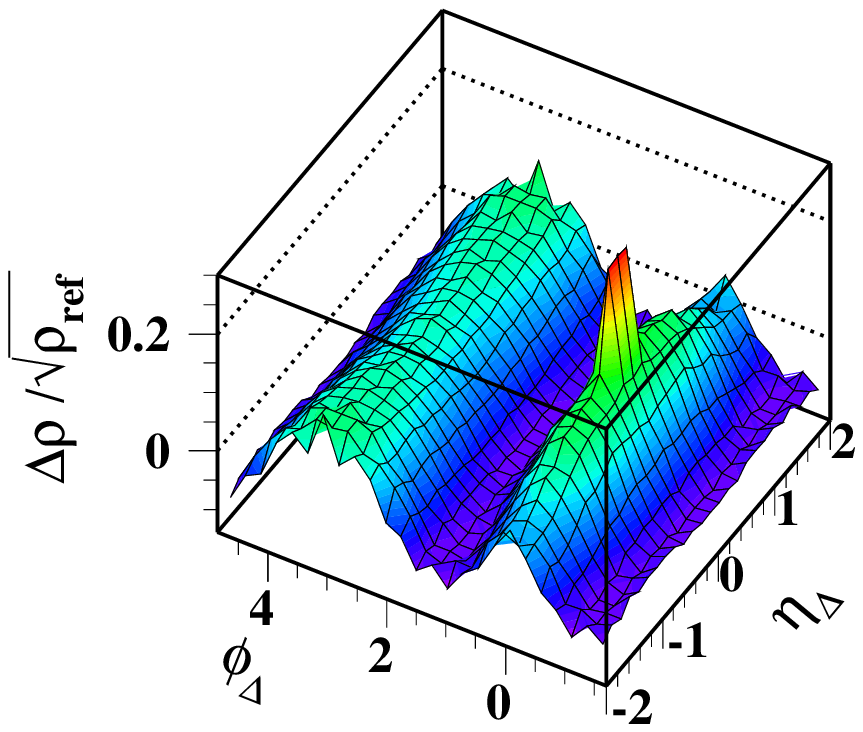}
\put(-100,85){\bf (h)}
\caption{\label{Figure1} 
(Color online) Perspective views of two-dimensional charge-independent angular correlations $\Delta\rho/\sqrt{\rho_{\rm ref}}$ on  $(\eta_{\Delta},\phi_{\Delta})$ 
for Au-Au collisions at $\sqrt{s_{\rm NN}}$ = 200 and 62~GeV (upper and lower rows respectively). Centrality increases left-to-right from most-peripheral to most-central.
Corrected total cross-section fractions are (left to right) 84-93\%, 55-64\%, 18-28\% and 0-5\% for the 200 GeV data and 84-95\%, 56-65\%, 18-28\% and 0-5\% for the 62 GeV data (see Tables~\ref{TableI} and \ref{TableII}). 
} 
\end{figure*}

\section{Data}
\label{Sec:Data}

Data for this analysis were obtained with the STAR detector~\cite{star}
using a 0.5~T uniform magnetic field parallel to the beam axis.
Minimum-bias triggered events for collision energies $\sqrt{s_{\rm NN}}$ = 62 and 200~GeV were obtained 
by requiring a coincidence of two Zero-Degree Calorimeters (ZDCs) and a minimum number of charged-particle hits in the Central Trigger (scintillator) Barrel (CTB). Charged-particle measurements with the Time Projection Chamber (TPC) and event triggering are described in \cite{star}. Primary vertices, defined using TPC tracks, were required to fall within 25~cm of the axial ($z$-axis) center of the TPC. The data accepted for this analysis included 6.7 million events at $\sqrt{s_{\rm NN}}$ = 62 GeV {(Run 4 - 2004)} and 1.2 million events at 200 GeV {(Run 2 - 2001)}. The present analysis is not limited by statistics; the 1.2M events from Run 2 are sufficient for all analysis requirements.

Accepted particle trajectories fell within the TPC acceptance defined by $p_t>0.15$~GeV/$c$, $|\eta| < 1.0$ and $2 \pi$ azimuth. Primary tracks in each event were required to have a Distance of Closest Approach (DCA) less than 3~cm from the reconstructed primary vertex, accepting a large fraction of true primary hadrons plus approximately 12\% background contamination~\cite{starspec200,molnarthesis} from weak decays and interactions with detector material. Conversion electron-positron backgrounds were reduced by excluding particles with dE/dx (specific energy loss in the TPC) within 1.5$\sigma$ of that expected for electrons in the momentum ranges $0.2 < p < 0.45$ GeV/c and $0.7 < p < 0.8$ GeV/c. Charge signs were determined, but particle identification was not otherwise implemented.  Further details of track definitions, efficiencies and quality cuts are described in Refs.~\cite{starspec200,ayathesis}.  

Event pileup results in tracks from an untriggered event coexisting with a triggered event in the TPC. Although the pileup rate for these data was typically less than 1\% such pileup can produce significant unwanted structure in angular correlations. A method to correct angular correlations for pileup is described in App.~\ref{pileupcorr}.


The pileup-corrected minimum-bias event sample at each energy was divided into eleven centrality bins: nine each with {\em nominally} 10\% of the total cross section and the most-central 10\% split into 5\% bins. 
The {\em corrected} centrality fractions reported in Tables~\ref{TableI} and \ref{TableII} were determined from the minimum-bias distribution plotted as $dN_{\rm event}/dN_{\rm ch}^{1/4}$ versus $N_{\rm ch}^{1/4}$ on {\em accepted} event multiplicity $N_{\rm ch}$ after adjustments for inefficiencies due to triggering, collision vertex finding and particle trajectory reconstruction. That distribution is nearly uniform because the minimum-bias distribution $dN_{event}/dN_{ch}$ is observed to approximate a  ``power-law'' trend $\propto N_{ch}^{-3/4}$~\cite{powerlaw}. The low-multiplicity end point of the distribution on $N_{\rm ch}^{1/4}$ was constrained by measured {\it p-p} minimum-bias collision multiplicities~\cite{ua5} normalized to the STAR TPC acceptance.

Multiplicity $N_{\rm ch}$ used to determine the centrality was integrated over the same pseudorapidity acceptance $|\eta| < 1$ used for the correlation analysis. Use of $N_{\rm ch}$ from a restricted interval (e.g. $|\eta| < 0.5$ as in~\cite{starraa}) to define the collision centrality results in artifacts in 2D histograms due to {\em canonical suppression}. Correlations (fluctuations) within the restricted pair acceptance are suppressed relative to those outside it, leading to substantial systematic errors in the angular correlations.

Centrality is represented in a Glauber context by parameter $\nu = 2\langle N_{\rm bin} \rangle / \langle N_{\rm part} \rangle$, the average number of \mbox{N-N} binary collisions per incident participant nucleon (in either nucleus) as obtained from Monte Carlo Glauber-model simulations~\cite{raycentral} related to 
{
62 and 200~GeV minimum-bias distributions on $dN_{\rm ch}/d\eta$ from~\cite{starspec200,molnarthesis} 
}
and denoted $\nu_{62}$ and $\nu_{200}$. Parameter $\nu$ is matched to observable $N_{\rm ch}$ through the integrated total cross section {\em via} the approximately rectangular power-law distribution on $N_{\rm ch}^{1/4}$ as described in Ref.~\cite{powerlaw}. At the lower-multiplicity end point (half-maximum point) $\nu \equiv 1$ while at the upper-multiplicity end point ($b = 0$) $\nu = 5.29 \pm 0.20$ and $6.17 \pm 0.23$ for 62 and 200~GeV data, respectively. The estimated mean value of $\nu$ for {\it p-p} (N-N) collisions is 1.25 (differing from 1 because of the skewness of the N-N multiplicity distribution~\cite{powerlaw}).

The Glauber parameters can also be viewed as purely geometric measures unrelated to a particular N-N process: $\nu$ can be thought of as the average participant path length. The 200 GeV parameters (assuming a 42 mb N-N cross section~\cite{siginel200}) are then adopted as default geometry measures for both energies. Centrality measure $\nu$ facilitates tests of the N-N linear-superposition hypothesis.

Estimates of ensemble-mean $\bar{N}_{\rm ch}$ for each centrality bin were obtained from minimum-bias multiplicity distributions~\cite{molnarthesis} and from Monte Carlo Glauber-model simulations assuming a two-component hadron production model~\cite{kn}.  
The two methods agreed within ~10\% (most peripheral) and 1\% (most central)
and were within 6\% and 3\% for the intervening centralities for the 200
and 62 GeV data, respectively. Average values were used for the corrected multiplicities $\bar N_{\rm ch}$, listed as angular density $d\bar N_{\rm ch} /d\eta \equiv \bar N_{\rm ch} / 2$ in Tables~\ref{TableI} and \ref{TableII}.


\section{2D angular autocorrelations}
\label{Sec:Dist}

Figure~\ref{Figure1} shows perspective views of data histograms $\Delta\rho/\sqrt{\rho_{\rm ref}}(\eta_{\Delta},\phi_{\Delta})$ for representative centralities obtained from Au-Au collisions at $\sqrt{s_{\rm NN}}$ = 62 and 200~GeV. The histograms show (within a constant factor) the event-wise mean number of correlated pairs per final-state particle in each $(\eta_{\Delta},\phi_{\Delta})$ bin.

The pair angular acceptances were divided into 25 bins on the $\eta_{\Delta}$ axis and 25 bins on $\phi_{\Delta}$, a compromise between statistical error magnitude and angular resolution. The histograms are by construction symmetric about $\eta_\Delta = 0$ and $\phi_\Delta = 0, \pi$. The 25 bins on $\phi_\Delta$ actually span $2\pi + \pi/12$ to insure centering of major peaks on azimuth bin centers. Statistical errors are $\sim \pm 0.002$ ($\pm 0.004$) for 62 (200)~GeV data near $|\eta_{\Delta}| = 0$. Because of the $\eta_\Delta$ dependence of the pair acceptance statistical errors increase with $|\eta_\Delta|$ as  $\sqrt{\Delta\eta/(\Delta\eta - |\eta_{\Delta}|)}$ for $\eta$ acceptance $\Delta\eta = 2$ but are uniform on $\phi_{\Delta}$. Statistical errors are approximately independent of centrality for this per-particle measure. 
Statistical errors are larger than the above trends by approximately $\sqrt{2}$ for angle bins with $\eta_{\Delta} = 0$, $\phi_{\Delta} = 0$ and $\pm \pi$ because of reflection symmetries. An additional overall $\sqrt{2}$ increase applies to the two most-central centrality bins which split the top 10\% of the total cross section.

Although the principal features of the correlations presented and discussed in the remainder of this article are consistent with those  reported in Ref.~\cite{axialci} the details are much clearer. 
The centrality dependence is accurately determined over the full range from N-N to $b = 0$ \mbox{Au-Au}, and the collision-energy dependence is measured for the first time. 
The most-peripheral Au-Au centrality bin corresponds approximately to minimum-bias N-N ($\sim${\it \mbox{p-p}}) collisions. The null hypothesis that A-A collisions are Glauber linear superpositions of N-N collisions is clearly valid for the more-peripheral Au-Au collisions, but strongly falsified for more-central collisions.

\begin{figure*}[t]
\includegraphics[width=.24\textwidth]{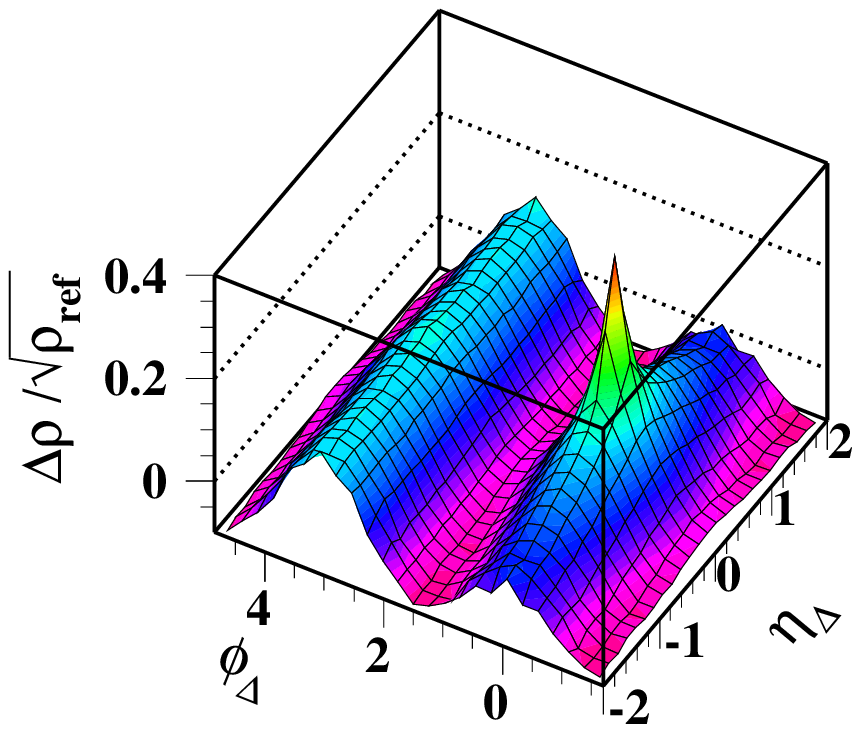}
\put(-100,90){\bf (a)}
\includegraphics[width=.24\textwidth]{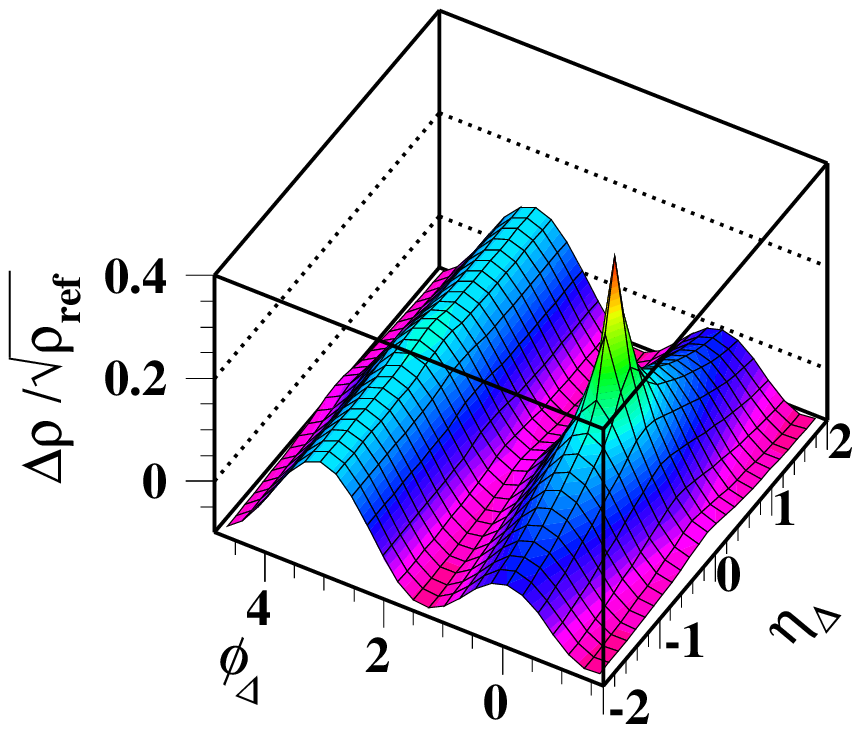}
\put(-100,90){\bf (b)}
\includegraphics[width=.24\textwidth]{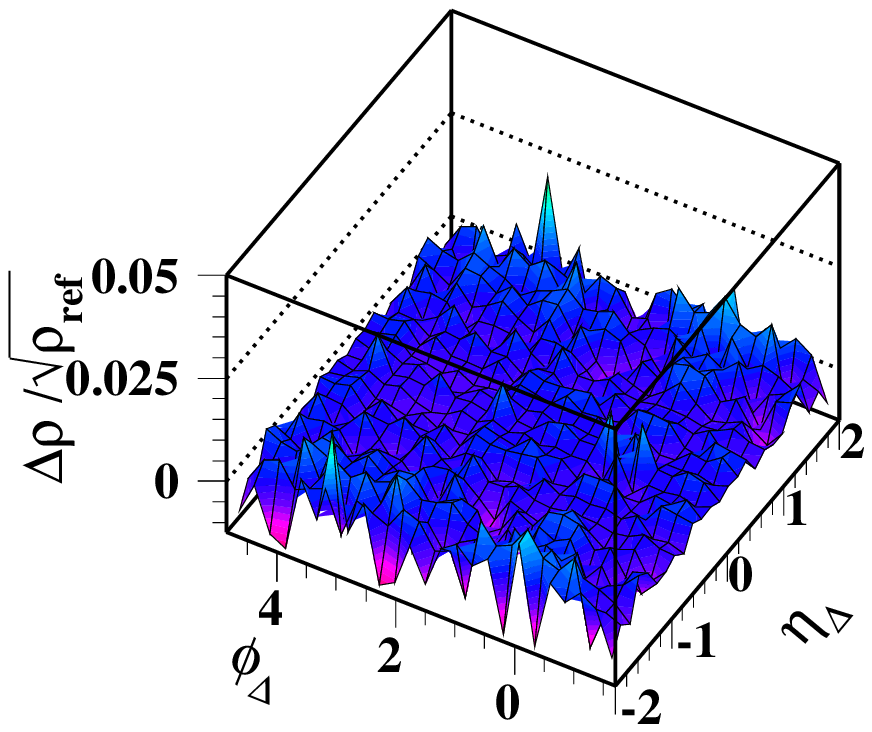}
\put(-100,90){\bf (c)}
\includegraphics[width=.24\textwidth]{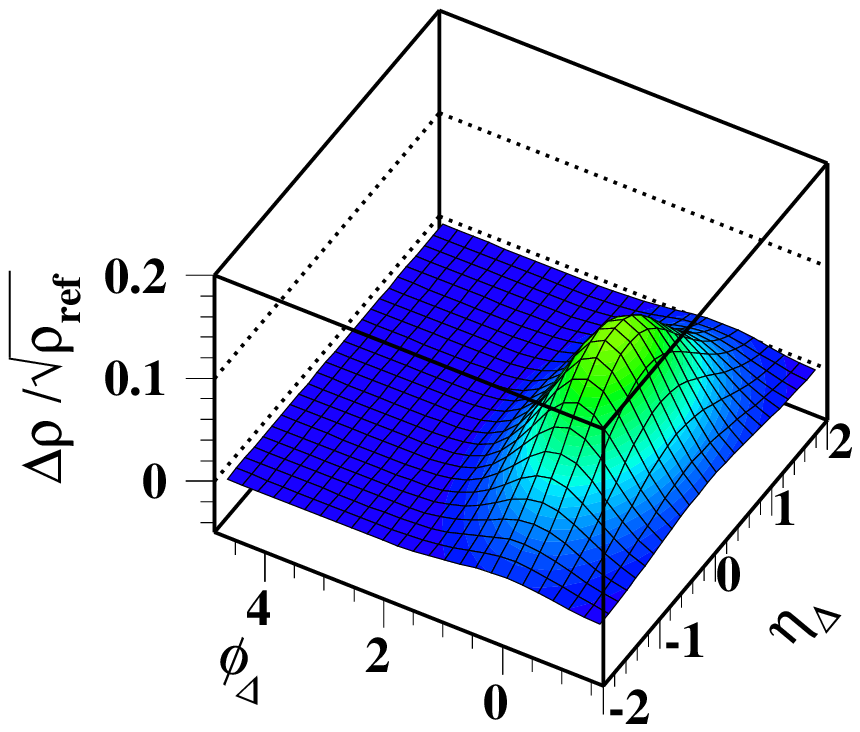}
\put(-100,90){\bf (d)} \\
\includegraphics[width=.24\textwidth]{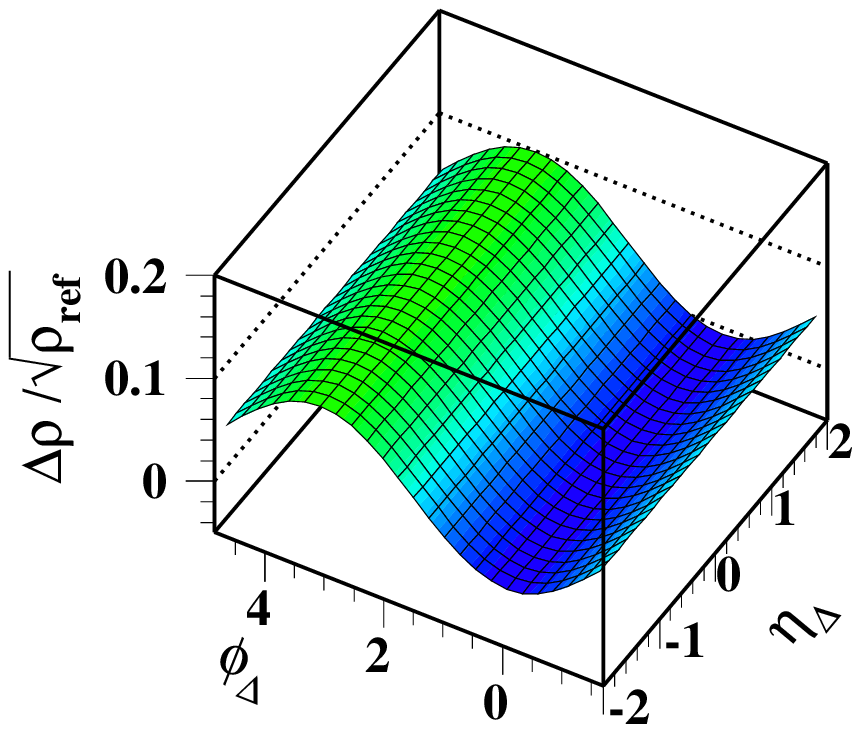}
\put(-100,90){\bf (e)}
\includegraphics[width=.24\textwidth]{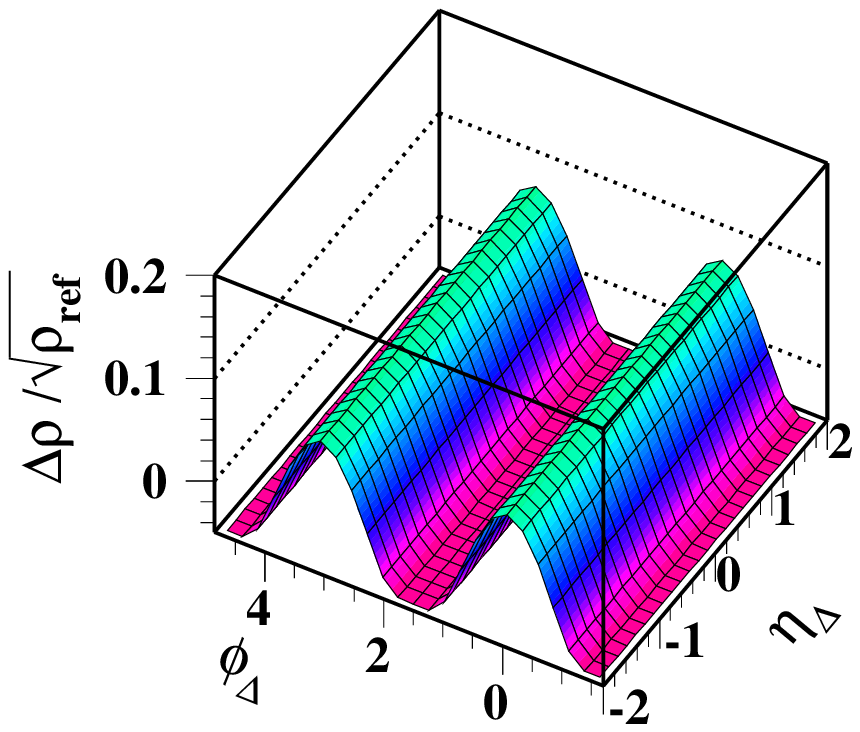}
\put(-100,90){\bf (f)}
\includegraphics[width=.24\textwidth]{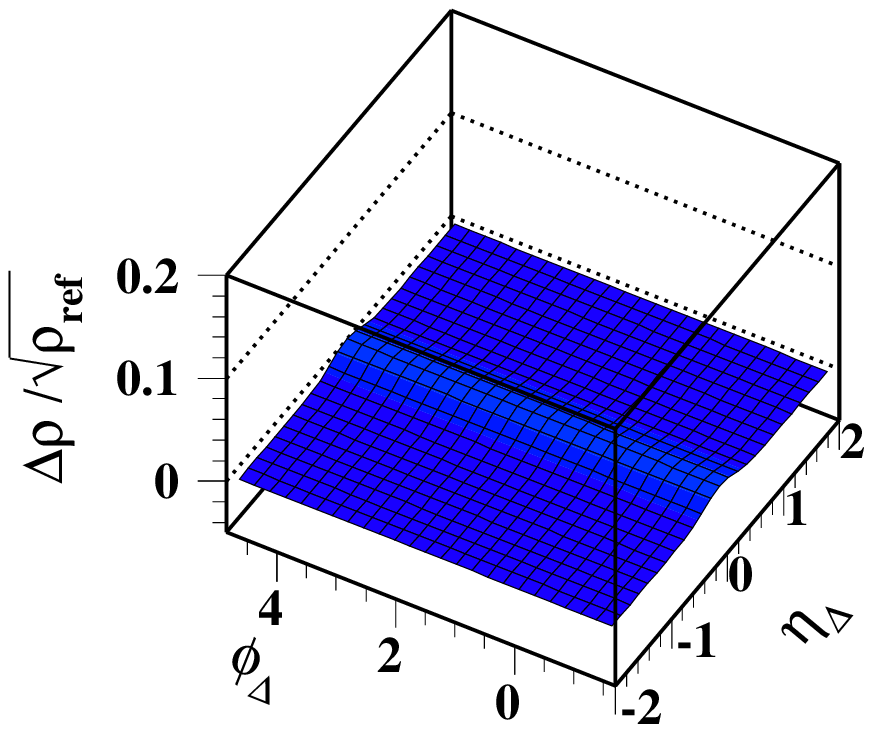}
\put(-100,90){\bf (g)}
\includegraphics[width=.24\textwidth]{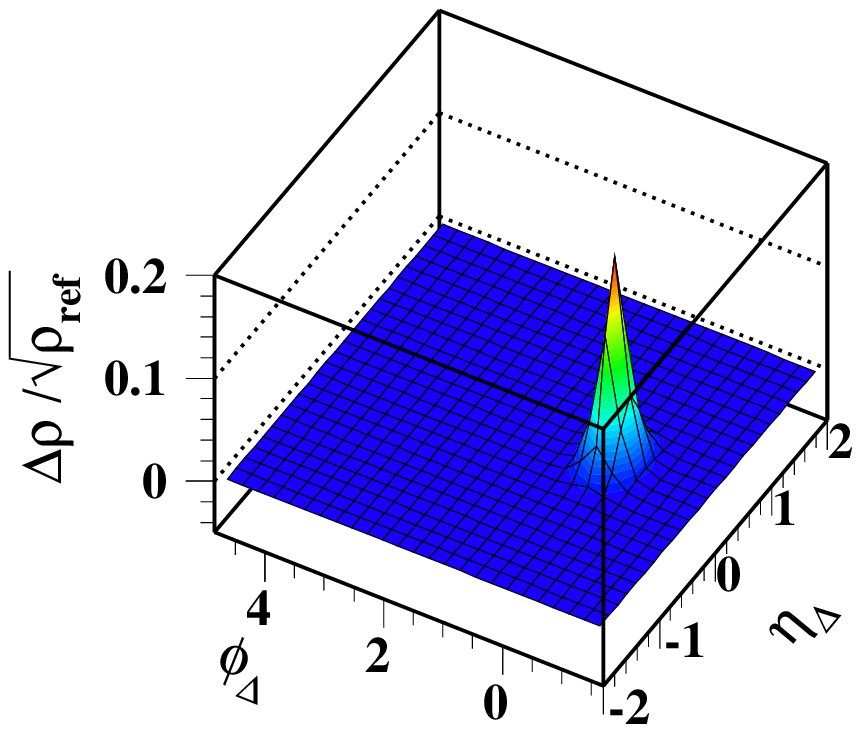}
\put(-100,90){\bf (h)}
\caption{\label{Figure2} 
(Color online) Fit decomposition of the 46-56\% centrality data for 62~GeV Au-Au collisions. 
The upper panels show from left to right the corrected data, model fit, fit residuals (data $-$ model) and same-side 2D Gaussian.  
The lower panels similarly show the away-side azimuth dipole,  the nonjet azimuth quadrupole, the 1D $\eta_\Delta$ Gaussian and the 2D exponential. This centrality is just below the sharp transition at $\nu_{trans} = 3$. Fit residuals (c) are scaled up eight-fold relative to the data.} 

\end{figure*}

Principal histogram features include (in the same order as Fig.~\ref{Figure2} panels after the fit residuals): 
(i) a same-side ($|\phi_{\Delta}| < \pi/2$)  2D  peak (approximately Gaussian) centered at $(\eta_{\Delta},\phi_{\Delta}) = (0,0)$ which increases in amplitude, narrows on $\phi_\Delta$ and dramatically broadens on $\eta_{\Delta}$ with increasing centrality;
(ii) an $\eta_{\Delta}$-independent away-side ($|\phi_{\Delta}| > \pi/2$) dipole (ridge) clearly apparent for the most peripheral bin and strongly increasing in amplitude with increasing centrality;
(iii) an $\eta_{\Delta}$-independent $\cos(2\, \phi_{\Delta})$ azimuth quadrupole
with maximum amplitude for mid-central collisions. The quadrupole feature has been conventionally identified with elliptic flow.
(iv) an approximately $\phi_{\Delta}$-independent 1D peak on $\eta_{\Delta}$ (approximately Gaussian) centered at $\eta_\Delta = 0$ (observed along the front edge of some panels and/or superposed on the away-side ridge), diminishing in amplitude to zero with increasing centrality; 
and 
(v) a narrow 2D  peak (approximately exponential) at $(\eta_{\Delta},\phi_{\Delta}) = (0,0)$ (due mainly to conversion
electrons and quantum correlations or HBT).

The features observed in peripheral 200 GeV \mbox{Au-Au} collisions agree well with those reported previously for 200 GeV {\it p-p} collisions~\cite{jeffpp1,aspect}. Based on systematic studies of two-particle angular and transverse-momentum correlations for {\it p-p} collisions~\cite{minijet,jeffpp1,aspect,starjets} we conclude that the same-side 2D peak [excluding the sharp spike at (0,0)] and away-side ridge represent semihard parton scattering and fragmentation (minijets). The visual features and fitting model components are discussed in the following section.

\section{Model Function and 2D Fits to Data}
\label{Sec:Model}

2D histograms have significant advantages over 1D projections and nongraphical numerical methods (e.g. some $v_2$ analysis).  Multiparameter fits to 2D histograms are generally less ambiguous than fits to their 1D projections because covariances among fit parameters are reduced by the additional information in the 2D histograms.

\subsection{2D model function} \label{2dmodel}

2D angular correlation histograms from Au-Au collisions for 22 energy and centrality combinations were fitted with a six-component model function. The  Au-Au model was adopted from one developed during analysis of 200 GeV {\it p-p} collisions~\cite{jeffpp1,aspect}. The fit model for {\it p-p} collisions was motivated by the simple geometrical forms apparent in the correlation data, not by an {\em a priori} physical model. A $\cos(2\,\phi_{\Delta})$ azimuth quadrupole component was added to the {\it p-p} model to describe the \mbox{Au-Au} data.

The model function applied to Au-Au correlation histograms includes (in the same order as panels in Fig.~\ref{Figure2} after the fit residuals)  (a) a same-side (SS) 2D Gaussian on $(\eta_{\Delta},\phi_{\Delta})$,  (b) an $\eta_{\Delta}$-independent away-side (AS) azimuth dipole $\cos(\phi_\Delta - \pi)$, (c) an  $\eta_{\Delta}$-independent azimuth quadrupole $\cos(2\, \phi_\Delta)$, (d) a $\phi_\Delta$-independent 1D Gaussian on $\eta_{\Delta}$, (e) a narrow SS 2D exponential on $(\eta_{\Delta},\phi_\Delta)$ and (f) a constant offset.  
The combined six-component model function in that order is
%
\bea \label{Eq4}
F   & = & 
A_1 \, \exp\left\{- \frac{1}{2} \left[ \left( \frac{\phi_{\Delta}}{ \sigma_{\phi_{\Delta}}} \right)^2  + \left( \frac{\eta_{\Delta}}{ \sigma_{\eta_{\Delta}}} \right)^2 \right] \right\} 
\\ \nonumber
&+& A_{\rm D}\, \cos(\phi_\Delta - \pi) \\ \nonumber
&+& A_{\rm Q}\, \cos(2\, \phi_\Delta)
+A_0\, \exp\left\{-\frac{1}{2} \left( \frac{\eta_{\Delta}}{ \sigma_{0}} \right)^2 \right\}  \\ \nonumber
&+&  
 A_2 \, \exp\left\{- \left[ \left( \frac{\phi_{\Delta}}{w_{\phi_{\Delta}}} \right)^2  + \left(\frac{\eta_{\Delta}}{ w_{\eta_{\Delta}}} \right)^2 \right]^{1/2} \right\} + A_3.
\eea

Given that mathematical description of the data the model elements can be interpreted physically. Terms (a) and (b) taken together are interpreted as a minijet contribution based on arguments in App.~\ref{minijets}, at least in peripheral \mbox{Au-Au} collisions. Term (c) is conventionally identified with elliptic flow~\cite{tomv2method1,v2cumul}. Term (d) is associated with participant-nucleon fragmentation (local charge conservation results in unlike-sign charged hadron pairs appearing nearby on $\eta$~\cite{jeffpp1,lund}). Term (e) models quantum correlations (HBT) and conversion-electron pairs.

The away-side ridge, attributed to $p_t$ conservation (e.g.\ back-to-back jets), can be modeled either by an AS azimuth dipole (better for low-$p_t$ fragments from minimum-bias or small-$Q$ partons) or by a 1D Gaussian at $\phi_\Delta = \pi$ with image peak at  $\phi_\Delta = -\pi$ (better for higher-$p_t$ fragments from more-energetic partons). With decreasing parton energy and increasing peak width the AS Gaussian periodic array approaches an AS azimuth dipole as a limiting case~\cite{tzyam}. The AS dipole then provides a more efficient description of the AS ridge. The effect of the AS ridge model choice on other fit parameters is included in the systematic uncertainties discussed in Sec. VII.

$\chi^2$ fits to the data were conducted by averaging the combined model function over a $5 \times 5$ grid within each $(\eta_\Delta,\phi_\Delta)$ bin rather than using function values at bin midpoints. The averaging technique becomes important in regions where the model function has large curvatures. In particular it affects the relation between the 2D exponential and 2D Gaussian near the angular origin.

Figure~\ref{Figure2} shows an example of fit decomposition and residuals using the 62~GeV 46-56\% corrected centrality bin (nominal 50-60\% bin). Similar results are obtained for each centrality bin and energy.  The upper panels show (left to right) data, model fit, residuals (data$-$model) and SS 2D Gaussian. 
The lower panels show the AS azimuth dipole $\cos(\phi_{\Delta}-\pi)$, azimuth quadrupole $\cos(2\phi_{\Delta})$, 1D Gaussian on $\eta_{\Delta}$ and 2D exponential. 
For this centrality, and for all other data except a few more-central bins, the residuals are comparable in magnitude to statistical errors and are negligible compared to the amplitudes of the principal correlation structures. 
In this example the 1D  Gaussian on $\eta_\Delta$ (g) describes a small artifact in 2D correlations at $\eta_\Delta = 0$

Absence of 
physically-significant structure in the fit residuals indicates that the 2D fit model of Eq.~(\ref{Eq4}) exhausts all statistical information in these data. The data do not require additional model components. 
%
Fit residuals for a few more-central bins at both energies include a small-amplitude non-statistical structure (AS dipole modulation on $\eta_\Delta$) discussed in Sec.~\ref{extracomp}.  
For minimum-bias ($p_t$-integral) angular correlations the SS 2D peak is well-described by a single 2D Gaussian. There is no systematically significant evidence for a separate non-Gaussian ``ridge'' in the SS 2D peak structure for angular correlations integrated over $p_t > 0.15$~GeV/c. Discussion of possible additional data structure and model components (e.g., $v_3$) is presented in Sec.~\ref{v3} and App.~\ref{multi}.

\subsection{Model-fit results}

Best-fit descriptions of data were based on a $\chi^2$ minimization procedure.  
For most centralities any 
substantial
excess contribution to total $\chi^2$ was confined to the acceptance edge $|\eta_{\Delta}| > 1.5$. Excluding those bins from the fitting procedure had a negligible effect on the best-fit model parameter values.
The resulting model parameters are presented in App.~\ref{taberrors} (Tables~\ref{TableI} and \ref{TableII}).
The columns of Tables~\ref{TableI} and \ref{TableII} correspond to the eleven centrality classes. The first eleven rows in both tables present the fit parameters from Eq.~(\ref{Eq4}) plus the statistical (fitting) and systematic uncertainties. The remaining rows report centrality and other derived parameters.
Centrality is measured by participant path length $\nu$ from a 200~GeV Au-Au Monte Carlo Glauber model used as a common geometry parameter for both energies. Most of the model fit parameters exhibit strong variations with centrality.  


The error matrix for the fit parameters revealed statistically significant covariances among some of the parameters, for example among the dipole, quadrupole and SS 2D Gaussian amplitudes for the more-central histograms. In order to account for covariances the corresponding statistical uncertainties were estimated by an iterative procedure. A given parameter was displaced from its optimum $\chi^2$ fit value, the other ten parameters were adjusted to minimize $\chi^2$, the selected parameter was further displaced and the data refit until the total $\chi^2$ increased by 1.  The reported uncertainties thus reflect covariances among the parameters. The incremental uncertainties (r.m.s.\ variances and covariances) are all observed to be small compared to the magnitudes of the parameters. Fitting errors for the model parameters in Eq.~(\ref{Eq4}) are listed in App.~\ref{taberrors}.


The fit model in Eq.~(\ref{Eq4}) includes non-orthogonal components which could lead to ambiguities in the best-fit solutions (e.g., multiple {\em local} $\chi^2$ minima). The possibility of ambiguities was studied in detail. Ambiguities were eliminated by conducting many independent $\chi^2$ fits assuming thousands of initial-value combinations for the 11 model parameters to locate the best-fit {\em global} minimum. 





For the more-central data at both energies a continuous fitting ambiguity developed when the 1D $\eta$ Gaussian amplitude was allowed to become negative. The concave-upward shape in the away-side ridge for the more-central data (see Fig.~\ref{Figure1} -- right-most panels) pulled the 1D Gaussian amplitude negative and forced the width to become large. In combination with the $\eta_\Delta$-broadened SS 2D Gaussian that lead to a 
continuous fitting instability: The offset, dipole, quadrupole, 2D Gaussian amplitude and 1D Gaussian amplitude and width could simultaneously co-vary over substantial intervals without significantly reducing the residuals (always less than 5\% of the SS 2D peak amplitude). 

The source of the ambiguity was identified as a statistically-significant residual structure not described by Eq.~(\ref{Eq4}). The ambiguity could be prevented by placing a (lower or upper) bound on the value of any one of the affected model components. Since the $m = 1$, 2 sinusoids and SS 2D peak parameters are the main focus of this paper we chose to remove the instability by requiring the 1D $\eta$ Gaussian to be non-negative. The impact of the imposed lower bound on the best-fit parameters was included in estimation of the systematic uncertainties. See Sec.~\ref{extracomp} for further discussion of the excess residuals.

\section{Anomalous centrality evolution}
\label{Sec:Anomalous}

\begin{figure*}[t]  
\includegraphics[width=.3\textwidth,height=.294\textwidth]{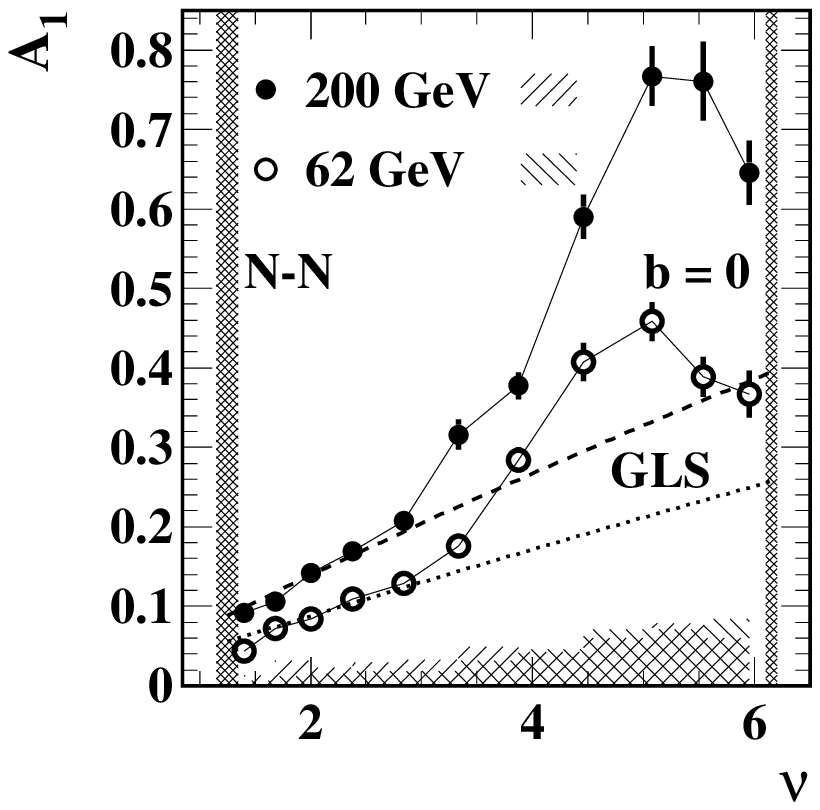} 
\includegraphics[width=.3\textwidth,height=.3\textwidth]{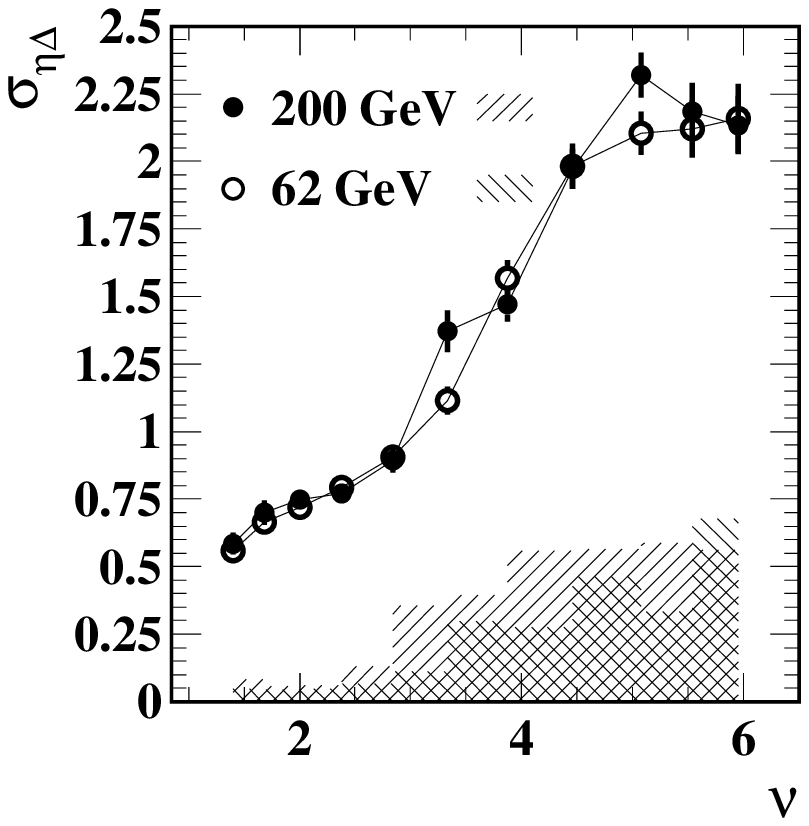}   
\includegraphics[width=.3\textwidth,height=.3\textwidth]{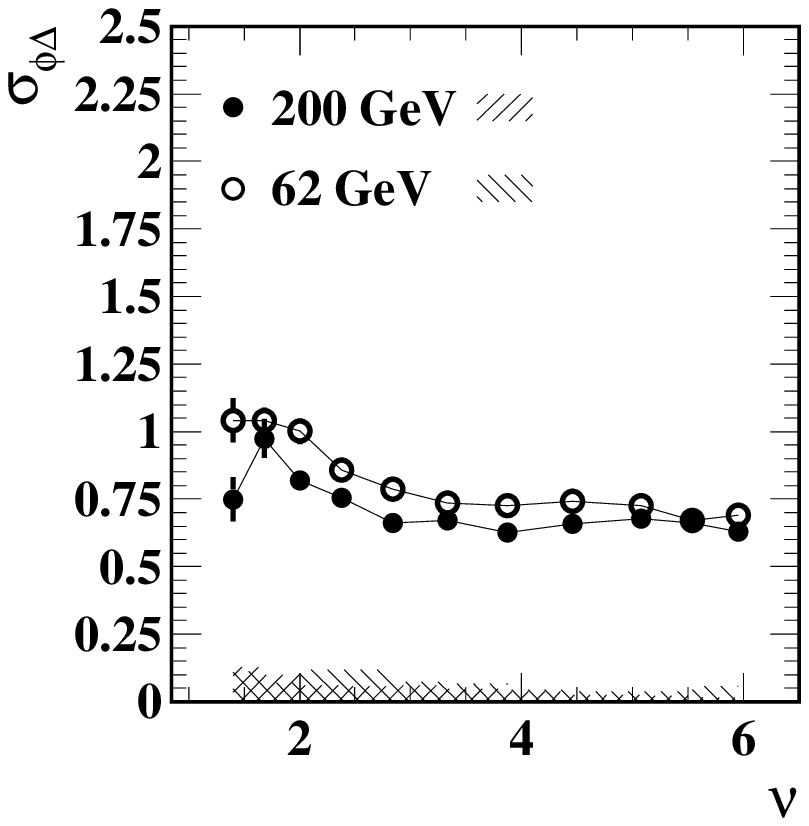} 
\includegraphics[width=.3\textwidth,height=.3\textwidth]{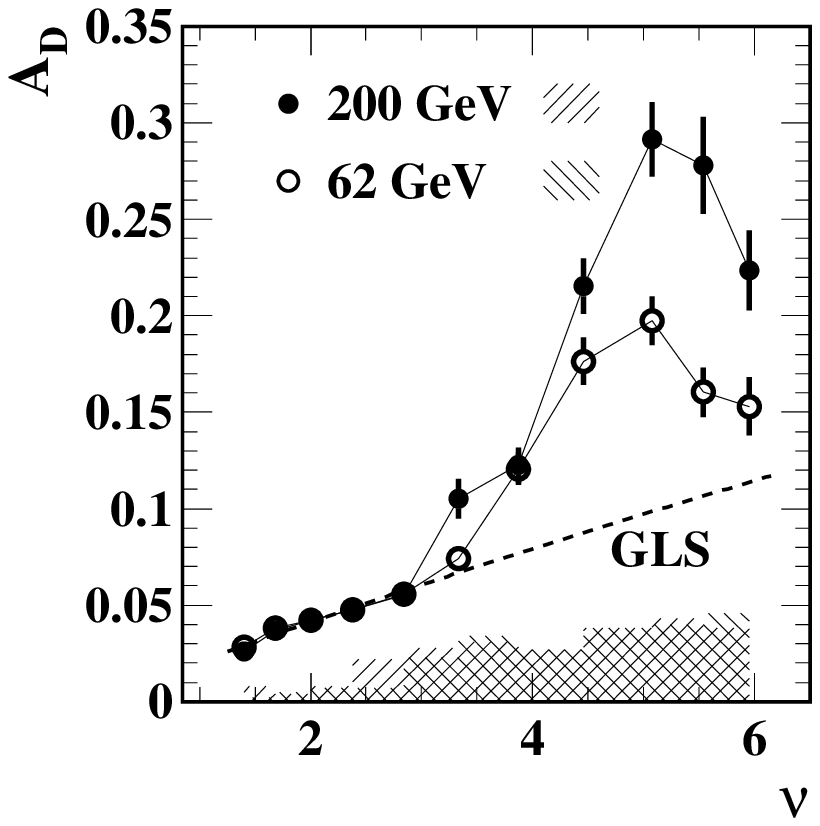}
\includegraphics[width=.3\textwidth,height=.3\textwidth]{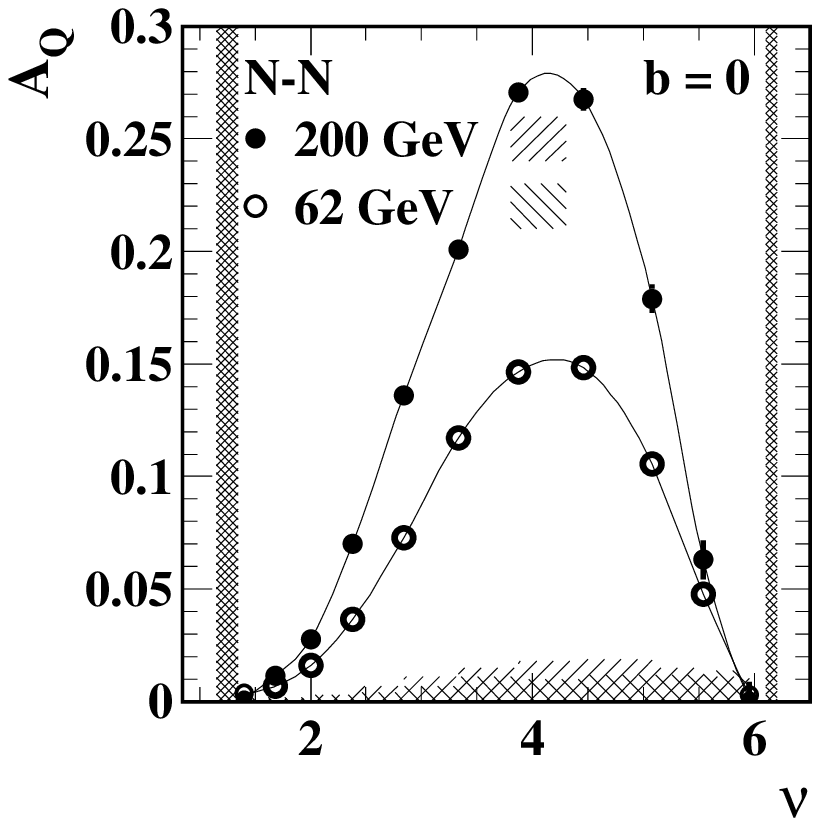} 
\includegraphics[width=.3\textwidth,height=.293\textwidth]{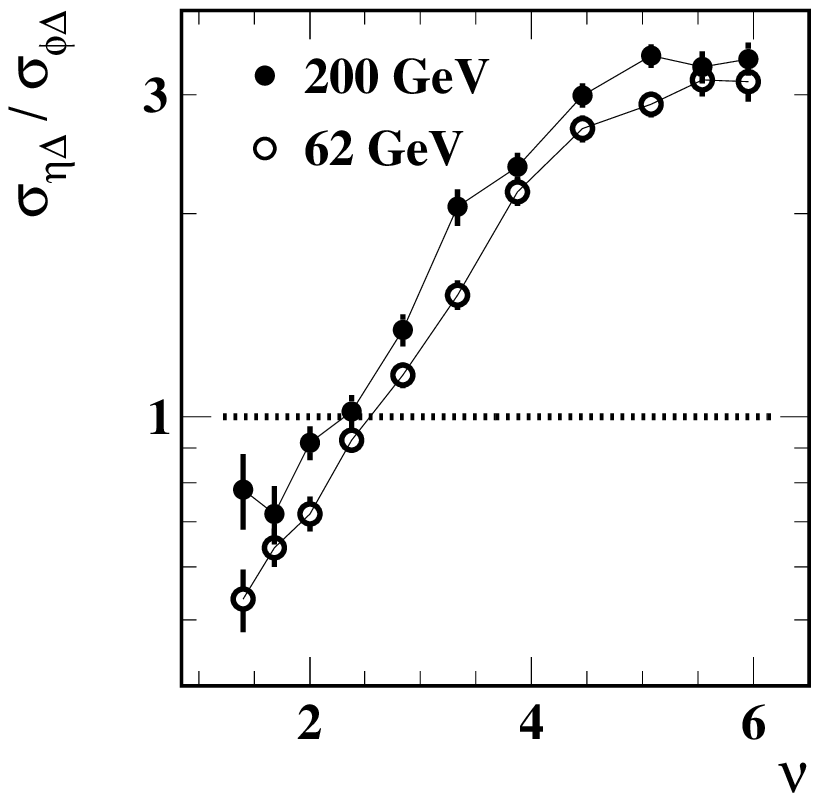}
\put(-127.3,106){\bf 2}
\caption{\label{Figure3}
Fit parameters for $(\eta_{\Delta},\phi_{\Delta})$ correlation data from Au-Au collisions at $\sqrt{s_{\rm NN}}= 62$ (open symbols) and 200~GeV (solid symbols) versus centrality measure $\nu$ computed at fixed energy (200~GeV). The same-side 2D Gaussian amplitudes, $\eta_{\Delta}$ widths, and $\phi_{\Delta}$ widths are shown in the left, center and right panels respectively of the upper row.  The lower row shows from left to right the amplitudes for the dipole, quadrupole, and same-side peak width aspect ratio $\sigma_{\eta_\Delta}/\sigma_{\phi_\Delta}$.
Fitting errors are indicated by error bars where larger than the symbols. Solid lines connect the points for clarity. The dotted and dashed curves indicate Glauber linear superposition estimates for 62 and 200~GeV peak amplitudes respectively, as discussed in the text. 
The quadrupole data are consistent with Ref.~\cite{quadpaper}.
 The hatched regions indicate the full range of systematic uncertainties listed in App.~\ref{taberrors}. The vertical dark bands indicate estimated $\nu$ equivalents for N-N collisions and $b = 0$ Au-Au collisions.
}
\end{figure*}

Figure~\ref{Figure3} shows the centrality and energy dependence of physically-relevant fit parameters reported in Tables~\ref{TableI} and \ref{TableII}. Two important trends emerge: (i) strong centrality variation tightly correlated between the SS 2D peak and AS dipole amplitudes and (ii) smooth variation of the azimuth quadrupole amplitude. In  this analysis we hypothesize that trend (i) is related to semihard parton scattering (minijets for more-peripheral collisions). The azimuth quadrupole [trend (ii)] is conventionally associated with elliptic flow. Comparisons with previous Au-Au 130~GeV results are discussed in App.~\ref{130}.

The term ``anomalous'' in the section title refers  to two aspects of centrality evolution: 
(a) the large increase in slope of centrality trends for the SS 2D peak amplitude, its width on $\eta_\Delta$ (represented by model parameters $A_1$ and $\sigma_{\eta_\Delta}$)  and AS 1D peak amplitude  $A_{\rm D}$ by factors 3.5, 5 and 3.5 respectively within one centrality bin  ({\em sharp transition}) and 
(b)  the large amplitude increase (up to twice the GLS trend) and significant azimuth width {\em decrease} of the SS 2D peak with increasing centrality, both trends contradicting conventional expectations for jet quenching in a strongly-coupled dense medium. Anomalous centrality evolution is discussed further in Sec.~\ref{anomdisc}.

\subsection{Centrality and energy trends} \label{centenergy}


With increasing centrality the SS 2D peak exhibits (a) a pronounced increase in the slope of the amplitude trend (i.e., a transition in the parameter trend with centrality) at {\em transition point}  $\nu_\text{trans} = 3.1 \pm 0.3$ (including statistical errors and bin-to-bin correlated and uncorrelated systematic uncertainties) accurately mirrored by the amplitude trend of the AS dipole, (b) a similar increase in the slope of the $\eta_\Delta$ width at the same transition points $\nu_{trans}$, and (c) a $\phi_\Delta$ width {\em decrease}. There is no significant difference in $\nu_{trans}$ for the two collision energies. $\sigma_{\phi_{\Delta}}$ for more-central collisions approaches a fixed value $\approx 0.7$.
Above the transition point the SS 2D peak and AS dipole amplitude trends for both energies increase uniformly on centrality to $\nu \approx 5$, beyond which they decrease. The correlated-pair yield decreases above $\nu = 5$ are intriguing, but are also comparable to the systematic uncertainties presented in App.~\ref{taberrors}. 

The SS 2D peak is actually strongly elongated on {\em azimuth} ($\sigma_{\phi_{\Delta}}$:$\sigma_{\eta_{\Delta}}$ = 2:1) in peripheral collisions. But with increasing centrality the angular asymmetry reverses and the 2D Gaussian becomes three times broader on \deta{} than on $\phi_{\Delta}$. The smooth shape evolution is shown by the aspect ratio plotted in Fig.~\ref{Figure3} (bottom-right panel).
It is notable that the SS 2D peaks for 55-64\% 200 GeV and 56-65\% 62 GeV histograms in Fig.~\ref{Figure1} have unit aspect ratio (equal r.m.s.\ widths on $\eta_\Delta$ and $\phi_\Delta$), but the peaks appear to be elongated on $\eta_\Delta$ because the histograms as plotted have an aspect ratio of $2\pi$:$4 \approx 3$:$2$. The SS 2D peak widths are further discussed in Sec.~\ref{anom}.

In contrast to the sharp transition in same-side 2D peak properties, the azimuth quadrupole amplitude varies smoothly with centrality, with no manifestation of the transition behavior observed in the SS 2D peak trends. The quadrupole amplitude depends only on {\em geometric} path length $\nu$ (estimated by $\nu_\text{200}$), with functional form independent of collision energy~\cite{quadpaper}. 

The energy dependence of the SS 2D peak in number angular correlations can be compared with that of the azimuth quadrupole and the SS peak in previously-measured $p_t$ angular correlations~\cite{ptedep}. In Ref.~\cite{quadpaper} an inferred energy factor of the form $\ln(\sqrt{s_{\rm NN}} / 13.5~ \text{GeV})$ was found to describe $v_2$ data measured by $A_{\rm Q} \equiv 2\rho_0(b)\, v_2^2\{{\rm 2D}\}(b)$ (defining 2D fit parameter $v_2\{{\rm 2D}\}$) above 17 GeV, where $\rho_0(b) =  dN_{\rm ch}/2\pi d\eta$ is the single-particle 2D angular density. The quadrupole amplitudes obtained in this analysis agree with those from Ref.~\cite{quadpaper}.  

In the present analysis we observe that the same-side 2D peak amplitudes for two energies and central \mbox{Au-Au} collisions are in the ratio $A_1(62) / A_1(200) = 0.57 \pm 0.06$~(stat.) which can be compared with the energy-factor ratio
$\ln(62.4 / 13.5~ \text{GeV}) / \ln(200 / 13.5~ \text{GeV}) = 0.57$. The energy dependence of the SS peak amplitude in number correlations is also consistent (within systematic uncertainties) with a $\ln(\sqrt{s_{\rm NN}} / 10~ \text{GeV})$ energy dependence of the SS peak amplitude inferred from $p_t$ angular correlations in Ref.~\cite{ptedep}. Given the uncertainties in the lower-energy SPS $p_t$ correlation measurements 10 GeV can be interpreted as a lower limit on the intercept consistent with 13.5 GeV from Ref.~\cite{quadpaper}. Thus, the SS 2D peak and azimuth quadrupole collision-energy trends agree above $17$ GeV and depend only on $\ln(\sqrt{s_{\rm NN}})$. The latter dependence is consistent with QCD processes.

The 1D peak on $\eta_{\Delta}$, interpreted to arise from participant-nucleon fragmentation~\cite{Whitmore,lund}, is small compared to the SS 2D  peak and falls monotonically to zero by mid-centrality ($\nu \sim 3.5$).




Centrality parameter $\nu = 2 N_{bin} / N_{part}$ is smoothly (not discontinuously) related to the fractional cross section $\sigma(b) / \sigma_{\rm tot}$ and to participant number $N_{\rm part}$. The sharp transition in SS 2D peak properties near $\nu = 3$ appears as well when fit parameters are plotted on other centrality measures. In the case of $N_{\rm part}$ the transition shifts to the extreme left end of the parameter range and is therefore visually obscured. Parameter $\nu$ presents the essential linear-superposition reference in a simple form: proportionality of the reference to binary-collision number $N_{\rm bin}$ as discussed in the next subsection.


\subsection{Testing the linear-superposition hypothesis}

Accurate measurement of centrality trends for \mbox{Au-Au} angular correlations down to the N-N limiting case makes possible a rigorous comparison of Au-Au correlations to N-N binary-collision scaling---the Glauber linear-superposition reference (baseline). In the GLS reference model of Au-Au collisions the SS 2D peak amplitudes (and volumes) from minimum-bias {\it p-p} ($\sim$N-N) collisions are linearly superposed (summed) at the angular-difference origin ($\eta_\Delta = \phi_\Delta = 0$) proportional to the Glauber-model number of N-N binary collisions $N_{\rm bin}$. 
In the GLS hypothesis the SS 2D peak \deta\, and \dphi\, widths retain fixed values characteristic of {\it p-p} collisions.


For per-particle measure $\Delta \rho / \sqrt{\rho_\text{\rm ref}}$ in Eq.~(\ref{Eq3}) binary-collision scaling of the SS 2D peak amplitude and volume translates to scaling as $N_{\rm bin}/N_{\rm ch}$. If $X_{pp}$ represents a correlation peak amplitude or volume in {\it p-p} ($\sim$N-N) collisions the GLS variation with A-A centrality should be
\bea
\label{Eq5}
X_{\rm AA}(\nu) \hspace{-.03in} & = & \hspace{-.03in} X_{pp} \frac{N_{\rm bin}\, N_{{\rm ch},pp}}{N_{\rm ch,AA}} =  X_{pp} \frac{\nu\, N_{{\rm ch},pp}}{(2/N_{\rm part}) N_{\rm ch,AA}} \nonumber \\
& \approx & X_{pp} \frac{\nu}{1 + x(\nu - 1)},
\eea
where the second line assumes the two-component hadron production model of Kharzeev and Nardi (K-N)~\cite{kn}. Amplitude or volume $X_{pp}$ can be estimated by direct \mbox{{\it p-p}} measurements or by extrapolation to N-N from several peripheral \mbox{A-A} centralities. 

Parameter $x$, the coefficient of the binary-collision scaling component, is held fixed in the K-N two-component model. Assuming $x$ to be independent of centrality provides a reasonable description of experimental probability distributions on multiplicity~\cite{raycentral}. More-differential spectrum analysis suggests that the {\em effective} $x$ increases substantially from {\it p-p} to central Au-Au collisions~\cite{ppspectra,TomAuAuspectra,jetyield}. For this GLS reference $x$ is held fixed at 0.02, the {\it p-p} value for acceptance $\Delta \eta = 2$ \cite{ppspectra,fragevo}.


GLS references for the SS 2D and AS peak amplitudes are shown as the dotted and dashed curves in Fig.~\ref{Figure3} for 62 and 200~GeV data respectively. The amplitude data closely follow  the GLS reference with increasing centrality 
(within small systematic uncertainties)
until the {\em transition point} $\nu_\text{trans}$, beyond which the data substantially exceed the reference trends. Peak widths on both $\eta$ and $\phi$ show significant deviations from GLS constant values $\sigma_{\eta_\Delta} = 0.55$ and $\sigma_{\phi_\Delta} = 1.10$ corresponding to N-N ({\it p-p}) collisions. SS peak width trends are further discussed in Sec.~\ref{anom}.

The aspect ratio trend in the lower-right panel is particularly interesting. It confirms the large eccentricity of the same-side 2D peak observed previously in {\it p-p} collisions with substantial elongation  on $\phi$ ($\sigma_{\phi_\Delta}$:$\sigma_{\eta_\Delta}$ = 2:1)~\cite{aspect} and shows the strong evolution with Au-Au centrality to large elongation on $\eta$ ($\sigma_{\eta_\Delta}$:$\sigma_{\phi_\Delta}$ = 3:1). In {\it p-p} collisions the elongation on $\phi$ was found to vary strongly with particle $p_t$, with larger $\phi$ elongation for smaller particle $p_t$ down to 0.5 GeV/c for each particle~\cite{aspect}.


{\sc hijing}~\cite{hijing} predictions for the SS 2D peak amplitude~\cite{daugherity} from 200~GeV Au-Au collisions with jets implemented but no jet quenching deviate strongly in more-central \mbox{Au-Au} collisions from the GLS trend extrapolated from {\it p-p} data. A discussion of the discrepancy is presented in Sec.~\ref{hijsec}. The {\sc hijing} SS 2D peak widths on $(\eta_{\Delta},\phi_{\Delta})$ are respectively 0.75 and 0.9 (radians) and remain constant with centrality, in marked contrast to the large angular asymmetries and strong centrality dependence observed in the data.




\section{Systematic uncertainties}
\label{Sec:Errors}

Systematic uncertainties in the parameters of the fitting function in Eq.~(\ref{Eq4}) are primarily due to secondary particle and other contamination backgrounds in the data, uncorrected detector and event reconstruction effects, ambiguities in the choice of fitting model function, and statistically significant residual structures not accounted for by the fitting model.  The specific sources of uncertainty and the method of error estimation are discussed in the following subsections. Systematic uncertainties for the fitting parameters are listed in Tables~\ref{TableI} and \ref{TableII} in App.~\ref{taberrors}.


\subsection{Uncertainties in the histogrammed data}
\label{Sec:ErrorsA}

The largest source of systematic uncertainty is a 12\% non-primary particle contamination~\cite{starspec200,molnarthesis} 
with unknown correlation structure in the particle sample used for the analysis. This background is primarily from weak-decay daughters from the collision and secondary particles produced in the detector material which were misidentified as primary particles, {i.e.} those emitted directly from the triggered collision. $e^+$-$e^-$ pair contamination produced by photon conversions in the detector material are discussed in the following subsection. Correlation measure $\Delta\rho/\sqrt{\rho_{\rm ref}}$ was computed assuming particle DCA $<3$~cm (standard cut admitting a 12\% secondary contamination) and DCA $<1$~cm (reduced contamination fraction) and the resulting histograms were compared. Any difference in correlation structure should be dominated by secondary particles preferentially removed by the modified DCA cut. Differences were found to be dominated by statistical fluctuations. Any systematic structures were less than 3\% of the primary correlation amplitudes, resulting in a $\pm 3$\% uncertainty estimate assigned to the five amplitude parameters in the fit model.

Pileup contamination was corrected as described in App.~\ref{pileupcorr}. We observe that pileup mainly affects the 1D $\eta$ Gaussian amplitude and mainly near mid-centrality
\footnote{The average pileup event superposed on a triggered event has a substantial average multiplicity (tens of particles), promoting the triggered event to a greater (and incorrect) centrality. The result is maximum pileup distortion near mid-centrality.},
 causing the amplitude to vary non-monotonically with centrality $\nu$. Comparing the centrality dependences of parameter $A_0$ before and after pileup correction suggests that about $\pm 15$\% of the full pileup effect may remain in the Au-Au 62 GeV data (with larger initial pileup fraction) after correction. Residual pileup contamination in the 62 and 200 GeV correlations was therefore estimated separately in each centrality bin as $\pm 0.15 \Delta\rho/\sqrt{\rho_{\rm ref}}$(pileup) from Eq.~(\ref{Eq4a}).

Pair reconstruction inefficiencies~\cite{axialci} induce depletion of $\Delta\rho/\sqrt{\rho_{\rm ref}}$ at small opening angles, visible in more-central collisions as grooves in uncorrected $\Delta\rho/\sqrt{\rho_{\rm ref}}$ near $(\eta_\Delta$,$\phi_\Delta) = (0,0)$ for $|\eta_\Delta| < 0.08$ and $|\phi_\Delta| < 1$. Although corrections (pair cuts to both sibling and reference pairs) remove most of this effect, close examination of the 2D histograms suggests that small artifacts remain which are approximated as a 2D Gaussian with amplitude $\sim$ $-$0.025 and $-$0.04 for 62 and 200 GeV 0-5\% centrality data respectively, and with $\eta_\Delta$ and $\phi_\Delta$ widths 0.08 and 0.5. Estimates for the other centrality bins were obtained by scaling the above amplitudes by the particle pair density, $(d^2N_{\rm ch}/d\eta d\phi)^2$. 


Other systematic effects considered include: intermittent electronics outages, pseudorapidity acceptance dependence on longitudinal ($z$-axis) collision vertex position in the TPC, collision-vertex position inaccuracy due to reconstruction error, particle momentum resolution, TPC central-membrane particle-trajectory crossing inefficiency, and residual dependence on the event-mixing bin sizes for collision vertex position in the TPC and event multiplicity. The overall contribution to 2D angular correlations from those sources was found to be insignificant compared to the reported correlation structure.

The effect of the above uncertainties in 2D correlation histograms on the fit-model parameters was estimated by separately adding each of the above representations [e.g. $0.15\Delta\rho/\sqrt{\rho_{\rm ref}}$(pileup), and small Gaussians for errors due to two-particle inefficiency, electronics outages and pseudorapidity acceptance dependence] to the data, refitting the data, increasing the amplitude of the added function, refitting the data again, and so on until a linear trend exceeding statistical fluctuations could be determined.

Parameter uncertainties due to secondary backgrounds were assumed to be Gaussian distributed. Those due to residual pileup were assumed to be uniformly distributed, implying that the ``true'' parameter value lies between fitted values obtained by adding or subtracting $0.15\Delta\rho/\sqrt{\rho_{\rm ref}}$(pileup) to the 2D correlations with uniform probability. The other parameter uncertainties were assumed to be uni-directional with uniform probability~\footnote{
The expression means that the systematic effect tends to shift the 
fitted parameter in one direction, either up or down. The uncertainty due to the 
actual systematic effect is assumed to be uniformly distributed 
between the stated limits. 
}. Mean shifts and variances estimated in this subsection were added linearly to those discussed in the next subsection. 

\begin{figure*}[t]  
\includegraphics[width=.24\textwidth]{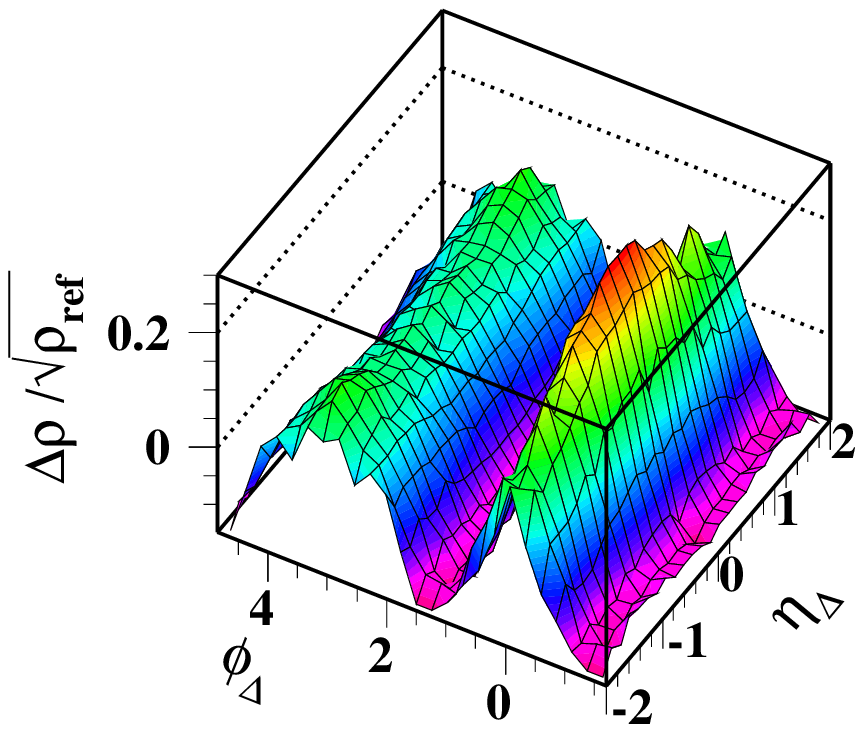} 
\put(-100,85) {\bf (a)}
\includegraphics[width=.24\textwidth]{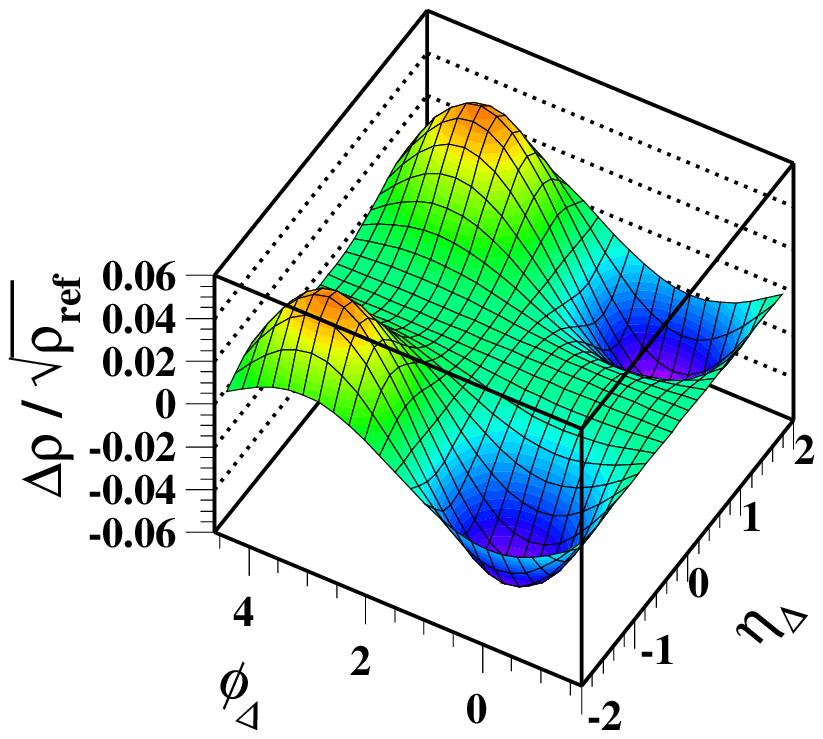}  
\put(-100,85) {\bf (b)}
\includegraphics[width=.24\textwidth]{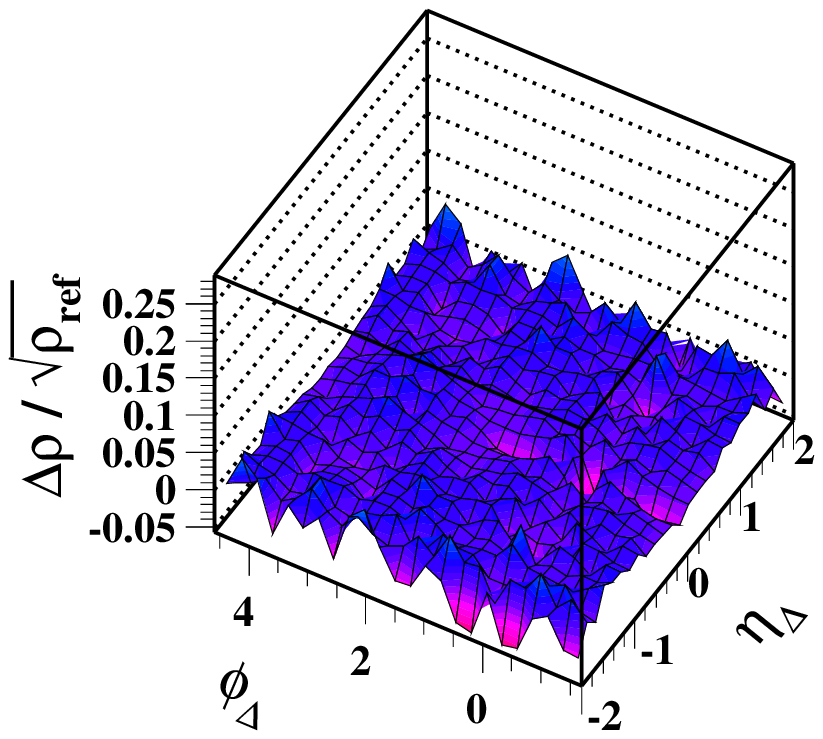} 
\put(-100,85) {\bf (c)}
\includegraphics[width=.24\textwidth]{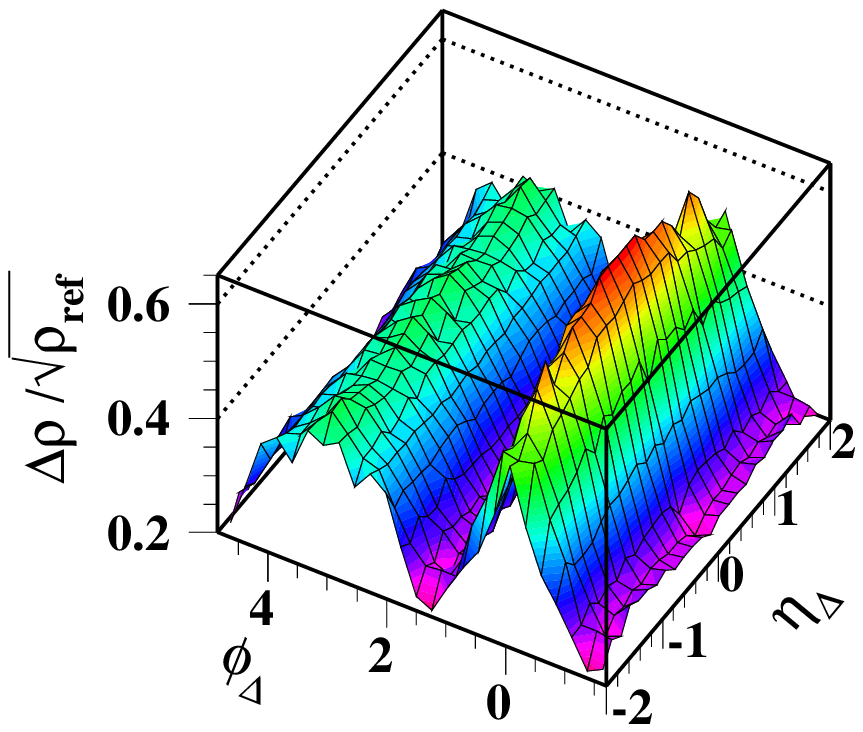}  
\put(-100,85) {\bf (d)}
\caption{\label{Figure4} (Color online)
(a) 2D data histogram from 0-5\% central 200 GeV Au-Au collisions [$(0,0)$ bin suppressed],
(b) Additional model component fitted to the 2D histogram and consisting of an AS azimuth dipole modulated by function $F(\eta_\Delta)$ described in the text, 
(c) Fit residuals including the new model component,
(d) Data histogram in the first panel minus the additional model component and fitted offset. Remaining structure is described accurately by an away-side dipole and same-side 2D Gaussian.
}  
\end{figure*}

The uncertainty in pre-factor $\sqrt{\rho^{\prime}_{\rm ref}}$ or $d^2\bar{N}_{\rm ch}/d\eta d\phi$ derived from spectrum analysis is $\pm 8$\% ($\pm 7$\%) for the 62 (200)~GeV data~\cite{molnarthesis,starspec200}. This overall normalization uncertainty is not included in the parameter uncertainties reported in Tables~\ref{TableI} and \ref{TableII} in App.~\ref{taberrors}.

%



\subsection{Uncertainties arising from the fitting model}

In Sec.~\ref{Sec:Model} we noted that the away-side ridge can be described by either an azimuth dipole or periodic array of 1D Gaussians with common width $\sigma_{\phi,{\rm AS}}$~\cite{tzyam}. In addition, for more-central collisions the away-side ridge amplitude is not constant on $\eta_\Delta$ but displays a concave-upward dependence, a feature not readily described by the fitting function adopted for this analysis. Also, similar model fits applied to 2D angular correlation data with $p_t$ cuts imposed suggest that the SS 2D peak for those data may be better described by a non-Gaussian function. 
Alternative model descriptions of the SS 2D peak for more-central collisions could include a SS 1D Gaussian ridge on azimuth (see Sec.~\ref{v3} and App.~\ref{multi}). 
We therefore estimate the extent to which the choice of fitting model function affects the accuracy of our description of the principal correlation structures.

To explore systematic uncertainties derived from the choice of model function the components in Eq.~(\ref{Eq4}) were modified. The data were refit and any changes in the parameters of the unmodified components in Eq.~(\ref{Eq4}) were recorded. The modifications included: a periodic series of 1D Gaussians replacing the AS azimuth dipole, 
additional $\eta_\Delta^2 \cos{\phi_\Delta}$ and $\eta_\Delta^2 \cos{2\phi_\Delta}$ terms (modeling alternative $\eta_\Delta$ dependence), 
modified SS 2D Gaussian [difference exponent $n$ allowed to deviate from 2 in e.g.\ $\exp\{-(x - \bar x)^n\}$], 
description of the SS 2D peak for more-central collisions as the sum of a 1D Gaussian on $\phi_\Delta$ plus an alternative 2D Gaussian,
modified 2D exponential (difference exponent allowed to deviate from 1), 
and similar exponent variation for the 1D $\eta$ Gaussian. Corresponding shifts in the parameters of unaltered components of Eq.~(\ref{Eq4}) for each modified fit model determined the uncertainties, assumed to be unidirectional with uniform probability. Variances were obtained from the parameter shifts.

The sharp spike at (0,0) apparent in Fig.~\ref{Figure2} is predominantly caused by $e^+$-$ e^-$ pairs produced by photon interactions in detector material which survive particle-identification (dE/dx)  
and primary-particle selection cuts. Quantum correlations also contribute near the angular origin, as shown by projecting HBT correlations onto $(\eta_\Delta , \phi_\Delta)$. Both background correlations are well described by the single 2D exponential in Eq.~(\ref{Eq4}). In addition to allowing the difference exponent to vary we also studied the impact of this component by removing the 2D exponential from the model and a few histogram bins near (0,0) from the fit and refitting the remaining data with a truncated 8-parameter model. The resulting changes in parameters were assumed to be uni-directional, with uniform probability distributions.


\subsection{Additional model element for central collisions} \label{extracomp}



2D histograms for the most-central Au-Au collisions exhibit a distinct $\eta_\Delta$ dependence in the AS dipole not observed in less-central Au-Au or {\it p-p} collisions, with a minimum at $\eta_\Delta = 0$. 
As an example, the data histogram for 0-5\% central 200 GeV Au-Au collisions is shown in Fig.~\ref{Figure4} (first panel). Attempts to model the visible AS minimum with a Gaussian negative on $\eta_\Delta$ and uniform on $\phi_\Delta$ were rejected because the fits were unstable and did not significantly reduce the residuals.

Detailed examination of the 2D residuals (data $-$ model) for the nominal fits in Sec.~\ref{Sec:Model} for the most-central collisions at both energies led to modification of the AS dipole model component shown in Fig.~\ref{Figure4} (second panel): adding an AS dipole times an $\eta_\Delta$-symmetric function with minimum value zero at the origin having the form $F(\eta_\Delta) =|\eta_\Delta|^m\, \exp\{-|\eta_\Delta / \sigma|^n/2\}$, with $m \sim 2$, $n\sim 5$ and $\sigma \sim 1.5$.  The fractional $\eta_\Delta$ modulation of the total AS dipole is about 15\% for 0-5\% central collisions and decreases to zero for mid-central collisions. The additional model component is obviously orthogonal to the azimuth quadrupole component and any other $v_m$ for $m > 2$. As with other 2D model elements the AS dipole modulation was introduced in response to observed data structure and is not motivated by a physical mechanism.

Residuals from a $\chi^2$ fit with the additional model component are shown in Fig.~\ref{Figure4} (third panel). Subtracting the AS dipole modulation term from the data histogram in the first panel leads to the fourth panel. What remains is a uniform AS dipole and an SS 2D peak well described by a 2D Gaussian. Those $0-5$\% data are consistent with zero quadrupole amplitude and with the corresponding entries in Table~\ref{TableI}.
This result suggests that in some cases {\em apparent} deviations of the SS 2D peak from an ideal  2D Gaussian shape may actually result from superposition of a small AS dipole modulation, with characteristic azimuth width ($\sim1.6$) much larger than the azimuth width of the SS 2D peak ($\sigma_{\phi_\Delta} \approx 0.65$). The SS 2D peak itself then does not deviate significantly from a 2D Gaussian.

The impact of this residual on parametrizations of the most central collisions at 62 and 200 GeV was estimated in two ways. First, the 1D Gaussian component in Eq.~(\ref{Eq4}) was replaced by the above residual model function $F(\eta_\Delta)\cos{(\phi_\Delta - \pi)}$, the data were re-fitted, and the shifts in the parameters of the remaining components recorded. Errors for the other centralities were estimated by scaling the preceding $0-5$\% errors by the approximate overall amplitudes of the residual in the centrality bin. Residuals above the transition centrality have shapes similar to the above model while those below the transition are dominated by statistical fluctuations.  Second, the residual histograms at each centrality were separately fitted with a dipole, quadrupole, SS 2D Gaussian and 1D Gaussian on $\eta_\Delta$. The resulting very small sinusoid amplitudes were included in the errors. The associated Gaussians were added to the nominal fitted Gaussians, and effective amplitudes and widths were computed from the volumes and second moments of the combined Gaussians. The differences between the effective amplitudes and widths and the nominal parameters were included in the systematic uncertainties. Each of the systematic uncertainties in this subsection was assumed to be uni-directional, with uniform probability distribution.

\subsection{Total systematic uncertainties}

The mean shifts (in Secs.~\ref{Sec:ErrorsA} and~\ref{extracomp}) and total variances of the systematic uncertainty contributions (15 in all) discussed in this section were summed for each of the eleven parameters in the model function, accounting for the symmetric or uni-directional nature of the uncertainty and its assumed probability distribution, Gaussian or uniform. If the mean shift exceeded the total systematic uncertainty r.m.s. value, then the error bar was extended to include the nominal fit value.

The principle sources of systematic uncertainty varied with centrality and parameter, but in general the secondary-particle contamination, the apparent $\eta_\Delta$ dependence of the AS dipole and the amplitude and shape (exponent) of the 2D exponential dominated. 
Systematic uncertainties due to some sources of systematic error tend to be correlated across parts of the centrality range. Total systematic uncertainties in adjacent centrality bins may therefore be partially correlated. 
The nominal fit parameter values and their statistical and systematic uncertainties are listed in Tables~\ref{TableI} and \ref{TableII} in App.~\ref{taberrors}. 


The total systematic uncertainties are also represented in Fig.~\ref{Figure3} by the hatched regions at the bottoms of the panels. The full range of uncertainty is represented. In most cases the uncertainties tend to be symmetric about the plotted values. The exception is 200 GeV $\sigma_{\eta_\Delta}$ where the uncertainties extend mainly  above the plotted values.

\subsection{Other possible correlation structures} \label{otherstruct}

Small variations near $|\eta_{\Delta}| = 2$ in the $\cos(n\phi_{\Delta})$ sinusoids result from residual effects of finite collision-vertex-position and event-multiplicity bin size. Random structures near $|\eta_{\Delta}| = 2$ reflect limited two-particle statistics near the $\eta$ acceptance boundary. These structures are found to have a negligible effect on fitted parameters.

Simulated angular correlation structures due to resonance decays (mainly $\rho^0$ and $\omega$ for the present $p_t$ acceptance) were found to contribute less than 10\% of the 2D same-side Gaussian peak within $|\eta_{\Delta}| < 0.5$ and $|\phi_{\Delta}| < 2$~\cite{axialci,mevsim} (and were negligible elsewhere). Such correlation structures were not observed in the fit residuals and thus were not included in the fitting model.

Global transverse-momentum conservation produces per-pair angular correlations measured by $\Delta \rho / \rho_{\rm ref}$ proportional to $\vec{p}_{t1} \cdot \vec{p}_{t2} / \bar N_{\rm ch} = (p_{t1}p_{t2} / \bar N_{\rm ch}) \, \cos(\phi_{\Delta})$~\cite{lisazibi} and is therefore included in the AS dipole amplitude. The magnitude of the corresponding per-particle dipole amplitude in $ \Delta \rho/ \sqrt{\rho_{\rm ref}}$ could be as large as 0.015 to 0.02 at either energy, but should be independent of centrality. Any global momentum-conservation contribution to the AS dipole should thus be relatively insignificant in more-central collisions. 
Additional energy- and momentum-conservation-induced correlations  (e.g.\ $p_{z1} \times  p_{z2}$~\cite{lisazibi}) would produce distinct $\eta_\Delta$ dependence (hyperbolic functions) in the 2D angular correlations which are evidently too small to be statistically significant in the residuals.


Any reduction of $v_2$ with increasing $|\eta|$ should produce a corresponding reduction in the quadrupole amplitude with $|\eta_{\Delta}|$. The systematic uncertainty estimated via the $\eta_{\Delta}^2 \cos{2\phi_\Delta}$ model component allows for this possible structure in the correlations. The $v_2(\eta)$ data~\cite{v2eta} for $|\eta| < 1$ do not require such a reduction, but would be consistent with reductions of a few percent. The present data do not require an $|\eta_{\Delta}|$-dependent quadrupole amplitude. Directed flow ($v_1$)~\cite{v1a,v1b} might contribute an $|\eta_{\Delta}|$ dependence to the dipole amplitude, but is estimated to be too small to observe in these data and is not apparent in the data or residuals. Higher-order azimuth sinusoid components were found to be negligible in the fit residuals projected onto $\phi_\Delta$ compared to the additional dipole component discussed in Sec.~\ref{extracomp}. Also, see Sec.~\ref{v3}.


It has been conjectured that a sextupole component referred to as $v_3$ may actually be present in angular correlation data. Alternatively, the SS 2D peak may include a SS 1D Gaussian component uniform on $\eta_\Delta$. Such structures are not observed in fit residuals from $p_t$-integral data using the standard model defined by Eq.~(\ref{Eq4}). The consequences of adding such model elements to the standard fit model are discussed in App.~\ref{multi}. In App.~\ref{multisys} related systematic changes in the fit results reported in Tables~\ref{TableI} and \ref{TableII} are estimated to be substantially smaller than the combined errors reported in those tables.

\begin{figure*}[t]  
\includegraphics[width=.3\textwidth,height=.293\textwidth]{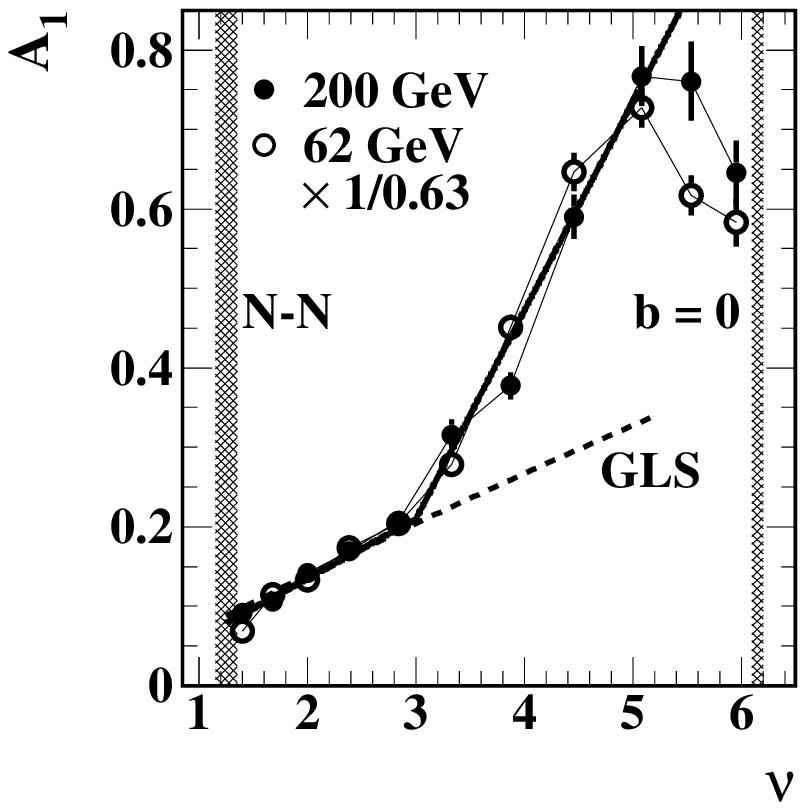}   
\put(-30,35) {\bf (a)}
\includegraphics[width=.3\textwidth,height=.3\textwidth]{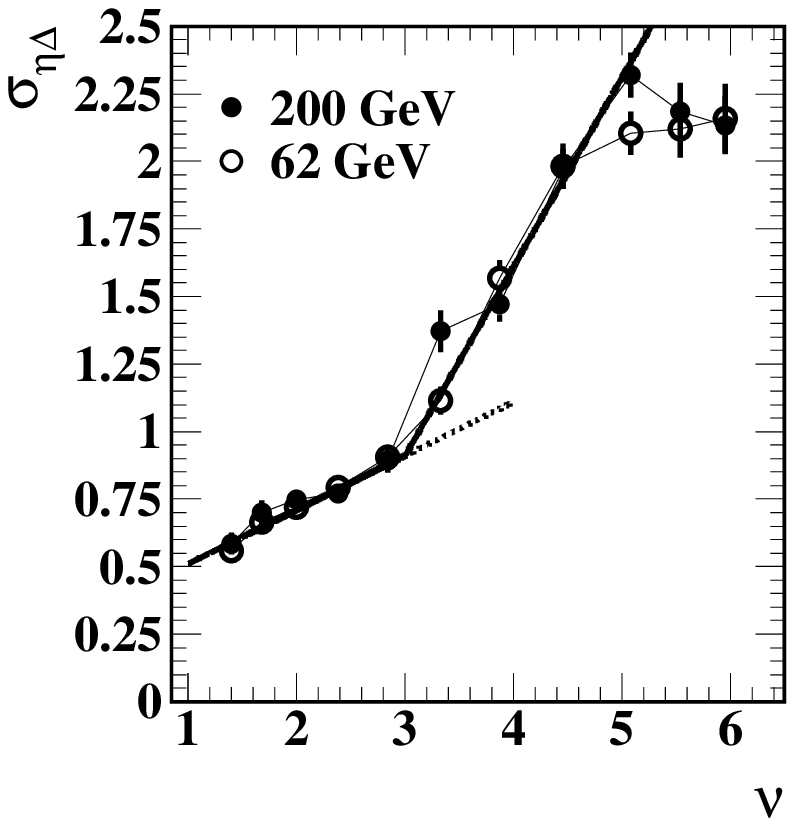} 
\put(-30,35) {\bf (b)}
\includegraphics[width=.3\textwidth,height=.3\textwidth]{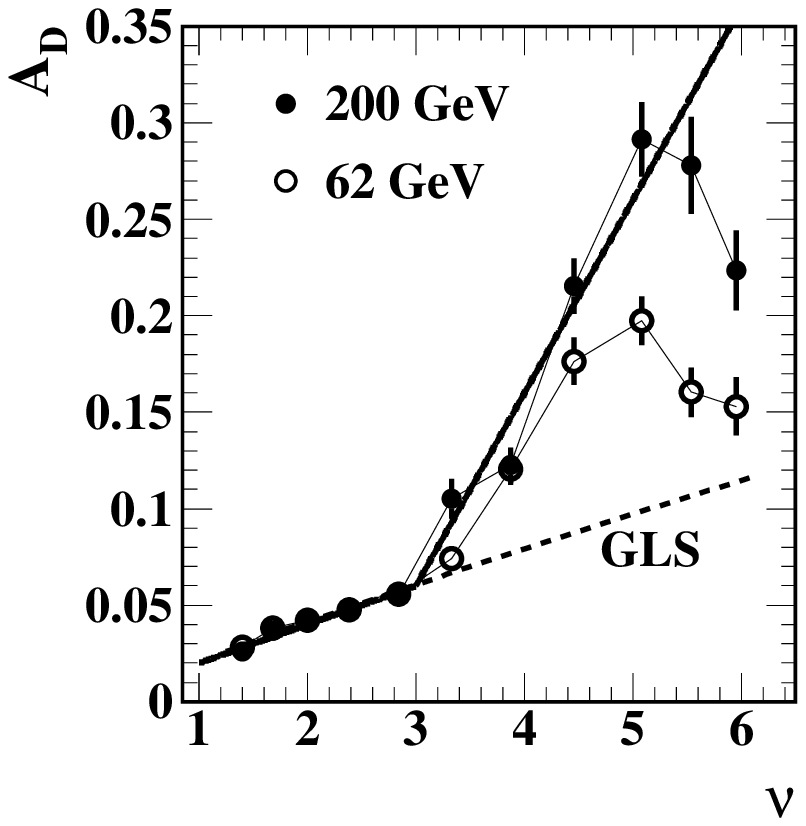}
\put(-30,35) {\bf (c)}
\caption{\label{slopes}
 Fit parameters for $(\eta_{\Delta},\phi_{\Delta})$ correlation data from Au-Au collisions at $\sqrt{s_{NN}}= 62$ (open symbols) and 200~GeV (solid symbols) versus centrality measure $\nu$ illustrating a {\em sharp transition} in centrality trends at $\nu_{trans} \approx 3$. The dashed curves labeled GLS are calculated predictions corresponding to linear superposition of N-N collisions (binary collision scaling). The bold solid lines illustrate slope changes by factors 3.5 in panels (a) and (b) and 5 in panel (c) within one centrality bin on $\nu$. The dotted line in panel (b) simply continues the slope trend from below the transition. Panel (a) shows that 62 and 200 GeV same-side 2D peak amplitudes $A_1$ are related by a $\log(\sqrt{s_{NN}})$ factor $\approx 0.6$~\cite{davidhq}, whereas panel (c) demonstrates that the away-side 1D peak amplitude $A_D$ is approximately independent of collision energy.
}\end{figure*}

\section{Discussion}
\label{Sec:Diss}

%

The sharp transition in jet-like angular correlation trends~\footnote{``Jet-like angular correlations'' refers to a generic same-side 2D peak and away-side 1D ridge structure in 2D angular correlations.} 
revealed by the present analysis introduces a surprising new aspect of RHIC Au-Au collisions. The transition occurs at a specific value of mean participant path length $\nu$ common to both collision energies ($\nu_{trans} \approx 3$). 
Such a transition in the SS 2D peak amplitude and width on $\eta_\Delta$ could mark the onset of a new correlation mechanism beyond semihard parton scattering and fragmentation.  However, any proposed theoretical description, including novel collision mechanisms, must describe accurately the smooth
centrality dependence of the SS 2D peak azimuth width and  $p_t$ angular
correlations [59] above {\em and below} $\nu_\text{trans}$. We now consider
further details.


\subsection{Goals and unique aspects of this analysis} \label{unique}

This analysis has as its primary goal accurate description of 2D angular correlations over the complete range of Au-Au centralities with a minimal complement of simple functional forms. We wish to determine under what circumstances and to what extent more-central Au-Au collisions deviate from a simple linear superposition of N-N collisions according to the Glauber model of A-A collisions extrapolated from peripheral collisions. A major issue for this study is the extent to which {\em minimum-bias} jet structure observed in \mbox{p-p} collisions is modified in more-central Au-Au collisions. Based on related studies we adopt a minijet hypothesis: that minimum-bias jet-like structure in more-peripheral A-A collisions corresponds approximately to that from 3 GeV jets (minijets). We then examine the extent to which that structure is modified in more-central Au-Au collisions. The extent and nature of such modifications is compared to several theoretical scenarios as a test of their validity.

The present analysis is unique in several aspects: (a) consideration of the full range of \mbox{A-A} centralities down to N-N collisions, (b) use of a statistically well-defined per-particle correlation measure, (c) definition of a Glauber linear superposition reference, (d) accurate model fits to 2D angular correlations and (e) distinct measurements of a same-side 2D peak, away-side 1D peak and nonjet quadrupole. Conventional $v_2$ analysis~\cite{2004} describes only 1D azimuth projections and only with one Fourier series term, disregarding the $\eta$ dependence critical for distinguishing separate structures and mechanisms. Analysis of ``dihadron'' azimuth correlations~\cite{trigger} also considers only 1D azimuth projections and subtracts a background based on possibly-biased $v_2$ data~\cite{tzyam}. Trigger-associated 1D and 2D angular correlations~\cite{joern} include only part of the jet structure (high-$p_t$ hadrons) and typically do not represent the 2D structure with model functions. 

Those characteristics can be compared with a 2D angular correlation analysis by the PHOBOS collaboration presented in Ref.~\cite{phoboscorr}. In the PHOBOS analysis correlations were measured over $2\pi$ azimuth and $\eta \in [-3,3]$. The correlation measure defined in Eq.~(1) of Ref.~\cite{phoboscorr} is that used in Ref.~\cite{axialci}. Particle $p_t$  could not be reconstructed over the full $(\eta,\phi)$ acceptance and was not used for the analysis in Ref.~\cite{phoboscorr}. Although 2D angular correlations were inferred, emphasis was placed on the projected 1D distribution on $\eta_\Delta$ which has a single peak feature (``short range'' correlations) fitted by a 1D Gaussian (cluster model).  No distinction was made between the SS 2D peak and 1D Gaussian on $\eta_\Delta$ isolated in the present analysis. The underlying physical mechanism for the projected 1D peak was not specified. Properties of the fitted 1D Gaussian  were inferred only for the upper 50\% of the total cross section due to systematic uncertainties in the event selection procedure  for low-multiplicity events. The sharp transition in SS 2D peak properties reported in the present study is then inaccessible. It was concluded that angular correlation structure in central Au-Au collisions is similar to that in p-p collisions, and an inferred cluster size decreases with increasing centrality, dramatically contradicting the large increases in jet-like structure observed in the present study.

\subsection{Anomalous  evolution of correlation structure} \label{anomdisc}

The reported anomalous evolution has two aspects (see Fig.~3): (a) Three correlation model parameters (SS peak amplitude, AS peak amplitude, SS peak $\eta$ width) undergo large slope changes in their centrality trends within a small interval on centrality at $\nu = \nu_{\rm trans}$ common to two energies. In addition, the azimuth width of the SS peak, which decreases significantly from p-p up to  $\nu_{\rm trans}$, maintains a fixed value above that point. 
(b) The large increase in the SS peak amplitude above $\nu_{\rm trans}$, interpreted as a minijet manifestation, is inconsistent with expectations of scattered-parton thermalization (strong jet quenching) in more-central A-A collisions, and the decrease in the azimuth width is inconsistent with parton multiple scattering and other jet-quenching scenarios~\cite{kll,nayak,mjshin,jethydro}.


\begin{figure*}[t]  
\includegraphics[width=.3\textwidth,height=.3\textwidth]{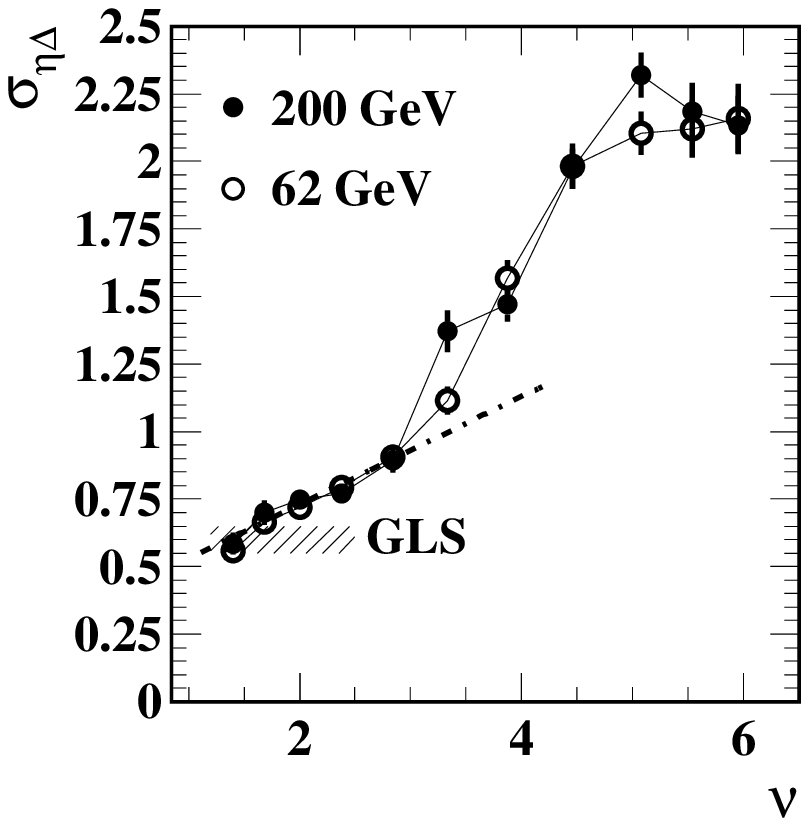}   
\put(-28,90) {\bf (a)}
\includegraphics[width=.3\textwidth,height=.3\textwidth]{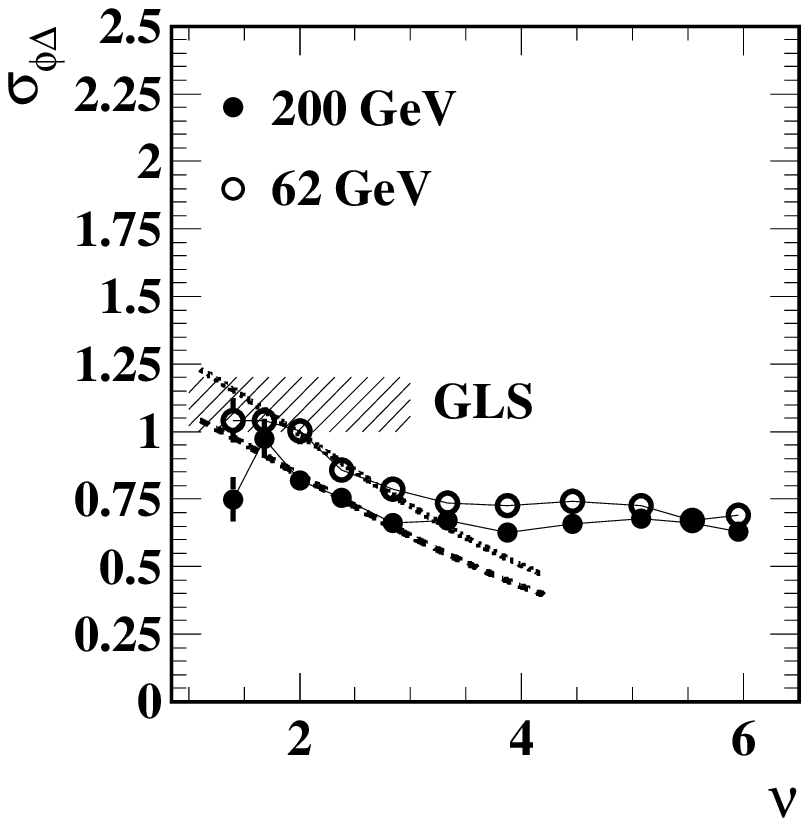} 
\put(-28,90) {\bf (b)}
\includegraphics[width=.3\textwidth,height=.294\textwidth]{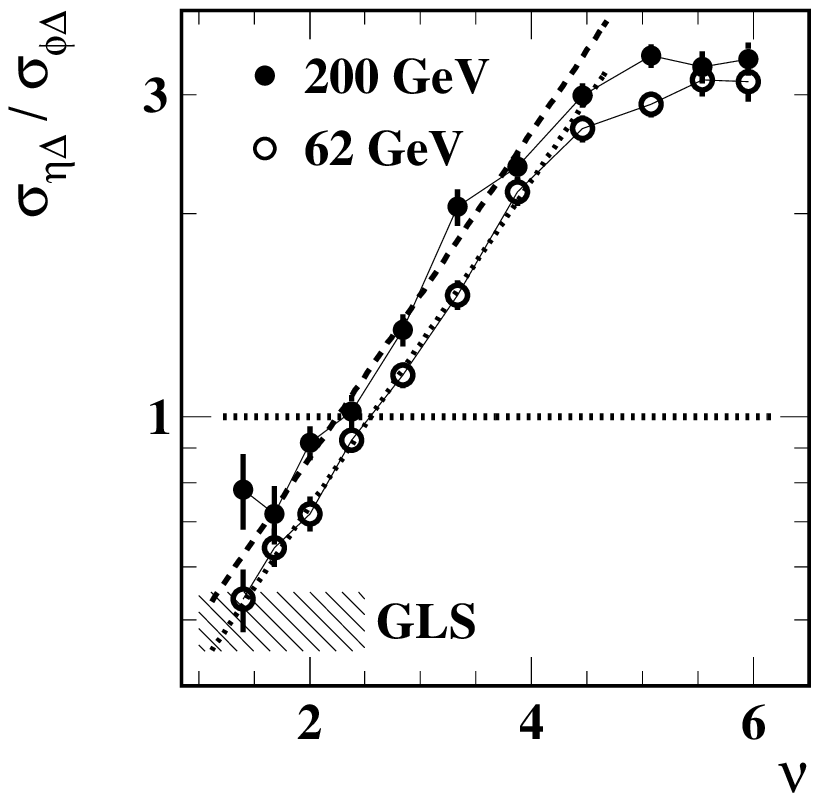}
\put(-28,90) {\bf (c)}
\caption{\label{pkwidths}
Same-side 2D peak width trends for $(\eta_{\Delta},\phi_{\Delta})$ correlation data from Au-Au collisions at $\sqrt{s_{NN}}= 62$ (open symbols) and 200~GeV (solid symbols) versus centrality measure $\nu$ computed at fixed energy (200~GeV). The same-side 2D Gaussian $\eta_{\Delta}$ and $\phi_{\Delta}$ widths are shown in panels (a) and (b) respectively.  Panel (c) shows the same-side peak width aspect ratio $\sigma_{\eta_\Delta}/\sigma_{\phi_\Delta}$. The hatched regions indicate GLS trends based on {\it p-p} collision data. The bold curves are explained in the text.
}
\end{figure*}

Figure~\ref{slopes} illustrates manifestation of anomalous centrality evolution in the form of feature (a) large slope change within a small centrality interval. In all three panels the {\em sharp transition} (rapid slope change) occurs near $\nu_{\rm trans} = 3$ (slope change of solid lines). In the first two panels the slope ratios of the solid lines  are fixed at 3.5. In the third panel the slope ratio is 5. Data deviations from the solid lines in the first two panels are consistent with fit uncertainties (error bars). In the third panel the 62 GeV transition may be slightly displaced to larger $\nu$. Nevertheless, three fit parameters at two energies (six instances) exhibit large slope changes in their centrality trends within the same small centrality interval $\nu_{trans} = 3.1 \pm 0.3$. The interval 0.6 about $\nu = 3$ corresponds to 10\% on fractional cross section $\sigma / \sigma_0$.

In the first panel the 62 GeV data are brought into coincidence with the 200 GeV data by a factor 1/0.63. That relation corresponds to the energy scaling quantity $R(\sqrt{s_{NN}} = \text{62 GeV}) = \log(62 / 13.5) / \log(200/13.5) = 0.57$ from Ref.~\cite{davidhq} which describes the azimuth quadrupole per-particle energy dependence (relative to values at 200 GeV) reported there. The correspondence is notable. Also, whereas the SS 2D peak amplitude does scale as $\log(\sqrt{s_{NN}})$ the AS dipole amplitude in panel (c) does not. That difference is expected for pQCD dijets, holds for most Au-Au centralities, and the same energy scaling trend continues up to LHC energies~\cite{ppcms}.

Centrality variation of the SS 2D peak amplitude and $\eta$ width in panel (b) indicates that the integrated number of SS peak correlated pairs per final-state charged particle (SS peak volume) exceeds binary-collision scaling by an order of magnitude in more-central Au-Au collisions. In the minijet context the same-side peak pair number corresponds to the product of the event-wise minijet number in the angular acceptance and the mean fragment pair number ($\approx$ mean jet fragment multiplicity squared)~\cite{jetyield}. Is the large correlated-pair increase due to excess production of minijets with N-N properties in central A-A collisions (relative to binary-collision scaling)? Or does the mean multiplicity associated with each minijet increase relative to that for N-N collisions, the minijet number remaining consistent with binary-collision scaling? 
The quantitative correspondence among correlations, spectra and pQCD discussed in Sec.~\ref{minispectra} seems to indicate that parton scattering changes little with increasing centrality, but the details of parton fragmentation to jets changes substantially.

While the sharp transition in SS 2D peak properties is itself notable, the fact that minijet correlations increase at all with increasing Au-Au centrality seems to conflict with the conventional expectation that most jets are ``quenched'' in the dense medium formed in central \mbox{A-A} collisions, therefore not appearing as correlation structures in the final state. Evidence from $R_{\rm AA}$ measurements (hadron suppression at high $p_t$)~\cite{starraa} and high-$p_t$ jet AS azimuth correlations (disappearance of the away-side jet)~\cite{staras} seemed consistent with that expectation. In contrast, we observe that $p_t$-integral minimum-bias jet-like structure increases dramatically with centrality.

\subsection{Anomalous SS 2D peak width trends}  \label{anom}

The most notable feature of the SS 2D peak in more-central Au-Au collisions, its large elongation on $\eta_\Delta$, is not predicted by present pQCD theory. But neither is the comparable elongation on $\phi_\Delta$ observed in {\it p-p} collisions. A possible $\eta_\Delta$ elongation mechanism arising from color connections between struck partons and their parent nucleons has recently been suggested~\cite{fragevo}.  The interplay between the SS peak widths is here described in more detail.

Figure~\ref{pkwidths} shows centrality evolution of the two angular widths of the SS 2D peak in Au-Au collisions. The third panel shows the width (or aspect) ratios.
The $\sigma_{\eta_\Delta}$ trend (first panel) is approximated below the transition point at $\nu \approx 3$  by
$\sigma_{\eta_\Delta} = 0.53 + 0.2(\nu - 1)$ (dash-dotted line).
The aspect ratio trends (third panel, dashed and dotted lines) are described over a larger centrality interval by $\sigma_{\eta_\Delta} / \sigma_{\phi_\Delta} = \exp\{(\nu - \nu_0)/1.8\}$, with $\nu_0 = 2.25,$ and 2.55 for 200 and 62 GeV respectively. The dashed and dotted curves describing $\sigma_{\phi_\Delta}$ below the transition point in the middle panel are simply derived from those two results and indicate the consistency of the description.

Nominal GLS trends assuming linear superposition of {\it p-p} collisions are indicated by the hatched regions. The predicted widths from {\sc hijing} are $\sigma_{\eta_\Delta} = 0.75$ and $\sigma_{\phi_\Delta} = 0.9$ (radians) independent of centrality, in marked contrast with the large angular asymmetries and strong centrality dependence of the SS 2D peak observed in the data. 
While the SS and AS peak amplitudes follow the GLS trend below the transition point (see Fig.~3) the SS peak widths do not. The individual width trends (the slopes) change substantially at the transition point, but the aspect ratio varies smoothly (exponentially) from {\it p-p} to more-central Au-Au collisions, exhibiting no sign of a slope change. Any viable theoretical description of angular correlations in Au-Au collisions must accommodate that complex phenomenology.

\subsection{Minijets and $\bf p_t$ angular correlations}

Are new (e.g., non-pQCD) collision mechanisms required to accommodate the observed anomalous centrality evolution? Information from $p_t$ angular correlations  and single-particle spectra may help to reduce the ambiguity.
$p_t$ angular correlations, complementary to number angular correlations from the present analysis, have been obtained by inversion of $\langle p_t \rangle$ fluctuation scale dependence from 200~GeV Au-Au collisions~\cite{auto,ptedep,ptscale}. The $p_t$ correlation structure is qualitatively similar to that presented here (e.g., same-side 2D peak, away-side ridge, quadrupole), but there are significant quantitative differences. 
The same-side 2D peak amplitude trend on centrality, while increasing at least as fast as binary-collision scaling until $\nu \approx 4$ where it starts to decrease, does not show a substantial change in slope at $\nu_{\rm trans}$.

In the minijet context $p_t$ angular correlations suggest that \mbox{N-N} semihard parton scattering continues to drive the SS 2D peak structure above $\nu_\text{\rm trans}$ where the number of minijets (parton scatters) increases at least as fast as $N_{\rm bin}$ through and above $\nu_{\rm trans}$ until $\nu \approx 4$.
In this picture it follows that the large increase of SS peak pairs in Fig.~\ref{Figure3} is then due to strong modification of parton fragmentation leading to a large increase in the number of jet-correlated hadrons per initial-state parton scatter. That conclusion is substantiated by comparisons among spectrum hard components, pQCD fragment distributions and minijet number angular correlations~\cite{jetyield,fragevo}.





\subsection{Minijets and single-particle spectra} \label{minispectra}

The large changes in the number of correlated particle pairs observed in this analysis should also be manifested in single-particle 
$p_t$ spectra. In Ref.~\cite{ppspectra} a differential two-component analysis was applied to unidentified charged hadron $p_t$ spectra from non-single-diffractive (NSD) \mbox{{\it p-p}} collisions at $\sqrt{s}$ = 200~GeV. 
Hard and soft spectrum components were isolated as limiting cases of spectrum evolution with event multiplicity (centrality). The soft spectrum component was estimated from the $n_{ch} \rightarrow 0$ spectrum limit. Subtracting the soft component from the full spectrum revealed the hard component centrality evolution.
The hard component is interpreted as hadron fragments from minimum-bias large-angle-scattered partons (minijets). The soft component is interpreted as fragments from dissociation of projectile nucleons. The hard component has been subsequently identified with jet-like {\it p-p} correlations~\cite{jeffpp1,aspect}.

In Ref.~\cite{TomAuAuspectra} a similar two-component analysis was applied to 200~GeV Au-Au spectra for identified pions and protons. Hard and soft spectrum components were identified for each species. The hard components evolve strongly with centrality, accounting for essentially all of the per-participant spectrum evolution. Strong suppression at larger $p_t$ is accompanied by much larger enhancement at smaller $p_t$ relative to binary-collision scaling of the \mbox{N-N} hard components. Suppression (at high $p_t$) and enhancement (at low $p_t$) variations with centrality are strongly correlated, implying the same underlying mechanism. The sharp transition on centrality for each hadron species matches the transition revealed in the present correlation analysis. 

In Ref.~\cite{fragevo} the spectra from Ref.~\cite{TomAuAuspectra} were compared with a pQCD calculation of fragment distributions based on measured jet fragmentation functions (LEP, CDF) and pQCD predicted parton spectra. Again the agreement was found to be very good, lending strong support to interpretation of both spectrum hard components and minimum-bias jet-like correlations as pQCD jets.

In Ref.~\cite{jetyield} preliminary data from the present analysis were combined with a pQCD prediction of jet number in A-A collisions to infer parton fragment yields corresponding to minijet production. The fragment yields were in turn compared with yields inferred from $p_t$ spectra for identified hadrons and  the agreement was found to be good~\cite{TomAuAuspectra}. From that exercise it was concluded that about one third of the final state in 200 GeV central Au-Au collisions is contained in resolved minijets, mainly from 3 GeV scattered partons.

\subsection{Interpretation of the away-side dipole}

In Fig.~\ref{Figure3} (left panels) the amplitudes of the {away-side dipole} structure and same-side 2D Gaussian follow the same centrality trend, strongly suggesting that they share a common mechanism. The mean energy $\sim 3$~GeV for minimum-bias scattered partons~\cite{minijet,ffprd}  is comparable to the mean intrinsic $k_t \sim 1$~GeV/$c$~\cite{nucleonkt} within projectile nucleons, implying large acoplanarities for semihard scattered parton pairs and a broad away-side azimuth ``ridge'' (back-to-back parton correlation). It is easy to demonstrate that the large width of the away-side peak ($\sim$ Gaussian on azimuth at $\phi = \pi$) plus peak periodicity on azimuth are equivalent to dipole trend $\cos(\phi_{\Delta} - \pi)$~\cite{tzyam}. 


An away-side dipole could also be produced by global transverse-momentum conservation (Sec.~\ref{otherstruct} and \cite{lisazibi}) which might account for part of the observed AS dipole amplitude in peripheral Au-Au (N-N) collisions. For the per-pair quantity $\Delta\rho/\rho_{\rm ref}$ that contribution to the dipole term would be proportional to $1/N_{\rm ch}$. For per-particle measure $\Delta\rho/\sqrt{\rho_{\rm ref}}$ used in this analysis that contribution should be independent of centrality. 
We observe that the AS dipole amplitude follows the same strongly-increasing  centrality trend as the SS 2D peak. The close correspondence implies that the AS dipole is indeed a manifestation of  transverse-momentum conservation, but at the parton-parton scale (dijets), not the nucleus-nucleus scale. 

\subsection{Other correlation structures}
The azimuth quadrupole component $\cos(2\,\phi_\Delta)$ has been conventionally identified with elliptic flow. The 2D quadrupole amplitude reported here  is related to conventional measure $v_2$ by $A_{\rm Q} = 2\rho_0(b)\, v^2_2\{{\rm 2D}\}$, where $\rho_0(b) = dN_{\rm ch} / 2\pi d\eta$ is the single-particle angular density and ``2D'' denotes inference of quadrupole amplitudes from model fits to 2D angular correlations as described in this article~\cite{tomv2method1,tomv2method2}. In Fig.~\ref{Figure3} (bottom center) the smooth centrality variation of the quadrupole amplitude is in marked contrast to the SS peak amplitude and $\eta$ width. The quadrupole amplitude shows no counterpart to the sharp transition in SS peak properties. Although the centrality trends for SS peak properties and azimuth quadrupole are very different, the two amplitudes, when measured by statistically equivalent quantities $A_{\rm Q}$ and $A_1$, share the same $\ln(\sqrt{s_{\rm NN}})$ dependence characteristic of QCD scattering processes, as discussed in Sec.~\ref{centenergy}. Detailed analysis of the quadrupole component and its relation to other $v_2$ methods is presented in Refs.~\cite{davidhq,quadpaper}.

\begin{figure}[t]  
\includegraphics[width=.23\textwidth,height=.23\textwidth]{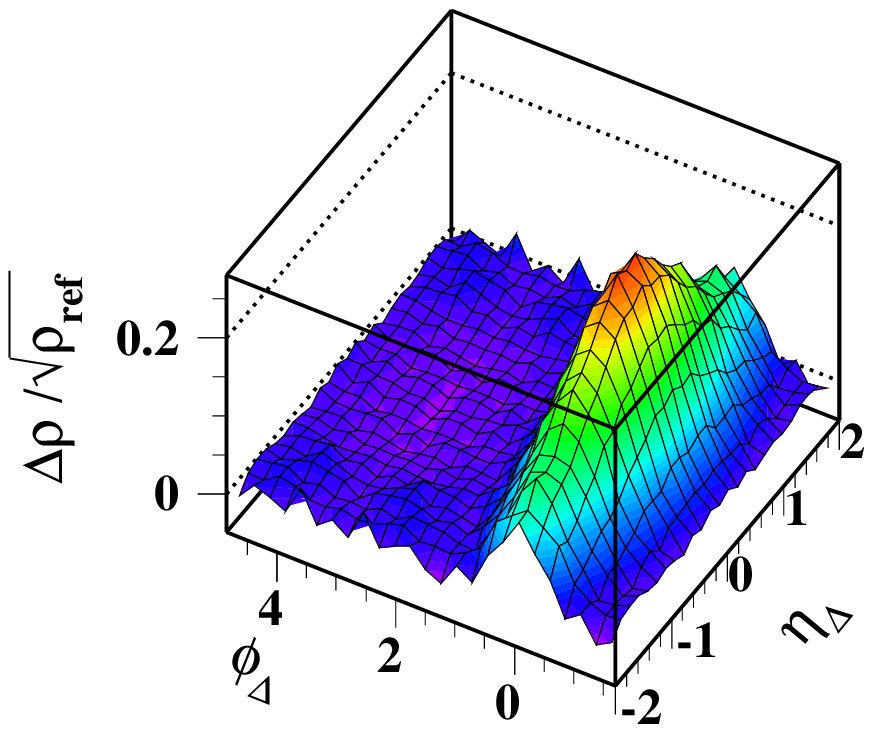}  
\includegraphics[width=.23\textwidth,height=.23\textwidth]{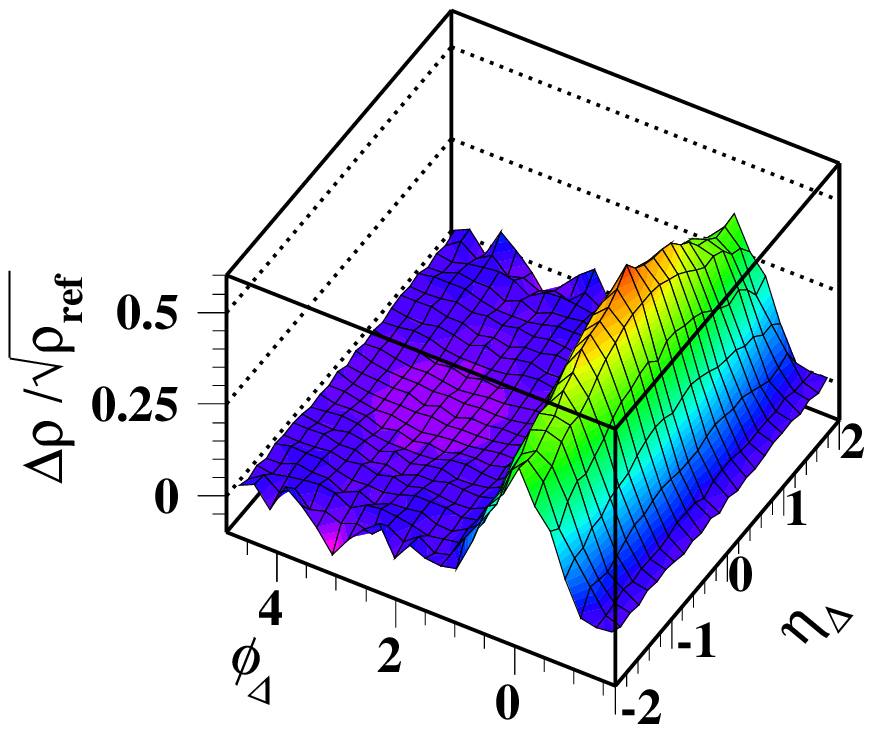} 
\caption{\label{ss2dsingles} Same-side 2D peaks from 38-46\% central (left panel) and 9-18\% central (right panel) 200 GeV Au-Au collisions. These histograms were obtained by subtracting the AS dipole ($A_D$), quadrupole ($A_Q$) and offset ($A_3$) elements of the 2D model fits from the data histograms.
}  
\end{figure}

The 1D peak on $\eta_\Delta$, interpreted to represent longitudinal projectile-nucleon fragmentation, has a simple centrality dependence. It is quite visible in \mbox{p-p} (N-N) collisions, especially for low-$p_t$ particles,~\cite{jeffpp1} but falls monotonically to zero by mid-central A-A collisions (See Fig.~\ref{Figure1}  and Tables~\ref{TableI} and \ref{TableII}) . The structure was first observed at the ISR~\cite{Whitmore} and has been compared to similar structure predicted by the Lund model (PYTHIA)~\cite{jeffpp0,lund}. In Ref.~\cite{jeffpp0} a comparison of charge-dependent (CD) structure dominated by that model component ($A_0$) was compared to PYTHIA data (Figs.~3 and 4 of that reference), and good agreement was observed.  In contrast to $A_{\rm Q}$ (nonjet quadrupole) and $A_1$ (minijets), the $A_0$ amplitude shows negligible energy dependence. 

\subsection{Presence of higher multipoles measured by $\bf v_m$} \label{v3}

\begin{figure}[t]  
\includegraphics[width=.23\textwidth,height=.235\textwidth]{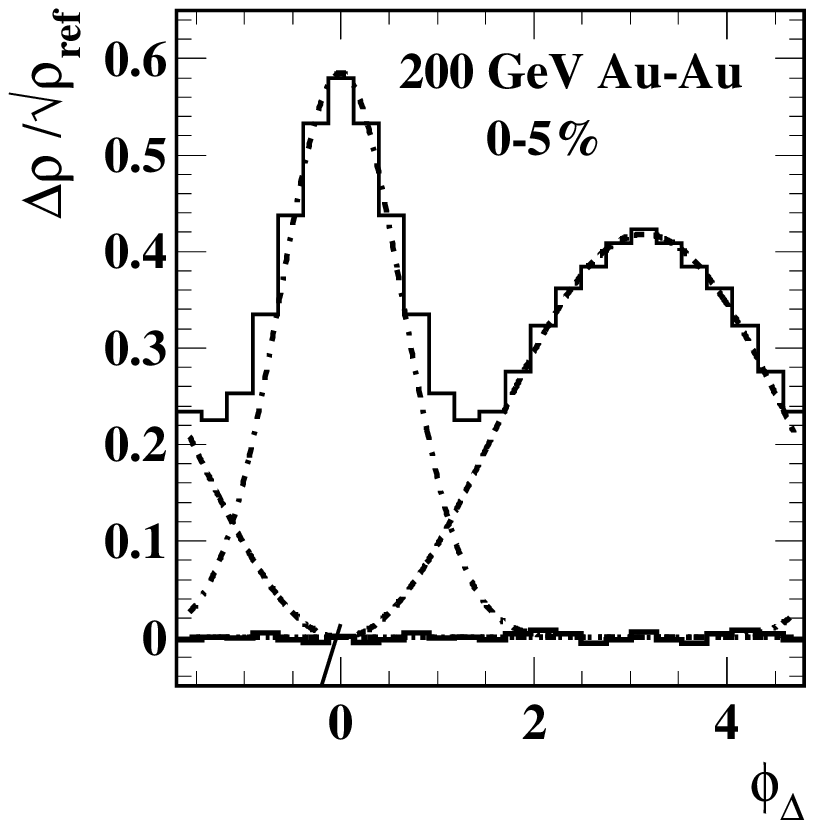}  
\includegraphics[width=.23\textwidth,height=.235\textwidth]{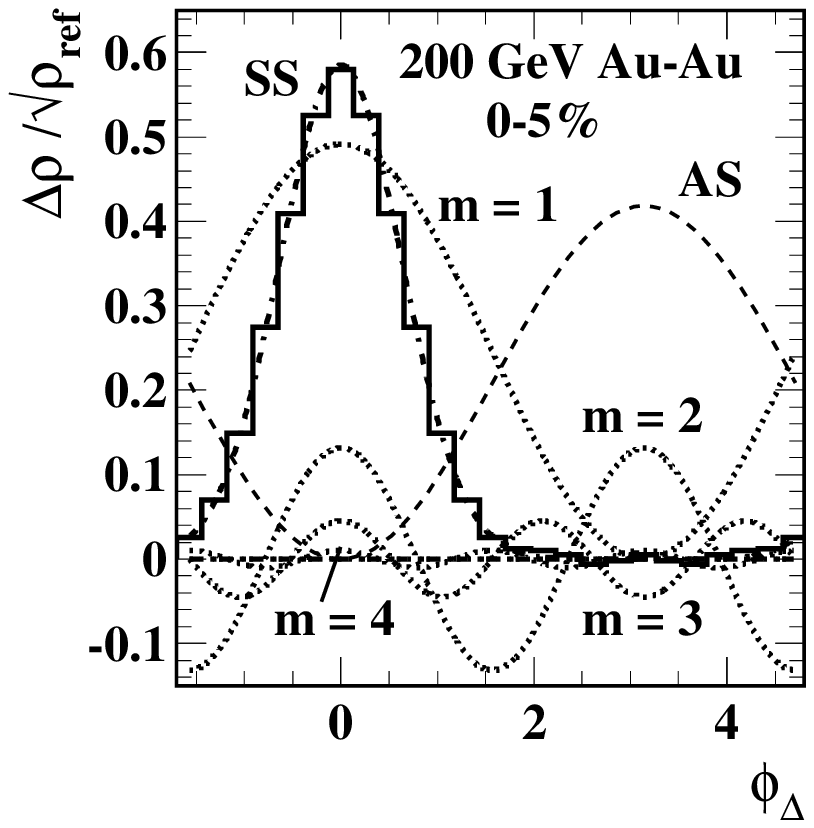} 
\caption{\label{vv3}
Left:  2D histogram from  Fig.~\ref{Figure4} (fourth panel) projected onto $\phi_\Delta$ (light histogram). The bold histogram with near-zero amplitude is a projection of 2D residuals from Fig.~\ref{Figure4} (third panel). The dash-dotted and dashed curves represent components of the model fit from the present analysis.
Right:  The bold histogram is the 1D data histogram in the left panel minus the fitted AS dipole component (dashed curve). The dash-dotted curve is the 1D projection of the fitted 2D Gaussian from this analysis. The bold dotted curves represent multipoles $m = 1\ldots4$ derived from the SS 1D Gaussian (see the text).
}  
\end{figure}

There are two modes in which higher-order ($m > 2$) azimuth multipoles may arise in model descriptions of the present data. First, the data may {\em require} additional multipoles for a statistically satisfactory description. That topic is mentioned below and discussed in more detail in App.~\ref{multi}. Second, the ``standard'' 2D model components defined in Eq.~(\ref{Eq4}) may be decomposed into $\eta_\Delta$-independent multipoles~\cite{multipoles}, based either on 1D projections onto azimuth or on the full 2D model elements. The former is discussed in this subsection, the latter in App.~\ref{multi}.

%
Figure~\ref{ss2dsingles} shows 2D angular correlations for two Au-Au centralities. The AS dipole ($m=1$) and quadrupole ($m=2$) terms of the 2D fit model have been subtracted from the data histograms. All that remains is the SS 2D peak, described by a 2D Gaussian. The 38-46\% centrality (left panel) is just above the sharp transition at $\nu_{trans} = 3$. The 9-18\% centrality (right panel) is at $\nu = 5$, the maximum of the SS 2D peak amplitude trend. 

As noted in Sec.~\ref{2dmodel} the present fit model exhausts all statistical information in most combinations of centrality and energy. In more-central collisions significant residuals structure does remain, but only associated with the AS dipole ($m = 1$) as discussed in Sec.~\ref{extracomp}. Thus, any higher multipoles $v_m$ for $m > 2$ must come from the SS 2D peak structure shown in Fig.~\ref{ss2dsingles}, since the conjectured $v_m$ are orthogonal to the subtracted model elements ($m=1$, 2). However, the SS 2D peak has a strong curvature on $\eta_\Delta$ which cannot be described by a 1D Fourier series on azimuth. Thus, the SS 2D Gaussian in $p_t$-integral angular correlations is unique. 

In this argument we focus on the 0-5\% centrality bin where the $A_Q$ amplitude is consistent with zero, providing a particularly simple example. However, the same arguments concerning higher $v_m$ apply to all Au-Au centralities, as demonstrated in Fig.~\ref{ss2dsingles}.

Figure~\ref{vv3} (left panel) shows the 2D histogram data in Fig.~\ref{Figure4} (fourth panel) projected onto azimuth difference $\phi_\Delta$ (light histogram). The dash-dotted and dashed curves represent the SS 1D Gaussian (projected SS 2D Gaussian) and AS dipole respectively, derived from the fit to 2D histogram data as reported in Table~\ref{TableI}. The bold histogram  with near-zero amplitude is the 1D projection of the residuals in Fig.~\ref{Figure4} (third panel). The r.m.s.\ residuals amplitude is about 0.5\% of the SS peak amplitude and consistent with statistical uncertainties. Thus, a model function consisting of SS 1D Gaussian (two parameters $A_{\rm 1D}$, $\sigma_{\phi_\Delta}$) plus AS dipole (one parameter $A_{\rm D}$) exhausts all statistical information in the 1D data histogram. There is no {\em necessity} for additional Fourier components to represent these 2D angular correlations, as shown in Fig.~\ref{vv3} (left panel). That is, the systematic uncertainty interval for such amplitudes includes zero.

\begin{figure*}[t]  
\includegraphics[width=.24\textwidth,height=.235\textwidth]{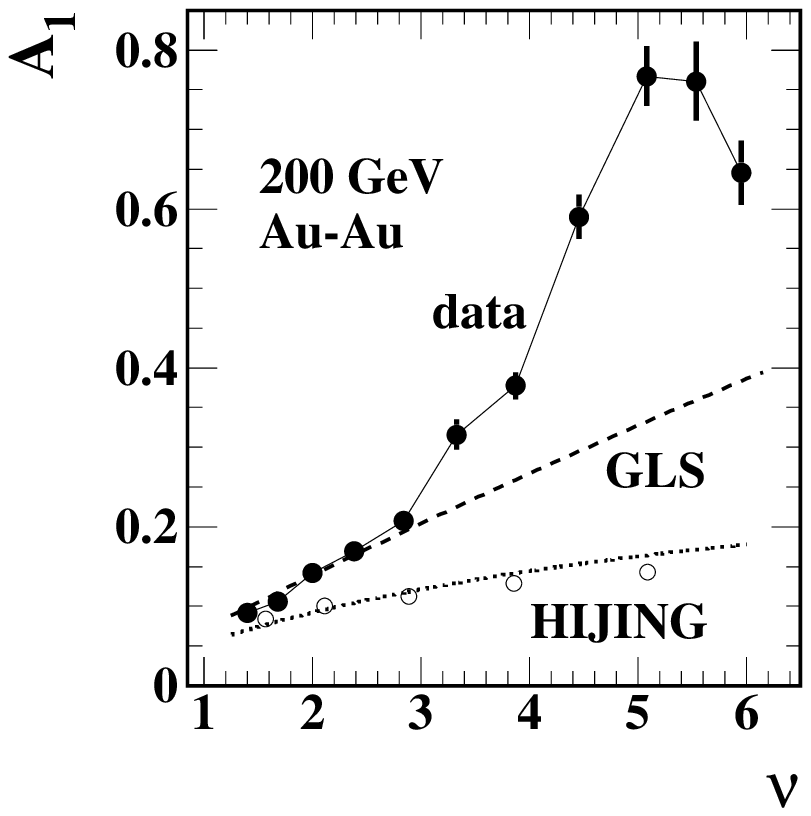} 
\put(-88,105) {\bf (a)}
\includegraphics[width=.24\textwidth,height=.235\textwidth]{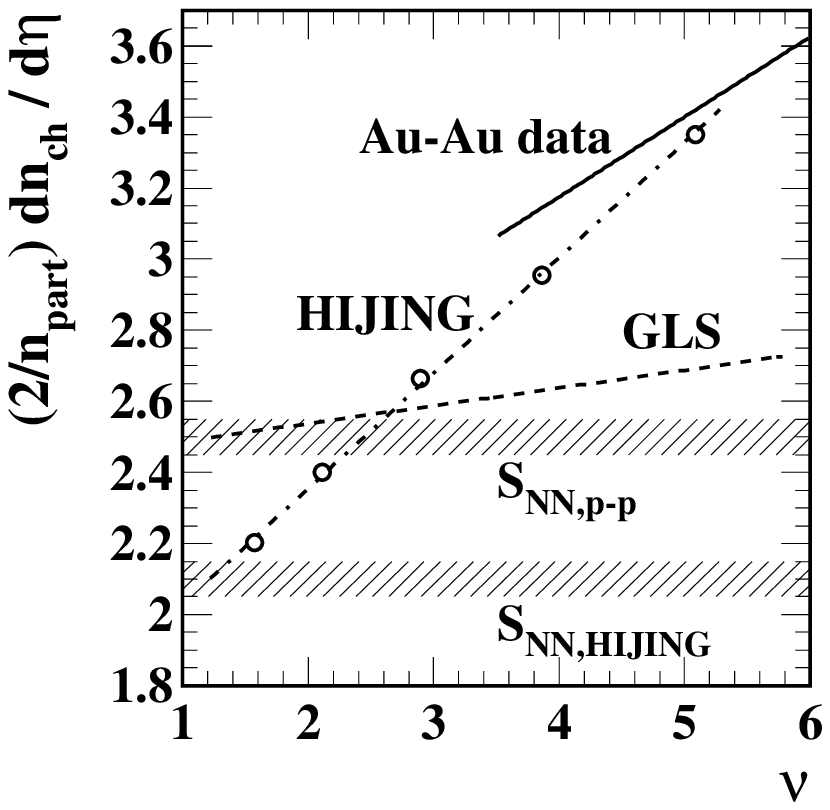}  
\put(-88,105) {\bf (b)}
\includegraphics[width=.24\textwidth,height=.24\textwidth]{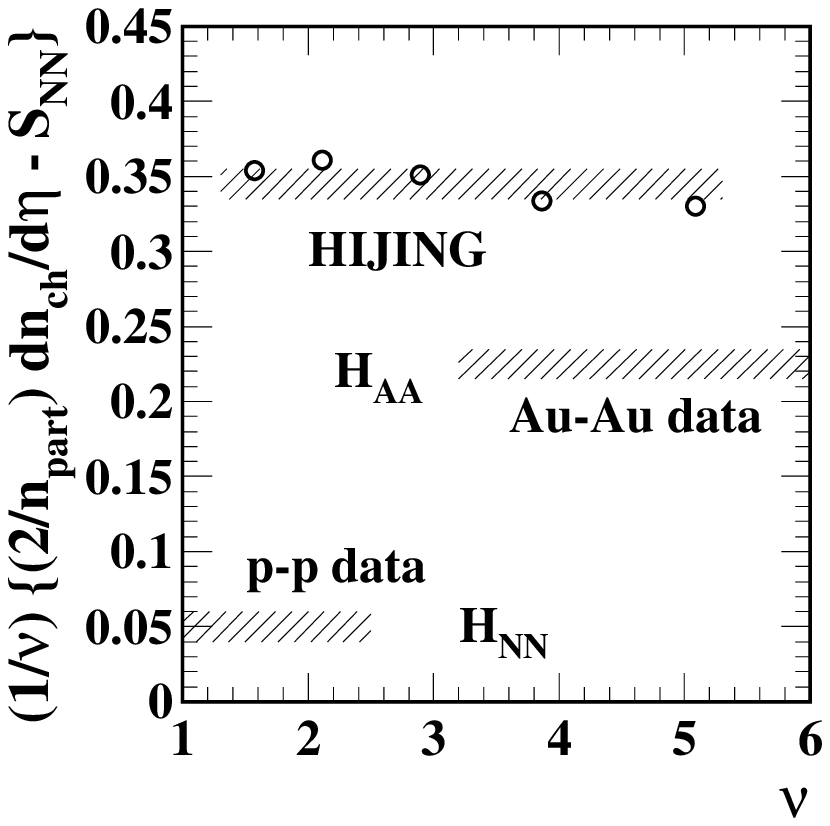}  
\put(-88,105) {\bf (c)}
\includegraphics[width=.24\textwidth,height=.24\textwidth]{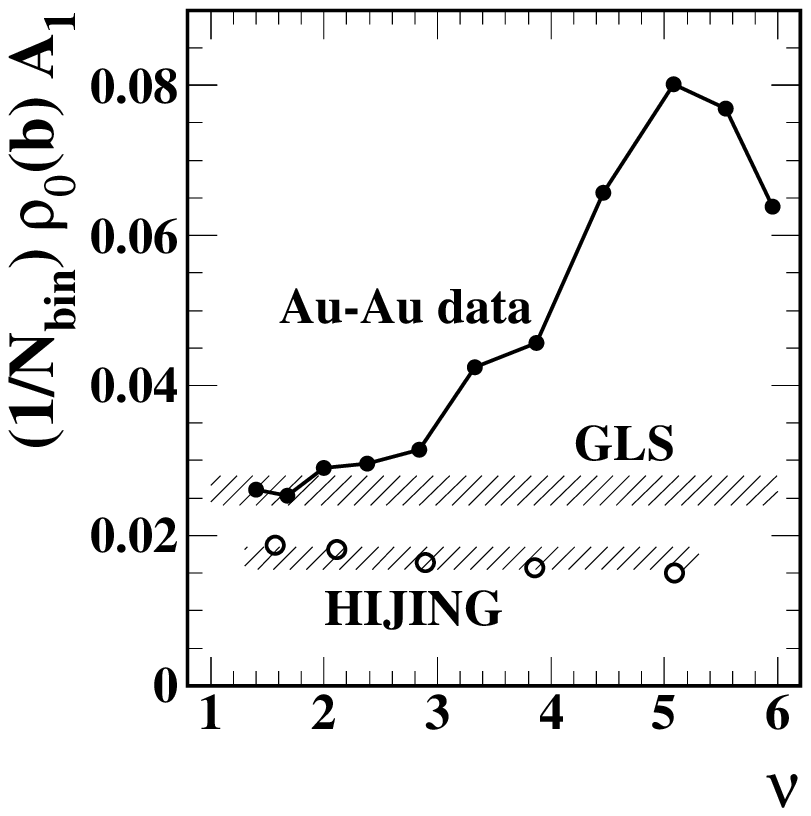} 
\put(-88,105) {\bf (d)}
\caption{\label{hijfig}
(a)  Amplitude of SS 2D peak (jet-correlated pairs per final-state hadron) for data (solid points), GLS extrapolation from {\it p-p} data (dashed curve) and from the {\sc hijing} Monte Carlo (open points, dotted curve).
(b)  Single-particle production extrapolated from {\it p-p} data (dashed curve, GLS), and from more-central Au-Au data (solid line) and {\sc hijing} (dash-dotted line, open points).
(c) Single-particle hard-component production per N-N binary collision from {\it p-p} data,  from more-central Au-Au data and from {\sc hijing} (open points).
(d) Jet-correlated pair production (SS 2D peak) per N-N binary collision extrapolated from {\it p-p} data (GLS), and from Au-Au data (solid points) and {\sc hijing} (open points).
}  
\end{figure*}

Figure~\ref{vv3} (right panel) shows the data histogram in the left panel minus the fitted AS dipole (dashed curve). The difference (bold histogram) is described to the statistical limits of data by a 1D Gaussian (dash-dotted curve) with amplitude $A_{1D}$, consistent with the residuals in the left panel.
The bold dotted sinusoids (SS-peak multipoles) in the right panel represent the first four Fourier components ($m = 1\ldots 4$) of the SS 1D Gaussian measured by $2\rho_0(b) v_m^2 = F_m(\sigma_{\phi_\Delta})\, A_{\rm 1D}$, with $F_m(\sigma_{\phi_\Delta}) = \sigma_{\phi_\Delta}\sqrt{2/\pi}  \exp(-m^2\, \sigma ^2_{\phi_\Delta} / 2)$. The $m = 2$ term estimates the ``nonflow'' contribution from the SS  2D peak to the total azimuth quadrupole.  The $m = 3$ term is the azimuth sextupole (``triangular flow''~\cite{triflow}), the $m = 4$ term (just visible) is the octupole component. 

For the 0-5\% central data in Fig.~\ref{vv3} we obtain  $A_{\rm 1D} = 0.585\pm0.06$, $\sigma_{\phi_\Delta} = 0.63$, $F_3(0.63) = 0.077$ and $\rho_0(\text{0-5\%}) = 107$, from which we predict $v_3 = 0.015\pm0.0008$. We also obtain $v_2 = 0.025\pm0.0013$ (nonflow) and $v_4 = 0.007\pm0.0004$.  The uncertainty estimates are based on the $\approx10$\% total uncertainty in $A_{\rm 1}$ which is the dominant source.
Similar results are obtained for other centralities, but the nonjet quadrupole measured by  $v_2^2\{{\rm 2D}\}$ then contributes substantially. In general, attempts to measure a $v_3$ (or higher multipole) component in 1D projections onto azimuth of 2D angular correlations can be anticipated accurately from the properties of the SS 2D peak determined by this analysis. 

The 2D structure of angular correlations imposes strong constraints on proposed data models. For the $p_t$-integral data presented in this analysis the SS 2D Gaussian is the most efficient description of same-side angular correlations (requiring the fewest independent parameters). 
Any Fourier-series representation of the SS 2D peak (or hybrid representation including both 2D Gaussian and Fourier components) would be less efficient, requiring more parameters. 
Whereas the nonjet quadrupole measured by $v_2^2\{{\rm 2D}\}$ and obtained from fits to 2D histograms has negligible curvature (is approximately uniform) on $\eta_\Delta$ within the STAR TPC acceptance any multipoles associated with the SS 2D peak must all exhibit the same large curvature corresponding to the $\eta_\Delta$ dependence of the SS 2D peak. 

%
Since 2D angular correlations projected onto 1D azimuth implicitly include a $v_3$ Fourier component, among others, {\em as part of the SS 2D Gaussian}, the SS 2D peak contribution to $v_3$ and any other inferred Fourier component must be acknowledged before any claim of an independent $v_m$ component 
is made. 
%
%
Any higher-$v_m$  conjecture should be presented in the context of the full 2D histograms, not just 1D projections onto azimuth. 
Interpretation of 1D Fourier components of the SS 2D peak as flow manifestations competes with jet production by fragmentation as an alternative mechanism. 
From this analysis we conclude that a sextupole Fourier component $v_3$ is not required by the 2D data.


Although the $\eta$ interval external to the STAR TPC acceptance is obviously excluded from this analysis, any correlation structure there should be considered in relation to 2D model fits within the TPC acceptance as well as 1D Fourier series. 


The observation of anomalous centrality trends reported in this paper applies to the SS 2D peak in the correlation data, as represented by the {\em required} model elements defined in Eq.~(\ref{Eq4}). For alternative model representations, for example those which invoke higher azimuth multipoles, the relevant parameter {\em combinations} remain those that describe the amplitude and widths of the SS 2D peak in the data, regardless of how that peak structure may be represented mathematically. See App.~\ref{multi} for further discussion.

\subsection{Centrality trends for {\sc hijing}} \label{hijsec}

The {\sc hijing} Monte Carlo with jet quenching off is nominally a linear superposition of N-N collisions modeled by {\sc pythia} within the context of a Glauber model of A-A collisions. {\sc hijing} should therefore provide a GLS reference for Au-Au collisions but fails to do so. In this subsection we discuss the discrepancy.

Figure~\ref{hijfig} (first panel) shows SS 2D peak amplitude $A_1$ for 
default {\sc hijing} v1.382~\cite{hijing} 
simulations of 200 GeV Au-Au collisions (open points) compared to 200 GeV data (solid points) from Fig.~\ref{Figure3} (first panel). {\sc hijing} is specifically formulated to describe minijets~\cite{hijing}. If hard scattering is disabled in \mbox{{\sc hijing}} the SS 2D peak disappears ($A_1 \rightarrow 0$). The GLS (N-N) extrapolation from {\it p-p} data [dashed curve, see Eq.~(\ref{Eq5})] is $A_1 = 0.065\nu / [1 + 0.02(\nu - 1)]$. 
The dotted curve (explained in the text below) is $0.05\nu /[1 + 0.16(\nu - 1)]$.
Apparently, {\sc hijing} deviates strongly from the GLS reference extrapolated from \mbox{{\it p-p}} data, whereas we expect {\sc hijing} (no jet quenching) to represent an equivalent \mbox{N-N} linear superposition. The source of the discrepancy is determined as follows. 

Figure~\ref{hijfig} (second panel) shows single-particle yields from {\sc hijing} (points) and those expected from a GLS superposition of N-N in Au-Au collisions inferred from {\it p-p} spectra~\cite{ppspectra}. The {\it p-p} extrapolated (GLS) per-participant-pair multiplicity in K-N two-component form~\cite{kn} is $2.5 [1 + 0.02(\nu - 1)]$. The {\sc hijing} equivalent is  $2.1 [1 + 0.16(\nu - 1)]$. We find a large difference in hard-component coefficient $x$: 0.02 vs 0.16. Also, {\sc hijing} soft component $S_{\rm NN}$ is 20\% lower than {\it p-p} data. The K-N trend for more-central Au-Au data  is $2.5 [1 + 0.1(\nu - 1)]$. 

Figure~\ref{hijfig} (third panel) shows the hard-component (jet-related) hadron angular density per N-N binary collision based on {\it p-p} data, inferred from more-central \mbox{Au-Au} data and extracted from {\sc hijing} (open points) in the second panel. Again we find that the {\sc hijing} hard-component yield per N-N binary collision is much larger than what is observed in {\it p-p} collisions and even larger than observed in central Au-Au collisions. 


Part of the explanation for the {\sc hijing} discrepancy comes from the underlying parton spectrum. The default parton spectrum cutoff (lower bound) in {\sc hijing} is $p_0 = 2$ GeV/$c$. The spectrum cutoff inferred from 200 GeV {\it p-p} spectrum data is 3 GeV/$c$~\cite{ppspectra,fragevo}. That difference has substantial consequences. The parton spectrum varies approximately as $p_t^{-6}$ near 3 GeV/$c$. A shift in the lower bound by factor 2/3 leads to an increased hard-scattering cross section per N-N collision by factor $(3/2)^6 \approx 10$. We observe in Fig.~\ref{hijfig} (third panel) that the {\sc hijing} hard-component single-particle yield is a factor 7 larger than that inferred from {\it p-p} data, consistent with the difference in effective spectrum cutoffs. Another part of the {\sc hijing} explanation comes from correlated-pair yields.

Figure~\ref{hijfig} (fourth panel) shows the 2D density of jet-correlated pairs per N-N binary collision $(1/N_{\rm bin}) \rho_0(b) A_1$, with $\rho_0(b) A_1 = \Delta \rho[\text{SS peak}]$ expressed in terms of an absolute pair-number density and $\rho_0(b) = dN_{\rm ch} /2\pi d\eta \approx \sqrt{\rho_{\rm ref}}$, the single-particle density. What is plotted then is a jet-related pair-number density per N-N binary collision. We find that the jet-correlated pair density from {\sc hijing} is about 2/3 that observed in {\it p-p} collisions, not seven times larger. From the pair densities in this plot we calculate coefficients $X_{pp}$ for reference curves in the first panel: $0.05 = 2\pi \times 0.017 / 2.1$ ({\sc hijing}) and $0.065 = 2\pi \times 0.026 / 2.5$ (GLS trend from {\it p-p} data).


We conclude that whereas {\sc hijing} binary-collision-scaled single-particle yields are much greater than those for \mbox{{\it p-p}} (or even Au-Au) data, consistent with its default parton spectrum cutoff, the corresponding number of jet-correlated particle pairs is substantially lower than for \mbox{{\it p-p}} data.
Spectrum hard components and jet angular correlations extracted from  {\em measured} \mbox{{\it p-p}} data are mutually consistent with measured parton fragmentation systematics and a pQCD parton spectrum with cutoff near 3 GeV~\cite{ppspectra,fragevo,jetyield}. When combined they accurately predict the centrality trend for more-peripheral Au-Au collisions (GLS extrapolation of {\it p-p} data). {\sc hijing} spectrum and correlations hard components are mutually consistent only if a very different parton fragmentation scheme is associated with the default 2 GeV parton spectrum cutoff. The combination does not predict the centrality trend for more-peripheral Au-Au collisions (deviates strongly from GLS extrapolation of {\it p-p} data).

Because of the parton spectrum power-law trend most scattered partons appear near the spectrum lower bound, which for default {\sc hijing} is 2 GeV. The parton fragmentation process is poorly known there 
(see Ref.~\cite{fragevo} for absence of accurate fragmentation function data below about 7 GeV parton energy). 
Fragmentation to charged-hadron pairs which could contribute to the SS jet peak, as modeled by {\sc hijing}, appears to be substantially less than for real N-N collisions. The apparent agreement between {\sc hijing} and minijet correlation data near $\nu = 1$ seems to be an accident: {\sc hijing} underestimates both \mbox{{\it p-p}} spectrum soft-component $S_{\rm NN}$ and the number of jet-correlated pairs, the errors nearly canceling in the ratio. With increasing Au-Au centrality  excess hard-component single-particle production in {\sc hijing} combines with too few jet-correlated pairs to produce the observed large deviations from the GLS trend extrapolated from {\it p-p} data.
 Implementing the {\sc hijing} jet quenching mode causes reduction of the same-side 2D peak amplitude with increasing centrality, further deviating from the \mbox{Au-Au} data.


\subsection{Centrality trends for other conventional models} \label{other}


We compare centrality trends of Au-Au angular correlations from the present analysis to those expected for three generic models of nucleus-nucleus collisions: a) Glauber linear superposition of N-N collisions, b) parton/hadron rescattering in a dissipative medium, and c) a locally-thermalized ``opaque'' medium.

\paragraph{Linear superposition of N-N collisions}

$-$ If \mbox{Au-Au} collisions were simply linear superpositions of \mbox{N-N} collisions (A-A transparency) then parton scattering and jet production in N-N collisions should be randomly superposed in A-A momentum space according to the Glauber-model number of binary collisions. The same-side 2D peak widths should remain constant with centrality while the amplitude should increase according to Eq.~(\ref{Eq5}). The away-side ridge should remain fixed in shape and follow the amplitude of the same-side peak.
The shape of the single-particle spectrum hard component should also remain unchanged, and its amplitude should follow binary-collision scaling~\cite{TomAuAuspectra}. The GLS model is in fact our baseline reference for the centrality evolution of A-A angular correlations and single-particle spectra. It describes minijet correlations and spectra well for $\nu < \nu_\text{trans}$, but fails dramatically above the transition and does not describe the azimuth quadrupole. It is worth noting that the quadrupole amplitude reaches at least half its maximum value within the centrality range where N-N linear superposition accurately describes the SS 2D peak and dipole (Fig.~\ref{Figure3}).

\paragraph{Parton/hadron rescattering in a medium}

$-$ Parton and/or hadron rescattering in a dissipative medium (e.g.\ as in cascade or transport models) implies that a primary scattered parton and its hadron (jet) fragments are randomly deflected in angle and lose energy/momentum to the medium, as in Brownian motion~\cite{brown}. 
The $\eta$ and $\phi$ widths of hadron-number and $p_t$ same-side 2D peaks should {\em both} increase. The same-side 2D peak amplitude for number correlations may be reduced compared to the binary-collision reference. The $p_t$-correlations peak amplitude should definitely decrease, and the spectrum hard component should be shifted to lower momentum and possibly reduced in amplitude (parton dissipation, ``jet quenching''). The data reveal that the $\eta$ width of the SS 2D peak increases, but the azimuth width decreases with increasing centrality, inconsistent with random multiple scattering. A reduction in the $p_t$ same-side peak amplitude is observed in data~\cite{ptscale}, but not until $\nu > 4$, well above $\nu_\text{trans}$. And the same-side peak amplitude for number-correlations {\em increases dramatically} above $\nu_\text{trans}$.

Calculations using the multi-phase transport Monte Carlo model {\sc ampt}~\cite{ampt} for trigger-associated particle distributions on $\eta_\Delta$ and
$\phi_\Delta$ from central Pb-Pb collisions at $\sqrt{s_{NN}}$ = 2.76 TeV and central Au-Au collisions at $\sqrt{s_{NN}}$ = 200 GeV were presented in
Refs.~\cite{xuko} and~\cite{mawang} respectively. Both papers include ``string melting,'' or a parton cascade, as well as fluctuating initial conditions and pQCD jets described by {\sc hijing}. With Lund-model~\cite{lund} fragmentation parameters and parton-cascade cross sections adjusted to fit $v_2$ values at mid-centrality, the model predicts an $\eta$-broadened SS 2D peak and a broad double-peaked AS 1D structure on azimuth. The authors attribute those structures to anisotropic flow from fluctuating initial conditions (initial-state densities) and medium response to jets but do not quantitatively distinguish the two sources. The predicted double-peaked AS structure is not observed in the present data, and neither is the predicted double-hump (on $\eta_\Delta$) SS peak in Ref.~\cite{mawang}. {\sc ampt} predictions for the measurement conditions reported in this analysis are not available.

\paragraph{Thermalized opaque medium}  

$-$ In some models of A-A collisions copious initial-state low-energy scattered partons are expected to contribute to formation of a locally-thermalized opaque medium with zero mean free path~\cite{kll,mjmuel,nayak,mjshin}. In those models low-energy scattered parton thermalization is assumed to occur almost immediately after impact ($<1$~fm/c), and the scattered-parton energy is converted into thermal energy within 1-4 fm/c~\cite{nayak,mjshin} which raises the local temperature (``hot spots''), while the parton momentum, being conserved
overall, is dissipated to many final-state hadrons whose angular
distributions are broadened. With increasing centrality same-side number and $p_t$ angular correlations from low-energy partons (minijets) should be broadened and greatly reduced in amplitude relative to the GLS reference, as estimated in~\cite{jethydro}. The opposite trends are observed for the SS 2D peak amplitude and azimuth width while the $\eta_\Delta$ broadening reported here far exceeds that expected from early-stage minijet interactions~\cite{nayak,mjshin,jethydro} and jet quenching~\cite{hijing}.

Predictions for per-pair ($\Delta\rho/\rho_{\rm ref}$) angular
correlations from Au-Au collisions at $\sqrt{s_{NN}}$ = 200 GeV for charged particles with 0.2 $\leq p_t \leq$ 2.0 GeV/c from the non-viscous 3+1 hydrodynamics event-wise model {\sc nexspherio}~\cite{nexsph} for four centralities from a 0-80\% cross section fraction are presented in Ref.~\cite{sharma}. The {\sc nexspherio} model assumes that initial conditions are described by the {\sc nexus} event generator~\cite{werner} which includes event-wise initial-state density fluctuations and soft, semi-hard and hard scattering processes. The generated 2D angular correlations reveal a SS 2D peak and AS 1D ridge, but the physical origin of those structures is not identified~\cite{sharma}. Dramatic increases in the SS 2D peak amplitude and $\eta_\Delta$ width in more-central Au-Au collisions reported in the present analysis are not predicted by the model. Instead, {\sc nexspherio} predicts that the $\eta_\Delta$ width should decrease. Differential study and identification of physical mechanisms for correlation structures predicted by the {\sc nexspherio} model are anticipated.




\section{Summary and Conclusions}
\label{Sec:Summ}

We have measured charged-particle number angular correlations from Au-Au collisions at $\sqrt{s_{\rm NN}}$ = 62 and 200~GeV projected onto relative azimuth $\phi_{\Delta}$ and relative pseudorapidity $\eta_{\Delta}$ for eleven centrality bins on 0-95\% of the total Au-Au cross section. The dominant features are a same-side 2D peak (approximately Gaussian), an away-side $\cos(\phi_{\Delta}-\pi)$ dipole and a $\cos(2\phi_{\Delta})$ quadrupole. The same-side 2D peak and away-side dipole in more-peripheral A-A collisions, first observed as correlation structures in minimum-bias {\it p-p} collisions and directly related to features of {\it p-p} single-particle spectra, can be reasonably interpreted as corresponding to {\em minijets} from minimum-bias parton scattering and fragmentation. 

{\it p-p} spectrum and correlation data combined with assumed Glauber-model linear superposition (binary-collision scaling) provide an essential baseline reference for heavy-ion collisions. The amplitude of the same-side 2D peak in Au-Au collisions approximately follows \mbox{N-N} binary-collision scaling from peripheral to mid-central collisions and the widths are slowly varying, extending the {\it p-p} minijet hypothesis across the centrality range $\nu \leq \nu_{\rm trans}$ in heavy ion collisions.

It is conventional practice in analysis of RHIC data to omit a substantial fraction of the complete A-A centrality range from observation (i.e., more-peripheral collisions). The essential GLS reference is then inaccessible, and it is not clear what is unique about \mbox{A-A} collisions relative to p-p or N-N collisions, or where exceptional behavior actually appears on centrality.  The present analysis establishes for the first time that jet-related correlation data from Au-Au collisions actually closely follow the GLS trend (binary-collision scaling) up to mid-centrality.  Strong deviations of 2D angular correlations from the GLS reference in more-central collisions are quantitatively determined relative to the GLS reference. Such information provides a more differential (hence rigorous) test for any proposed theoretical model.

At an  intermediate centrality denoted by $\nu_\text{trans} \approx 3$ the same-side 2D peak amplitude and pseudorapidity width $\sigma_{\eta_{\Delta}}$ and away-side 1D peak amplitude transition to a qualitatively new centrality trend. The slopes of the trends on centrality measure $\nu$ increase within one centrality bin by factors 3.5 to 5. The transitions  for 62 and 200~GeV Au-Au data are located at similar centralities as measured by mean participant path length $\nu$ employed as an A-A geometry parameter. {The underlying mechanism for the transition and its relation to other collision parameters is presently under study.} 
Within the minijet context the large increase in jet-like correlations in more-central Au-Au collisions is inconsistent with expectations of strong minijet quenching as described in relativistic Boltzmann transport and diffusion theories.




Angular correlations from the present analysis combined with pQCD predictions of mean jet number per \mbox{A-A} collision were used to estimate jet fragment yields which agree quantitatively with pQCD calculations of parton fragment yields and with measured spectrum hard-component yields,
implying that about one third of the hadronic final state in 200 GeV central \mbox{Au-Au} collisions is contained in resolved jets.


A scaling analysis of event-wise mean-$p_t$ fluctuations revealed that the same-side 2D peak structure in $p_t$ (rather than number) angular correlations increases smoothly with centrality (i.e. no sharp change in slope)
from below $\nu_{\rm trans}$ to well above it. In the minijet context the
$p_t$ angular correlations trend suggests that the SS 2D number-correlation
peak, although strongly modified above the transition, is still initiated
by semihard parton scattering which increases as least as fast as the
binary-collision trend. A successful theoretical description of the
correlation structures presented here must not only account for the
rapid slope and magnitude changes in the SS 2D peak amplitude and
$\eta_\Delta$ width, but also the relatively smooth evolution of the 2D peak
azimuth width and same-side $p_t$ angular correlations above and below
$\nu_{\rm trans}$.



The A-A collision-energy dependence of the SS 2D peak amplitude from the present number-correlation analysis is consistent with that inferred for the corresponding structure in $p_t$ angular correlations and with the energy dependence of the nonjet azimuth quadrupole (as measured by equivalent statistical quantities). All follow the $\log(\sqrt{s_{\rm NN}})$ trend that might be expected for a QCD scattering process.



Two-dimensional  histograms were examined for {\em independent} ``higher harmonics'' (multipoles) represented by $v_m$ with $m \geq 3$. Relative to the fit model of Eq.~(\ref{Eq4}) no such structure was observed to the statistical limits of the data. 
The same-side 2D peak can be decomposed into a Fourier series of azimuth multipoles with a common {\em large curvature} on pseudorapidity. $v_m$ values inferred from 1D analysis of projected 2D angular correlations can therefore be predicted from the fitted same-side 2D Gaussian parameters. 
{Fits including an additional sextupole model element for the more-central collision data were shown to be equivalent to an alternative model of the same-side 2D peak consisting of a 2D Gaussian (with modified parameters) plus an additional 1D Gaussian on azimuth.  The sum of those two model elements is systematically equivalent to the fitted single same-side 2D Gaussian.
}
Addition of higher multipoles to the 2D model of  Eq.~(\ref{Eq4}) results in model redundancy and ambiguous fit results.

Results from the present analysis were compared with the expected trends from three scenarios for nuclear collisions at RHIC :  (i) Glauber linear superposition of N-N collisions (the A-A baseline reference), (ii) a locally-equilibrated opaque medium with zero mean free path, and (iii) an intermediate scenario where partons and their hadron fragments randomly scatter in a medium. 
The present analysis reveals substantial deviations from (i). 
The large same-side peak amplitude increase and azimuth width decrease with centrality apparent in  Figs.~\ref{Figure1} and~\ref{Figure3}, coupled with the smoothly increasing mean-$p_t$ angular correlation amplitude for  $\nu > \nu_{\rm trans}$ and corresponding $p_t$ spectrum hard-component centrality evolution above the transition,
represent the essential experimental constraints summarized here. 

We are unable to reconcile an interpretation in which RHIC
nuclear collisions are dominated by minijet structure and
the observations reported here are described by pQCD with
modified fragmentation, with that of a strongly-absorptive,
even opaque (zero viscosity), collision system as described
in scenarios (ii) and (iii). 
If the collision system turns out to be effectively opaque to
few-GeV partons the present observations would be inconsistent 
with the minijet picture presented here.

\vspace{0.1in}

We thank the RHIC Operations Group and RCF at BNL, the NERSC Center at LBNL and the Open Science Grid consortium for providing resources and support. This work was supported in part by the Offices of NP and HEP within the U.S.\ DOE Office of Science, the U.S.\ NSF, the Sloan Foundation, the DFG cluster of excellence `Origin and Structure of the Universe' of Germany, CNRS/IN2P3, FAPESP CNPq of Brazil, Ministry of Ed.\ and Sci.\ of the Russian Federation, NNSFC, CAS, MoST, and MoE of China, GA and MSMT of the Czech Republic, FOM and NWO of the Netherlands, DAE, DST, and CSIR of India, Polish Ministry of Sci.\ and Higher Ed., Korea Research Foundation, Ministry of Sci., Ed.\ and Sports of the Rep.\ Of Croatia, and RosAtom of Russia.

\begin{appendix}

\section{Minijets} \label{minijets}


Minijets have an experimental and theoretical history over more than twenty years~\cite{ua1,sarc,kll}. 
A growing number of experimental results indicate that minijets make significant contributions to the transverse dynamics of nuclear collisions above $\sqrt{s_{\rm NN}} \sim$ 13 GeV~\cite{axialci,fragevo,ptedep,jetyield,TomAuAuspectra,ptscale}. However, the phenomenological definition of minijets and their interpretation as true pQCD jets is not widely recognized within the heavy ion community. We review the experimental properties of minijets, their relation to QCD theory and their manifestations in combinatoric correlation analysis as in the present study.




\subsection{Minijets and calorimeter experiments}

The minijet concept emerged experimentally at the  S{\it p\=p}\,S from a UA1 analysis of $E_t$ structure with an event-wise cone jet finder down to exceptionally small integrated $E_t$ (5 GeV)~\cite{ua1}.  The resulting $E_t$ ``clusters'' were compared with pQCD predictions to test how low in energy a pQCD jet description is applicable to $E_t$ structure~\cite{sarc,kll}.

The UA1 analysis determined that $E_t$ clusters follow an approximate pQCD power-law parton spectrum down to 5 GeV. Azimuth correlations {\em between} clusters exhibit a peak at $\pi$ radians expected for back-to-back parton scattering. The 5 GeV cutoff in calorimeter data was related to a 3-4 GeV parton energy equivalent, the difference contributed by the underlying event~\cite{sarc}. UA1 concluded: ``...one can usefully define jets down to $E_{t,{\rm min}} \simeq 5$ GeV [3-4 GeV parton energy]. ... The agreement of the inclusive [minijet] cross section with [p]QCD over several orders of magnitude is quite remarkable''~\cite{ua1}.

More recently, the UA1 calorimeter analysis was repeated by the STAR collaboration, confirming a jet spectrum consistent with NLO QCD predictions for event-wise reconstructed jets down to 5 GeV energy (background corrected to 3-4 GeV)~\cite{starjets}. The same pQCD parton spectrum was used to describe spectrum hard components as fragment distributions in {\it p-p} and Au-Au collisions~\cite{fragevo}.

\subsection{Minijets and theory}

In Ref.~\cite{kll} minijet production was considered in a pQCD context for anticipated RHIC U-U collisions based on UA1 observations. ``The observed [UA1 minijet] rate is in agreement with [p]QCD and is quite large.'' Applicability of pQCD to minijets (low-$p_t$ jets, $E_t \sim$ few GeV) was studied in detail down to parton $p_{t}^{\rm min} = 3$ GeV/c in Ref.~\cite{sarc}. ``...a theoretical cutoff of $p_t^{\rm min} \sim 3$ GeV seems to describe the observed total minijet cross section with $E_T^{\rm jet} (E_T^{\rm raw}) \geq 5$ GeV [3 GeV parton energy].'' Minijets and cross sections in {\it p-p} and {\it p-\=p} collisions were also considered in~\cite{durand}. 

The minijet-based Monte Carlo {\sc hijing} was developed specifically to study the role of minijets in {\it p-p} and \mbox{A-A} collisions. The parton spectrum is given a lower cutoff $p_0$ with default value 2 GeV/c. {\sc hijing} predictions are compared to {\it p-p} collision data in~\cite{hijingpp,minijet} where they quantitatively match measured minijet correlations (same-side amplitude, widths, away-side ridge)~\cite{jeffpp1,aspect}. {\sc hijing} with ``jet quenching'' disabled applied to A-A collisions is equivalent to a Glauber linear-superposition reference [Eq.~(\ref{Eq5})]  as shown by the dash-dotted curve in the left-most panel of Fig.~\ref{hijfig}.
{\sc hijing} with hard parton scattering disabled shows no minijet correlations.


In a recent study of pQCD applied to Au-Au collisions, measured $p_t$ spectrum hard components identified with minijets were described quantitatively by pQCD calculations~\cite{fragevo}. The pQCD predictions are compatible with measured minijet transverse rapidity correlations on $(y_{t1},y_{t2})$~\cite{aspect}, especially a parton fragment spectrum mode at $p_t = 1$ GeV/c in both spectra and correlations. The direct comparison between pQCD and spectrum data confirms a 3 GeV/$c$ parton spectrum cutoff.

Theoretical descriptions widely assume that minijets are rapidly thermalized in RHIC collisions, contributing to QGP formation. For example,  ``minijets...will be reprocessed by the system and not emerge from it''~\cite{kll}. Thermalization by transport processes is studied in~\cite{nayak,cooper}. Thermalization time is estimated as 4-5 fm/c with $T \sim 200$ MeV~\cite{nayak}. 
Within those models minijet angular correlations should be strongly suppressed in amplitude, broadened on both $\eta$ and $\phi$, and the scattered parton $p_t$ strongly dissipated among final-state hadrons. The correlation structures reported here and in Refs.~\cite{axialci,ptedep,jeffpp1,aspect,daugherity,ptscale}, if generated by semihard parton scattering and fragmentation, contradict the assumption of minijet thermalization.


\subsection{Minijets and combinatoric analysis}

Minijet structure has also been observed directly in minimum-bias two-particle angular correlations from 200 GeV \mbox{{\it p-p}} collisions~\cite{jeffpp1,aspect}. Angular correlations with no ``jet'' minimum-$p_t$ requirement exhibit just the structure expected from pQCD jets: a narrow {\em intrajet} same-side 2D peak at the angular origin (parton fragmentation) with most-probable $p_t \sim 1$ GeV/c and an {\em interjet} away-side ridge at $\pi$ radians (back-to-back parton scattering). 

The intrajet 2D peak is interpreted to reflect the angular consequences of a parton fragment momentum distribution along the jet axis with mode near 1 GeV/c~\cite{ffprd,fragevo} and a soft momentum spectrum transverse to the jet axis~\cite{aleph}. The resulting 2D ``jet cone'' typically falls within one radian, with single-particle r.m.s.\ width $\sim 0.5$. The interjet 1D peak on azimuth is uniform over a large interval on $\eta_\Delta$ because the parton-parton CM is broadly distributed relative to the nucleon-nucleon CM. The away-side azimuth width is determined by a combination of the  intrajet width $\sim 0.5$ and acoplanarity due to parton intrinsic $k_t$ in the projectile nucleons~\cite{jeffpp1,aspect}. The observed systematics and angular correlations are qualitatively consistent with the {\sc pythia} Monte Carlo~\cite{pythia}.



Preliminary results from the present analysis~\cite{daugherity} were used to derive absolute spectrum hard-component yields from jet-like angular correlations~\cite{jetyield}. That analysis combined pQCD estimates of jet number in Au-Au collisions with factorization of jet-correlated pair numbers to obtain parton fragment yields. The results are in excellent agreement with fragment yields inferred from a two-component analysis of identified-hadron spectra~\cite{TomAuAuspectra}. 

A combination of several related analyses has established a quantitative connection among spectrum hard components, pQCD-predicted fragment distributions and jet-like correlations, providing strong support for a minijet interpretation. The combined analysis~\cite{jetyield} demonstrates that {\em one third} of the hadronic final state in central 200 GeV Au-Au collisions lies within resolved jet correlations.





\section{2D Fit-model ambiguities} \label{multi}

In Sec.~\ref{v3} we emphasized multipoles inferred by projecting the entire SS 2D peak onto 1D azimuth and evaluating the SS peak Fourier components. Those multipole amplitudes may correspond to ``nonflow'' bias in $v_m$ data inferred from nongraphical numerical methods~\cite{multipoles}. In this Appendix we address a more subtle problem: what is the consequence if a sextupole ($v_3$) or higher multipole element is added to the standard 2D fit model used for the present analysis. Is the extended model superior to the original? Is more information extracted from the data? To answer such questions we must distinguish between necessary and sufficient conditions, lower bounds vs upper bounds on fit parameters and algebraic equivalence between seemingly different fit models.

\subsection{Jet-related vs nonjet structure}

We first establish a distinction between ``nonjet'' and ``jet-related'' structure in 2D angular correlations. Based on correlation systematics below the sharp transition at $\nu_{trans}$ we interpret two elements of the data model in terms of (mini)jets: the SS 2D peak and the AS dipole. Whether a jet interpretation for those structures is appropriate in more-central Au-Au collisions is an open question. To provide consistent terminology for this analysis we refer to those two model elements as ``jet-related'' structure for all collision centralities. All complementary structure is by definition ``nonjet,'' including the soft-component 1D Gaussian on $\eta_\Delta$ and the BEC-electron-pair peak, but also the nonjet quadrupole component $A_Q$ which is {\em observed} to have negligible curvature on $\eta_\Delta$ within the TPC angular acceptance.  

In describing multipole elements $v_m$ we distinguish between (a)  $v_m$ obtained from 1D Fourier fits to all angular correlations projected onto 1D $\phi_\Delta$ and (b) $v_m$ obtained from model fits to 2D angular correlations which may include one or more multipole terms $A_X$ (letters $X = D,\, Q,\, S,\, O$ denote azmuth multipoles: dipole, quadrupole, sextupole, octupole, with pole number $2m$). Multipoles in case (a) must mix jet-related and nonjet structure as discussed in Sec.~\ref{v3}.  Multipoles in case (b) may represent jet-related and/or nonjet contributions, as discussed in this Appendix.

Given those definitions and no significant sextupole amplitude in fit residuals from our standard 2D model fits we conclude no {\em detectable} nonjet ($\eta_\Delta$-uniform) $v_3$ structure in the data. A test for ``false negatives'' can be performed by adding a known sextupole component to the data and refitting with the standard 2D model. We then observe a sextupole component in the residuals with about 50\% of the input amplitude. We therefore conclude from the data that the upper limit on any nonjet sextupole is about  10\% of the SS 2D peak sextupole component, which is negligible compared to other correlation structure. Thus, the {\em only source} for any inferred sextupole component is the jet-related SS 2D peak, as noted in Sec.~\ref{v3}.  We now consider the significance of possible jet-related sextupole amplitudes in 2D data histograms inferred by adding a $\cos(3\phi_\Delta)$ term to the standard 2D data model.



\subsection{SS 2D peak models vs peak properties}


It is important to distinguish between {\em model-independent} peak properties and peak-related model parameters. Model parameters should be accurately constrained by peak properties. But parameters from some models may be {\em poorly constrained} by data obtained within a {\em limited acceptance}.  
In more-peripheral collisions the entire SS 2D peak is resolved within the STAR TPC angular acceptance. The peak model is then unambiguous: a 2D Gaussian. The SS peak continues to be well resolved until well above the sharp transition in peak properties at $\nu_{trans}$. However, in more-central collisions the SS 2D peak extends sufficiently far outside the TPC $\eta$ acceptance that its modeling on $\eta_\Delta$ may become ambiguous.  When a fraction of the peak is observed some peak properties (e.g., amplitude and curvature at the mode) are more reliably determined than others (e.g., higher moments, tail structure). Model parameters relating to the latter may suffer from large systematic uncertainties.

The curvature of a peak near its mode is a well-defined algebraic quantity obtained directly from data as the second derivative (second difference) of histogram data. The curvature of a Gaussian function $A\exp(-x^2/2\sigma^2)$ is $A/\sigma^2$.  For a single Gaussian fitted to a single 1D peak the Gaussian amplitude and width parameters are directly and accurately related to the peak amplitude and curvature at the mode, the two most well-determined peak properties in all cases. Introduction of additional model parameters may result in large systematic uncertainties if corresponding peak properties (e.g., higher-order moments) are not well-determined by the data.


\subsection{Model equivalence: sextupole vs 1D Gaussian} \label{equiv}

The model elements $A_D$ (dipole) and $A_Q$ (quadrupole) already present in the standard 2D data model are in effect parts of a (truncated) Fourier series. Adding an $A_S$ (sextupole) term to the data model extends the truncated Fourier series. The next term $A_O$ (octupole) is typically at the level of statistical fluctuations~\cite{multipoles,triflow}. The separate sinusoids may serve both as representatives of distinct (nonjet) data multipoles and as parts of a Fourier series representing a localized (possibly jet-related peak) structure on azimuth~\cite{tzyam,multipoles}.

If a sextupole component uniform on $\eta_\Delta$ is added to the standard 2D fit model nonzero sextupole amplitudes are indeed inferred from model fits to more-central Au-Au data, and some other model parameters are shifted substantially from their original values. We demonstrate below that addition of the $A_S$ term is equivalent to modifying the SS 2D peak model by adding a SS 1D Gaussian narrow on $\phi_\Delta$ and uniform on $\eta_\Delta$. An illustration is provided in Table~\ref{params} for 9-18\% central 200 GeV data. The argument follows those presented in Refs.~\cite{tzyam,multipoles} and in text relating to Fig.~\ref{vv3} that equate a truncated Fourier series to a periodic peak array.  Other mid-central data give similar results. Fit parameters that do not change significantly and are not relevant to this discussion are omitted. The parameter labels are as defined in Eq.~(\ref{Eq4}).

\begin{table}[h]
  \caption{Model parameters for three fit models: (a)  standard model for this analysis, (b) standard model plus additional sextupole term $A_S$ and (c)  standard model plus additional SS 1D Gaussian on $\phi_\Delta$ with amplitude $A_{1D}$ as part of the SS 2D peak model. The fit parameters are as defined in Eq.~(\ref{Eq4}). Uncertainties in the second column illustrate typical fit uncertainties (statistical plus systematic) for each parameter. $A_X$ denotes the additional model parameter $A_S$ or $A_{1D}$.}
  \label{params}
\begin{center}
\begin{tabular}{|c|c|c|c|} \hline
 parameter &standard & std + $A_S$ & std + $A_{1D}$   \\ \hline
$A_{1}$  & $0.76 \pm 0.04$ & 0.51 & 0.47  \\ \hline
 $\sigma_{\eta_\Delta}$ & 2.3$\pm 0.3$ & 1.78 & 1.72  \\ \hline
$A_Q$ &0.18$\pm 0.008$ & 0.23 & 0.18   \\ \hline
$A_D$ &0.29$\pm 0.02$ & 0.175 & 0.28   \\ \hline
$A_X$ & -- & 0.014 & 0.29   \\ \hline
\end{tabular}
\end{center}
\end{table}

The second column (standard) shows fit results obtained with the standard 2D model. The third column (std + $A_S$) shows fit results when a sextupole term $A_S \cos(3 \phi_\Delta)$ is added to the 2D model and all else remains the same. Proof that addition of a sextupole term to the standard 2D model is equivalent to adding a 1D Gaussian on azimuth to the SS 2D peak model  proceeds as follows. 

\begin{table}[h]
  \caption{Comparison of {\em changes} in multipole amplitudes when an additional sextupole element is added to the standard model with Fourier coefficients of a SS 1D Gaussian on azimuth. The comparison demonstrates an equivalence between the changes in multipole amplitudes (columns 2 and 3 of Table~\ref{params}) and the Fourier coefficients of a 1D Gaussian on azimuth (column 5 of this Table). Column 5 is obtained by multiplying Fourier coefficients $F_m$ by the value $A_{1D} = 0.25$ that best matches the third column. That value also corresponds to the difference $\Delta A_1 = 0.76-0.51$ from Table~\ref{params}.}
  \label{fourierr}
\begin{center}
\begin{tabular}{|c|c|c|c|c|} \hline
 $m$ &parameter & data & $F_m$ & $F_m\, A_{1D}$   \\ \hline
1 &$-\Delta A_D$ &0.115 & 0.44 & 0.11   \\ \hline
2 & $\Delta A_Q$ &0.05 & 0.21 & 0.052   \\ \hline
3 & $\Delta A_S$ & 0.014 & 0.062 & 0.0155   \\ \hline
\end{tabular}
\end{center}
\end{table}

Table~\ref{fourierr} (third column, data) shows the {\em changes} in fitted multipole amplitudes between the standard model and that including the sextupole term. The third column ($F_m$ defined in Sec.~\ref{v3}) shows the calculated Fourier coefficients for a unit-amplitude 1D Gaussian on azimuth with width $\sigma_{\phi_\Delta} = 0.7$~\cite{multipoles}. The width value was adjusted to achieve the best match between columns 3 and 5. The fifth column ($F_m A_{1D}$) shows the Fourier coefficients for a SS 1D Gaussian with amplitude $A_{1D} = 0.25$. Comparison of the third and fifth columns reveals equivalence within small data uncertainties. 
The inferred amplitude $A_{1D}$ of the added SS 1D Gaussian on azimuth corresponds to the difference $\Delta A_1$ between SS 2D peak amplitudes for the two fit models. And the inferred width corresponds to the azimuth width of the SS 2D peak in the standard fit model. Thus, the parameter changes resulting from inclusion of a sextupole element in the 2D model are equivalent to adding a constant offset to the $\eta_\Delta$ factor of the SS 2D peak model.

Table~\ref{params} (fourth column) confirms that equivalence by presenting the result of a 2D fit replacing the added sextupole element by an offset in the $\eta_\Delta$ factor of the SS 2D peak model [$A_1\exp(-\eta_\Delta^2 / 2 \sigma^2_{\eta_\Delta}) \rightarrow A'_1\exp(-\eta_\Delta^2 / 2 \sigma^{\prime2}_{\eta_\Delta}) + A_{1D}$]. The added constant in the $\eta_\Delta$ factor then represents a SS 1D Gaussian on $\phi_\Delta$ in the 2D model [see Eq.~(\ref{Eq4})]. 
The fit results are systematically consistent with the previous exercise: the 1D Gaussian amplitude is $A_{1D} = 0.29$ compared to 0.25, and the modified 2D amplitude $A'_1 = 0.47$ is consistent with $A_{1D}$ (i.e., $A'_1 + A_{1D} = A_1 = 0.76$).  Modification of $\eta$ widths in the revised models is discussed below. Note that $A_D$ and $A_Q$ {\em have reverted to values for the standard fit model}, since the added 1D Gaussian on azimuth is modeled explicitly as a Gaussian rather than as a truncated Fourier series.
Equivalence of a 1D periodic peak array and a truncated Fourier series on azimuth is discussed in Refs.~\cite{tzyam,multipoles}.

\subsection{Model ambiguities and parameter significance}

We now return to the issue of SS 2D peak modeling and the significance of model-fit results. The standard model of the SS 2D peak for this analysis in Eq.~(\ref{Eq4}) is factorized to 1D functions on $\eta_\Delta$ and $\phi_\Delta$. Whatever the actual peak structure on $\eta_\Delta$ we do observe that the narrow 1D Gaussian on $\phi_\Delta$ is independent of $\eta_\Delta$. Thus, factorization accurately describes the SS peak in all cases, and we can simplify the model choices on $\eta_\Delta$ to a single 1D Gaussian with parameters $(A,\sigma_\eta)$ vs a 1D Gaussian plus constant offset with parameters $(A',\sigma'_\eta,A_{1D})$.

\begin{figure}[h]  
\includegraphics[width=.23\textwidth,height=.235\textwidth]{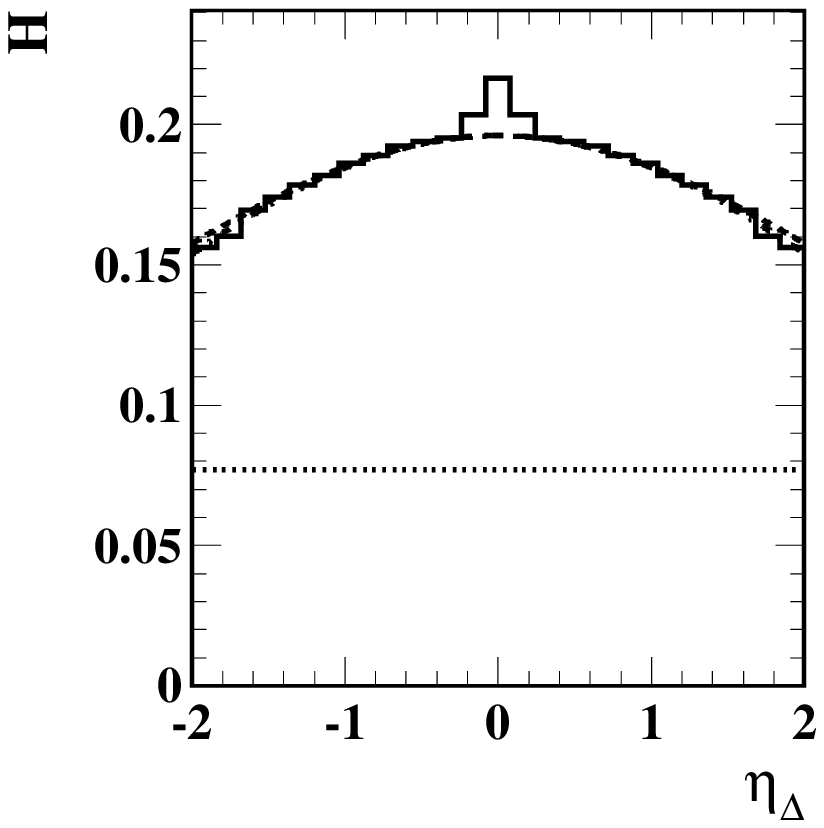}  
\includegraphics[width=.23\textwidth,height=.235\textwidth]{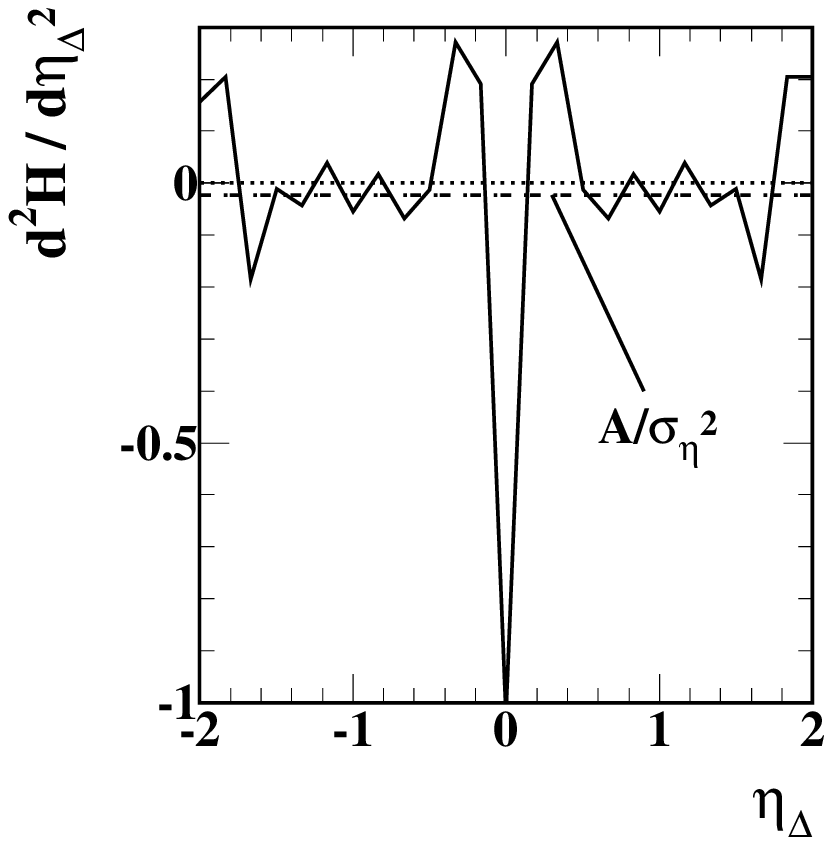} 
\caption{\label{modell}
Left: The data histogram is the 2D histogram in Fig.~\ref{ss2dsingles} (right panel) projected onto 1D $\eta_\Delta$. The narrow peak at the center is BEC + electron pairs. The dash-dotted curve through the data histogram is a single 1D Gaussian with parameters $(A,\sigma_\eta)$. The dashed curve is the sum of a 1D Gaussian $(A',\sigma'_\eta)$ plus constant offset $A_{1D}$. The two curves are barely resolved at the $\eta$ accepance limits. The dotted line is the constant offset.
Right:  The second difference of the data histogram in the left panel corresponds to its local curvature. The calculated curvature $A / \sigma^2_\eta$ of the 1D Gaussian in the left panel (dash-dotted curve) is the dash-dotted line in this panel. 
}  
\end{figure}



Figure~\ref{modell} (left panel) shows the histogram in Fig.~\ref{ss2dsingles} (right panel) projected onto 1D $\eta_\Delta$. The projection over $2\pi$ azimuth eliminates the small $\eta_\Delta$ modulation of the AS dipole discussed in Sec.~\ref{extracomp}. Those data can be described in a model-independent way with a Taylor series. Two series terms are systematically significant within the $\eta$ acceptance: the amplitude (constant) and the curvature (quadratic). The quartic term is not significant. The data are thus accurately described by a 1D Gaussian (dash-dotted curve) with amplitude $A = 0.195$ and width $\sigma_\eta = 2.9$. The Gaussian curvature is $A / \sigma_{\eta}^2 = 0.023 $. In the right panel the Gaussian curvature value (dash-dotted line) is plotted together with the second difference of the data histogram, comparing local  statistical noise to the curvature signal. The fourth difference (quartic) would be overwhelmed by statistical fluctuations.

If a constant offset is added to the 1D fit model the data can constrain {\em only certain parameter combinations}. The constraints are $A = A' + A_{1D}$ for the common peak amplitude and $A/\sigma_\eta^2 = A'/\sigma^{\prime2}_\eta$ for the common peak curvature at the mode. The parameter combination  $(A',\sigma'_\eta,A_{1D})$ is otherwise {\em free to vary} within those constraints. If we choose the value $A_{1D} = 0.077$ (dotted line, equivalent to $A_{1D} = 0.29$ for the unprojected 2D peak model) the data constraints determine that $A' = 0.12$ and $\sigma'_\eta = 2.27$. A dashed curve representing the sum of the modified 1D Gaussian and the constant offset is just visible in the left panel above the dash-dotted curve at the $\eta$ acceptance limits. If $\sigma'_\eta$ is reduced by 5\% the curves are not visually distinguishable. The same exercise could be carried out with a large range of $A_{1D}$ choices {\em including zero} (the standard data model). Returning to the 2D fit parameters in Table~\ref{params} we have $A = 0.76$, $A' = 0.47$, $A_{1D} = 0.29$ and $\sigma_\eta = 2.3$. The calculated $\sigma'_\eta = 1.82$ when reduced by 5\% is 1.73, consistent with the free-fit value 1.72 (see Table~\ref{params}).

SS 2D peak $\eta$ structure in data obtained within a limited $\eta$ acceptance, as in the present case, can only constrain two model parameters. Introduction of the offset  $A_{1D}$ (i.e., the 1D Gaussian on azimuth) as an additional model element leads to large fit instabilities because the data within the limited $\eta$ acceptance cannot constrain that parameter. The actual $A_{1D}$ values from model fits depend on small statistical fluctuations and systematic distortions at the $\eta$ acceptance boundary, as in Fig.~\ref{modell} (left panel). The additional model parameter acts to {\em amplify} relatively small and systematically insignificant variations in the data to appear as relatively large {\em but still systematically insignificant} variations in an extraneous model parameter. If the unnecessary element is introduced as a Fourier component ($v_3$) the definitions of other multipole amplitudes change substantially, and those parameters may then be misinterpreted. 



If a larger fraction of the SS 2D peak is accessible within a detector angular acceptance more peak properties are determined, such as higher peak moments (e.g., kurtosis) and possible non-Gaussian tail structure at larger $\eta_\Delta$. In more-peripheral Au-Au and p-p collisions essentially the entire SS 2D peak is resolved within the STAR TPC acceptance, and the 2D peak model is fully constrained. If the detector $\eta$ acceptance were extended  then even in more-central  Au-Au collisions peak structure at large $\eta_\Delta$ might be accurately determined. 

\subsection{Related systematic uncertainties} \label{multisys}

Appendix~\ref{equiv} describes the effects of adding  unnecessary model elements to a 2D fit model, and demonstrates the equivalence of adding either a $v_3$ or SS 1D Gaussian element to the model presented in this paper. The numerical analysis in that subsection demonstrates that any changes in the actual data description are less than 3\% of the reported parameter values. For instance, $A_1$ and the sum $A'_1 + A_{1D}$ agree to better than 3\%. Also, when a SS 1D Gaussian element is added to the standard model the fitted $A_D$ and $A_Q$ amplitudes change by less than 2\% of their values.

We further supplemented the systematic uncertainties analysis in Sec.~\ref{otherstruct} with a consideration of the possible effects of an undetected $v_3$ component or SS 1D Gaussian in the data. Data from representative centrality bins 38-46\% and 9-18\% for 200 GeV collisions were studied. 2D fit residuals from the standard model were fitted with either a $v_3$ or SS 1D Gaussian model component. Inferred amplitudes $A_S$ were less than 0.0003 and inferred amplitudes $A_{1D}$ were less than 0.01. Corresponding additional model elements with the inferred amplitudes held fixed were subtracted from the standard data model and the data were refitted with the revised model. Changes in the standard-model parameters were in all cases substantially less than the statistical fitting errors reported in this paper.

\subsection{Summary}

Ideally, data models should be composed of elements that are both necessary and sufficient. Removal of a necessary element from a data model results by definition in appearance of equivalent structure in the fit residuals. The model is incorrect without that element. A combination of {\em necessary} elements that leaves no significant structure in the fit residuals is {\em sufficient}. A model composed of a sufficient assembly of necessary elements accurately and efficiently describes the data. The model defined by Eq.~(\ref{Eq4}) is a necessary and sufficient data model for 2D histograms obtained within the STAR TPC angular acceptance, except as noted in Sec.~\ref{extracomp}.

{\em Within the STAR TPC acceptance} we find that
offset $A_{1D}$ values inferred by adding a SS 1D Gaussian on azimuth to the standard 2D fit model, {\em or equivalently} $v_3$ values inferred by adding a sextupole element, are not systematically significant. The results of this analysis have no implications for SS 2D peak structure at larger $\eta_\Delta$ outside the STAR TPC angular acceptance.

\section{Constructing ~${\bf \hat{r}_{ab}}$} \label{ratio}

Pair ratio $r = \rho_{\rm sib}/\rho_{\rm ref}$ is measured most accurately for a single charge combination ({i.e.}, $++$, $--$, or $+-$), event multiplicity $ N_{\rm ch}$ and vertex position $z_{\rm vtx}$ within the STAR TPC~\cite{star} because in that case charge-dependent tracking inefficiencies and acceptance effects most closely cancel in the ratio.  In a practical analysis particle pairs from small bins on $ N_{\rm ch}$ (subdivisions of a centrality bin) and $z_{\rm vtx}$ are combined into individual ratios denoted by 
\bea
\label{EqA1}
\hat{r}_{\alpha,ab} & = & \hat{n}_{\alpha,ab,{\rm sib}} / \hat{n}_{\alpha,ab,{\rm ref}},
\eea
where subscript $\alpha$ represents bin indices for charge-pair combination, event-multiplicity and vertex-position bins. 
%
Indices $(a,b)$ denote $(\eta_{\Delta},\phi_{\Delta})$ bins. 
We define normalized pair numbers
$\hat{n}_{\alpha,ab} = n_{\alpha,ab} / \sum_{ab} n_{\alpha,ab}$ for sibling or mixed (ref) pairs, where $n_{\alpha,ab}$ is the ensemble-averaged event-wise number of pairs in 2D angle bin $(a,b)$, with additional bin indices $\alpha$. We obtain  stable (optimized) angular correlations when  $ N_{\rm ch}$ and $z_{\rm vtx}$ bins are restricted to bin widths $\Delta N_{\rm ch} \leq 50$ and $\Delta z_{\rm vtx}\leq 5$~cm respectively.

Corrections were applied to each ratio $\hat{r}_{\alpha,ab}$ for two-particle reconstruction inefficiencies due to overlapping space points in the TPC
(two trajectories merged into one reconstructed track) and intersecting trajectories which cross paths within the TPC and are reconstructed as more than two tracks (splitting)~\cite{ayathesis}. The track-merging cuts rejected track pairs if both the longitudinal (along the TPC drift direction) and transverse (along the pad row direction) separation distances in the TPC were less than 5~cm at any of three radial positions in the TPC~\cite{star} (inner, middle and outer radii). The crossing cut rejected track pairs if their longitudinal separations at the middle and outer radii in the TPC were both less than 5~cm and if they cross in the transverse plane, {\em i.e.} $(\phi_{1,{\rm inner}} - \phi_{2,{\rm inner}})(\phi_{1,{\rm outer}} - \phi_{2,{\rm outer}}) < 0$, where angles $\phi$ for tracks 1 and 2 locate trajectory azimuth positions in the TPC at the inner or outer radii. The same track-pair cuts were applied to sibling and mixed pairs. Cuts have small overall effect except near  $(\eta_{\Delta},\phi_{\Delta}) = (0,0)$ where artifacts (depressions in $\hat{r}_{ab} - 1$) are significantly reduced.


The grand ratio for each angle and centrality bin was obtained by constructing a sibling-pair-weighted average over $\hat{r}_{\alpha,ab}$ for the relevant charge combinations (all four are summed), multiplicity bins, and vertex-position
bins indexed by $\alpha$. The final normalized ratio is defined by
\bea
\label{EqA2}
\hat{r}_{ab}(\nu) & = & \sum_{\alpha(\nu)} N_{\alpha,{\rm sib}}\, \hat{r}_{\alpha,ab} /
\sum_{\alpha(\nu)} N_{\alpha,{\rm sib}}
\eea
for centrality bin $\nu$, where $N_{\alpha,{\rm sib}} = \sum_{ab} n_{\alpha,ab,{\rm sib}}$.

\section{Event pileup corrections} \label{pileupcorr}

Event pileup can contribute spurious structure to angular correlations.
Ideally, each triggered collision event would appear in isolation in the TPC and be digitized over a readout time interval of 40~$\mu$s (TPC drift time~\cite{star}). At sufficiently high beam luminosity particle tracks from an untriggered A-A collision in a different beam-beam crossing (every 120 ns) may appear in the TPC tracking volume during readout of a triggered collision event, a source of contamination known as ``pileup.'' 
The extraneous particle tracks may be added to the triggered event or may be misidentified as the triggered collision event by the reconstruction code. We estimate that approximately 0.5\% and 0.05\% of the minimum-bias events in the 62 and 200 GeV Au-Au data sets respectively are contaminated by pileup. 

Although relatively few in number, pileup events can significantly affect angular correlations, particularly the $\eta_{\Delta}$ dependence. We observe that event pileup produces a characteristic ``W'' correlation shape on $\eta_{\Delta}$ due to the mis-match in $\eta$ range of particle distributions from out-of-time partial events ($\sim$1 unit) and those for intact reference events (2 units). Due to the $\eta$-range mismatch all particles in pileup events appear as correlated pairs relative to the normal (two units in $\eta$) reference-pair distribution, thus greatly amplifying the relative contribution of pileup events to the total angular correlations. For example, a 0.5\% pileup event rate contributes a few parts permil to sibling-to-mixed pair ratio $r$, which is comparable to the true correlation amplitude. The pileup structure has no significant $\phi_\Delta$ dependence.

Pileup events were identified and removed (with estimated
75\% efficiency) by exploiting the bi-directional drift of the STAR TPC
which causes reconstructed particle tracks from pre- and post-triggered
collisions to be either split at the TPC high-voltage central membrane
or truncated before the TPC readout plane.
These tracking artifacts produce distinctive patterns
in the event-wise track end-point distributions in the longitudinal drift direction
for trajectories within the active volume of the TPC, leading to an efficient pileup filter algorithm.

Pileup-free correlations $\Delta\rho/\sqrt{\rho_{\rm ref}}$ were constructed from 2D histograms obtained with and without the pileup filter by solving the equations
\bea
\label{Eq4a}
\frac{\Delta\rho}{\sqrt{\rho_{\rm ref}}}{\rm (no~filter)} & = &
\frac{\Delta\rho}{\sqrt{\rho_{\rm ref}}} +
\frac{\Delta\rho}{\sqrt{\rho_{\rm ref}}}{\rm (pileup)} \nonumber \\
\frac{\Delta\rho}{\sqrt{\rho_{\rm ref}}}{\rm (with~filter)} & = &
\frac{\Delta\rho}{\sqrt{\rho_{\rm ref}}} +
(1-f)\frac{\Delta\rho}{\sqrt{\rho_{\rm ref}}}{\rm (pileup)}, \nonumber \\
\eea
where $\Delta\rho/\sqrt{\rho_{\rm ref}}$(pileup) represents
the pileup contribution. 
The centrality-independent pileup detection efficiency $(1-f) = 0.25\pm 0.1$ assumed for this analysis is based on an estimate of the fraction of pileup events which had too few tracks crossing the central membrane to be identified by the adopted pileup signature. The estimated uncertainty in $(1-f)$ is propagated to the total uncertainties for the analysis results.

\begin{figure}[h]  
\includegraphics[width=.23\textwidth]{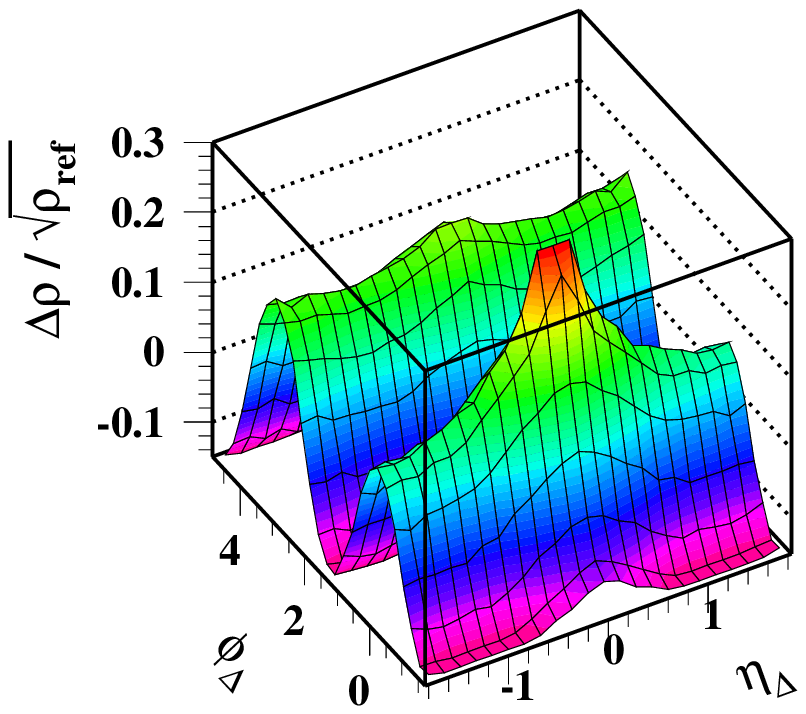} 
\includegraphics[width=.23\textwidth]{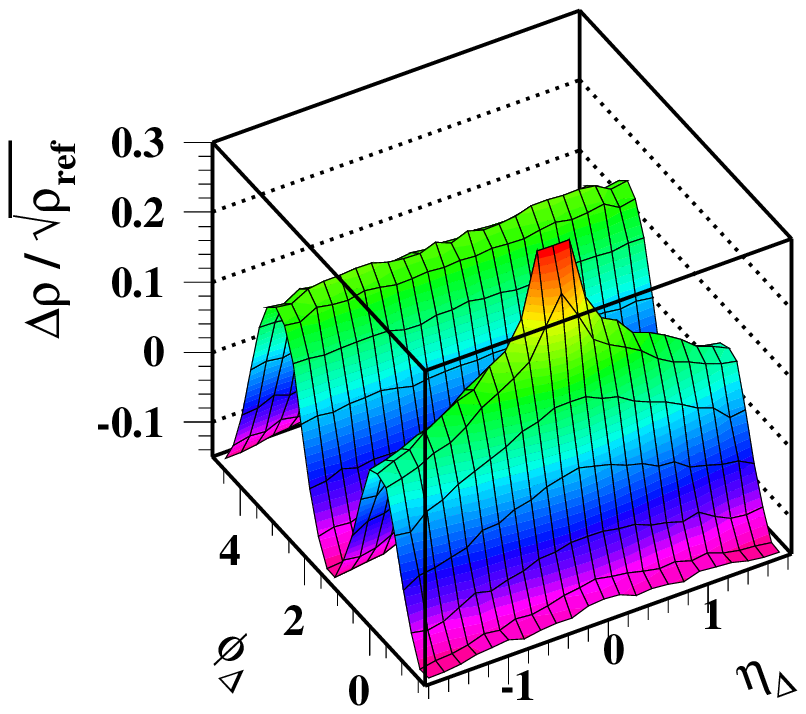}  
\caption{\label{pileup} (Color online)
Left: Uncorrected angular correlations from 62 GeV 37-46\% central Au-Au collisions showing pileup distortions, especially evident as the W-shaped nonuniformity of the away-side ridge on $\eta_\Delta$.
Right: The same data with pileup correction applied.
} 
\end{figure}

Figure~\ref{pileup} shows uncorrected (left panel) and corrected (right panel) histograms for mid-central 62 GeV Au-Au data where pileup distortions are most severe. Although pileup may affect a small fraction of the total event number (e.g. $< 1$\%), its effect on angular correlations can be substantial, as illustrated in the figure.

\section{130~GeV $\bf Au$-$\bf Au$ data} \label{130}

The 2D angular autocorrelation method used in this paper was first applied to unidentified charged hadrons from Au-Au collisions at 130 GeV~\cite{axialci}. Similar correlation structures were observed in the 130 GeV analysis, but with poorer statistics and coarser centrality binning. Here we compare 130 GeV data to the present analysis.

The 130 GeV centrality bins were defined by approximate cross-section fractions 40-70\%, 17-40\%, 5-17\% and 0-5\%. After conversion to measure  $\Delta\rho / \sqrt{\rho_{\rm ref}}$ using $\log{\sqrt{s_{\rm NN}}}$ interpolated prefactor $d^2\bar{N}_{\rm ch}/d\eta d\phi$ the same-side 2D Gaussian peak amplitude = 0.14, 0.22, 0.26, 0.24; $\sigma_{\eta_{\Delta}}$ = 0.58, 1.05, 1.34, 1.36; $\sigma_{\phi_{\Delta}}$ = 0.61, 0.55, 0.54, 0.53; dipole amplitude = 0.023, 0.052, 0.067, 0.057; and quadrupole amplitude = 0.079, 0.186, 0.090, 0.025 for fit parameters in Table~I of Ref.~\cite{axialci} for the four respective centralities. The quadrupole amplitude at 130 GeV interpolates the present results at 62 and 200 GeV. However, the dipole amplitude and the SS 2D Gaussian parameters do not, each being smaller than expected from linear interpolation of the 62 and 200 GeV parameters.

The 130~GeV correlation analysis differed from the present analysis in several respects: (a) In Ref.~\cite{axialci} a $p_t$ upper limit of 2.0 GeV/$c$ was imposed, whereas in this analysis there is no imposed upper limit. (b) Pair cuts were applied to minimize quantum correlations (HBT) and conversion-electron contamination near the angular origin. Those cuts reduced the correlation amplitude near $(\eta_{\Delta},\phi_{\Delta}) = (0,0)$ by about 20\%, including the same-side 2D Gaussian peak.  In the present analysis no such cuts were applied---the HBT/electron contribution is explicitly accounted for in the model function (the $A_2$ term).  (c) The 2D exponential term in the present fitting model was not included in Ref.~\cite{axialci}. (d) TPC two-track merging near the angular origin extended to twice the angular range of more recent data because the 130 GeV data were obtained with half the nominal magnetic field strength of the STAR detector.
(e) The 130~GeV analysis accepted a  larger pseudorapidity range ($|\eta|<1.3$) and included collision vertices distributed over a much larger distance interval ($|z_{\rm vtx}|<75$~cm).

It was not feasible to repeat the 130 GeV analysis starting from the raw data. However, the above $p_t < 2$~GeV/c restriction and extended pair cuts were applied to the present 200 GeV data in order to estimate their effects on the fitting results. The additional $p_t$ and pair cuts caused the 2D Gaussian amplitude and widths to decrease as did omitting the 2D exponential term in the fitting model. The dipole amplitude was similarly reduced, however the quadrupole amplitude was weakly affected except for the most-central 0-5\% bin where it increased. Omission of the 2D exponential had greater effect on the 2D Gaussian widths than on the amplitudes. The additional pseudorapidity range, increased track merging and extended collision vertex range in the 130 GeV data had negligible effects on the fit parameters.


Reversing the above fractional changes in amplitudes and shifts in widths and applying those changes to the 130 GeV fit parameters results in better agreement with expected energy-dependent trends, where: $A_1$ = 0.14, 0.32, 0.43, 0.31; $\sigma_{\eta_{\Delta}}$   = 1.5, 1.8, 2.3, 1.9; $\sigma_{\phi_{\Delta}}$   = 0.81, 0.67, 0.65, 0.58; $A_{\rm D}$ = 0.07, 0.11, 0.14, 0.08; and $A_{\rm Q}$ = 0.07, 0.17, 0.05, 0.01. The fractional amplitude and width increases introduce additional $\pm 10$\% uncertainties beyond that in Ref.~\cite{axialci}.

This estimation procedure is not intended as a substitute for a re-analysis of the 130 GeV raw data or even new measurements at 130 GeV. It does however provide a reasonable explanation for the discrepancy between the 130 GeV parameters in Ref.~\cite{axialci} and what would be expected given the present results at 62 and 200 GeV.

\section{{Tabulated data}} \label{taberrors}

Model fit parameters, statistical errors (in parentheses), and asymmetric systematic uncertainties (subscripts and superscripts) for 200 GeV and 62 GeV
Au-Au charged-particle angular correlation data for eleven centrality bins from most-peripheral (left column) to most-central (right column) are listed in Tables~\ref{TableI} and \ref{TableII}.  The volume of the
  same-side 2D Gaussian peak within the acceptance, corrected multiplicities, and Monte Carlo Glauber centrality
  measures $\nu$ and $N_{\rm part}$ are also listed. The full volume of the same-side 2D Gaussian peak extrapolated to $4\pi$ acceptance is given by $2\pi A_1 \sigma_{\eta_{\Delta}} \sigma_{\phi_{\Delta}}$. $\chi^2$/DoF values are for 158 degrees of freedom.

\end{appendix}

\setlength{\textheight}{10.0truein}
\begin{sidewaystable*}[t]
\caption{Model fit parameters, statistical errors (in parentheses), and asymmetric systematic uncertainties (subscripts and superscripts) for 200 GeV
Au-Au charged-particle angular correlation data for eleven centrality bins from
  most-peripheral (left column) to most-central (right column).  Prefactor ($d^2\bar N_\text{ch}/d\eta d\phi$) uncertainty (Sec.~\ref{Sec:ErrorsA}) is not included. The volume of the
  same-side 2D Gaussian peak within the acceptance, corrected multiplicities, and Monte Carlo Glauber centrality
  measures $\nu$ and $N_{\rm part}$ are also listed. $\chi^2$ values are for 158 degrees of freedom.
} 
  \label{TableI}
\begin{tabular}{ccccccccccccc}
 \multicolumn{12}{c}{\bfseries 200 GeV Au-Au} \\ \hline
$\sigma_{\rm AA}$(\%) & 84-93 & 74-84 & 64-74 & 55-64 & 46-55 & 38-46 & 28-38 & 18-28 & 9-18 & 5-9 & 0-5 \\
\hline
\vspace{0.05in}
$A_{\rm D}$ & 0.026(2)$^{+.004}_{-.004}$ & 0.037(2)$^{+.002}_{-.002}$ & 0.042(2)$^{+.004}_{-.004}$ & 0.047(3)$^{+.011}_{-.011}$ & 0.055(4)$^{+.013}_{-.015}$ & 0.105(10)$^{+.015}_{-.016}$ & 0.123(9)$^{+.012}_{-.015}$ & 0.215(14)$^{+.017}_{-.021}$ & 0.291(19)$^{+.020}_{-.023}$ & 0.278(25)$^{+.010}_{-.028}$ & 0.224(21)$^{+.014}_{-.017}$  \\
\vspace{0.05in}
$A_{\rm Q}$ & 0.002(1)$^{+.0005}_{-.001}$ & 0.011(1)$^{+.0005}_{-.001}$ & 0.028(1)$^{+.001}_{-.002}$ & 0.070(1)$^{+.003}_{-.004}$ & 0.136(2)$^{+.004}_{-.007}$ & 0.201(4)$^{+.006}_{-.009}$ & 0.270(4)$^{+.006}_{-.012}$ & 0.268(5)$^{+.007}_{-.012}$ & 0.179(6)$^{+.005}_{-.010}$ & 0.063(9)$^{+.001}_{-.010}$ & 0.001(8)$^{+.002}_{-.007}$  \\
\vspace{0.05in}
$A_1$  & 0.091(4)$^{+.007}_{-.006}$ & 0.105(10)$^{+.016}_{-.015}$ & 0.142(6)$^{+.012}_{-.011}$ & 0.169(6)$^{+.016}_{-.013}$ & 0.207(7)$^{+.022}_{-.011}$ & 0.316(19)$^{+.029}_{-.020}$ & 0.378(17)$^{+.033}_{-.009}$ & 0.590(28)$^{+.039}_{-.021}$ & 0.767(38)$^{+.048}_{-.031}$ & 0.760(50)$^{+.054}_{-.006}$ & 0.646(40)$^{+.036}_{-.012}$ \\
\vspace{0.05in}
$\sigma_{\deta}$ & 0.585(40)$^{+.053}_{-.031}$ & 0.699(45)$^{+.054}_{-.002}$ & 0.748(34)$^{+.057}_{-.005}$ & 0.771(35)$^{+.131}_{-0}$ & 0.892(45)$^{+.355}_{-0}$ & 1.372(78)$^{+.364}_{-.032}$ & 1.470(64)$^{+.562}_{-0}$ & 1.972(74)$^{+.566}_{-0}$ & 2.320(85)$^{+.589}_{-0}$ & 2.185(107)$^{+.561}_{-0}$ & 2.134(108)$^{+.656}_{-0}$ \\
\vspace{0.05in}
$\sigma_{\dphi}$ & 0.750(41)$^{+.064}_{-.064}$ & 0.974(72)$^{+.050}_{-.049}$ & 0.818(28)$^{+.031}_{-.030}$ & 0.756(23)$^{+.029}_{-.028}$ & 0.662(14)$^{+.012}_{-.049}$ & 0.670(16)$^{+.016}_{-.032}$ & 0.626(13)$^{+.009}_{-.029}$ & 0.659(11)$^{+.015}_{-.010}$ & 0.677(11)$^{+.015}_{-.007}$ & 0.663(15)$^{+.015}_{-.001}$ & 0.629(15)$^{+.009}_{-.006}$ \\
\vspace{0.05in}
$A_2$ & 0.184(43)$^{+.009}_{-.009}$ & 0.198(10)$^{+.008}_{-.008}$ & 0.335(13)$^{+.009}_{-.009}$ & 0.466(19)$^{+.012}_{-.011}$ & 0.656(21)$^{+.016}_{-.013}$ & 0.848(24)$^{+.018}_{-.014}$ & 1.120(33)$^{+.024}_{-.012}$ & 1.541(48)$^{+.024}_{-.012}$ & 2.180(97)$^{+.031}_{-.011}$ & 4.3$\pm$1.4$^{+.036}_{-.008}$ & 3.8(8)$^{+.043}_{-.007}$ \\
\vspace{0.05in}
$w_{\deta}$ & 0.059(21)$^{+.054}_{-.054}$ & 0.218(33)$^{+.012}_{-.012}$ & 0.164(12)$^{+.001}_{-.001}$ & 0.141(10)$^{+.008}_{-.008}$ & 0.113(5)$^{+.003}_{-.003}$ & 0.106(4)$^{+.004}_{-.005}$ & 0.090(3)$^{+.004}_{-.004}$ & 0.081(2)$^{+.002}_{-.002}$ & 0.069(2)$^{+.001}_{-.002}$ & 0.058(2)$^{+0}_{-.003}$ & 0.056(2)$^{+0}_{-.004}$ \\
\vspace{0.05in}
$w_{\dphi}$  & 0.126(27)$^{+.018}_{-.018}$ & 0.296(45)$^{+.009}_{-.009}$ & 0.183(14)$^{+.008}_{-.007}$ & 0.157(10)$^{+.007}_{-.007}$ & 0.133(7)$^{+.013}_{-.012}$ & 0.119(5)$^{+.012}_{-.011}$ & 0.093(4)$^{+.011}_{-.008}$ & 0.074(3)$^{+.004}_{-.001}$ & 0.056(3)$^{+.006}_{-0}$ & 0.029(11)$^{+.002}_{-.005}$ & 0.036(7)$^{+.002}_{-.008}$ \\
\vspace{0.05in}
$A_{0}$ & 0.034(2)$^{+.003}_{-.003}$ & 0.015(2)$^{+.001}_{-.001}$ & 0.013(2)$^{+.001}_{-.001}$ & 0.008(2)$^{+.002}_{-.002}$ & 0.004(1)$^{+.002}_{-.003}$ &
  $-$ & $-$ & $-$ & $-$ & $-$ & $-$ \\
\vspace{0.05in}
$\sigma_0$ & 0.593(44)$^{+.023}_{-.023}$ & 0.398(56)$^{+.019}_{-.019}$ & 0.309(42)$^{+.027}_{-.028}$ & 0.118(72)$^{+.074}_{-.076}$ & 0.210(66)$^{+.269}_{-.271}$ & $-$ & $-$ & $-$ & $-$ & $-$ & $-$ \\
\vspace{0.05in}
$A_3$ & -0.023(1)$^{+.001}_{-.001}$ & -0.025(2)$^{+.001}_{-.002}$ & -0.025(2)$^{+.002}_{-.003}$ & -0.025(2)$^{+.006}_{-.008}$ & -0.028(3)$^{+.006}_{-.012}$ & -0.057(7)$^{+.008}_{-.013}$ & -0.065(6)$^{+.005}_{-.014}$ & -0.124(10)$^{+.009}_{-.019}$ & -0.178(14)$^{+.011}_{-.020}$ & -0.172(18)$^{+.006}_{-.019}$ & -0.142(14)$^{+.007}_{-.015}$ \\

\hline\hline
 volume &  0.251 &  0.447 &  0.542 &  0.613 &  0.749 &  1.561 &  1.806 &  3.322 &  4.627 &  4.427 &  3.549 \\
$d\bar N_{\rm ch}/d\eta$ &   5.2  &  13.9 &  28.8  &  52.8  &  89  & 139   & 209  & 307  & 440 & 564  & 671 \\
       $\nu_{200}$ &   1.40 &   1.68 &   2.00 &   2.38 &   2.84 &   3.33 &   3.87 &   4.46 &   5.08 &   5.54 &   5.95 \\
     $N_{\rm part}$ &   4.6 &  10.5 &  20.5 &  36.0 &  58.1 &  86.4 & 124.6 & 176.8 & 244.4 & 304.1 & 350.3 \\
     $\chi^2$/DoF &  1.00  & 1.00    & 0.96   &  1.22  &  1.71  &  1.92  &  2.96  &  3.11  & 3.68   & 2.00   &  2.57  \\
\hline
  \end{tabular}
\end{sidewaystable*}


\begin{sidewaystable*}[t]
\caption{Same as Table~\ref{TableI} except for 62 GeV Au-Au charged-particle angular correlation data.}
  \label{TableII}
\begin{tabular}{ccccccccccccc}
 \multicolumn{12}{c}{\bfseries 62 GeV Au-Au} \\ \hline
$\sigma_{\rm AA}$(\%) & 84-95 & 75-84 & 65-75 & 56-65 & 46-56 & 37-46 & 28-37 & 18-28 & 9-18 & 5-9 & 0-5 \\
\hline
\vspace{0.05in}
$A_{\rm D}$ & 0.028(1)$^{+.002}_{-0}$ & 0.038(1)$^{+.004}_{-.001}$ & 0.042(1)$^{+.005}_{-.002}$ & 0.048(2)$^{+.005}_{-.002}$ & 0.056(2)$^{+.014}_{-.009}$ & 0.074(4)$^{+.022}_{-.012}$ & 0.121(8)$^{+.021}_{-.006}$ & 0.176(12)$^{+.025}_{-.013}$ & 0.197(13)$^{+.032}_{-.008}$ & 0.160(13)$^{+.046}_{-0}$ & 0.153(15)$^{+.054}_{-0}$  \\
\vspace{0.05in}
$A_{\rm Q}$ & 0.004(1)$^{+.001}_{-.001}$ & 0.007(1)$^{+.001}_{-.001}$ & 0.016(1)$^{+.002}_{-.001}$ & 0.037(1)$^{+.002}_{-.002}$ & 0.073(1)$^{+.004}_{-.003}$ & 0.117(1)$^{+.007}_{-.004}$ & 0.147(2)$^{+.007}_{-.003}$ & 0.148(4)$^{+.008}_{-.004}$ & 0.106(4)$^{+.010}_{-.002}$ & 0.048(4)$^{+.014}_{-0}$ & 0.003(5)$^{+.016}_{-0}$  \\
\vspace{0.05in}
$A_1$  & 0.043(3)$^{+.001}_{-.005}$ & 0.072(3)$^{+.004}_{-.008}$ & 0.084(3)$^{+.005}_{-.010}$ & 0.109(3)$^{+.007}_{-.011}$ & 0.128(4)$^{+.004}_{-.014}$ & 0.176(8)$^{+.006}_{-.025}$ & 0.284(16)$^{+.009}_{-.037}$ & 0.407(24)$^{+.024}_{-.047}$ & 0.458(25)$^{+.013}_{-.058}$ & 0.389(26)$^{+0}_{-.084}$ & 0.367(30)$^{+0}_{-.102}$ \\
\vspace{0.05in}
$\sigma_{\deta}$ & 0.559(40)$^{+.028}_{-.019}$ & 0.666(28)$^{+.027}_{-.018}$ & 0.720(27)$^{+.029}_{-.020}$ & 0.794(26)$^{+.038}_{-.029}$ & 0.907(36)$^{+.066}_{-.047}$ & 1.115(52)$^{+.170}_{-.128}$ & 1.567(69)$^{+.168}_{-.108}$ & 1.981(84)$^{+.237}_{-.126}$ & 2.104(80)$^{+.246}_{-.088}$ & 2.120(105)$^{+.417}_{-.261}$ & 2.159(128)$^{+.600}_{-.429}$ \\
\vspace{0.05in}
$\sigma_{\dphi}$ & 1.042(82)$^{+.054}_{-.057}$ & 1.039(50)$^{+.033}_{-.036}$ & 1.001(46)$^{+.040}_{-.040}$ & 0.859(22)$^{+.035}_{-.040}$ & 0.787(15)$^{+.034}_{-.041}$ & 0.736(12)$^{+.028}_{-.038}$ & 0.727(12)$^{+.015}_{-.030}$ & 0.741(12)$^{+.012}_{-.026}$ & 0.725(11)$^{+.005}_{-.032}$ & 0.672(15)$^{+0}_{-.058}$ & 0.689(18)$^{+0}_{-.064}$ \\
\vspace{0.05in}
$A_2$ & 0.086(9)$^{+.005}_{-.005}$ & 0.152(6)$^{+.007}_{-.006}$ & 0.251(7)$^{+.007}_{-.007}$ & 0.375(9)$^{+.008}_{-.008}$ & 0.495(10)$^{+.010}_{-.009}$ & 0.689(13)$^{+.013}_{-.011}$ & 0.935(16)$^{+.015}_{-.012}$ & 1.223(24)$^{+.020}_{-.010}$ & 1.676(40)$^{+.022}_{-.011}$ & 2.360(170)$^{+.031}_{-.006}$ & 4.2$\pm$1.0$^{+.033}_{-.007}$ \\
\vspace{0.05in}
$w_{\deta}$ & 0.100(19)$^{+.030}_{-.030}$ & 0.175(14)$^{+.022}_{-.022}$ & 0.169(9)$^{+.006}_{-.006}$ & 0.136(5)$^{+.004}_{-.005}$ & 0.133(4)$^{+.002}_{-.002}$ & 0.113(3)$^{+.001}_{-.002}$ & 0.098(2)$^{+.001}_{-.002}$ & 0.086(2)$^{+0}_{-.003}$ & 0.075(1)$^{+0}_{-.002}$ & 0.072(2)$^{+0}_{-.005}$ & 0.063(2)$^{+0}_{-.005}$ \\
\vspace{0.05in}
$w_{\dphi}$  & 0.227(35)$^{+.050}_{-.050}$ & 0.223(18)$^{+.022}_{-.022}$ & 0.195(10)$^{+.012}_{-.012}$ & 0.152(5)$^{+.004}_{-.004}$ & 0.134(4)$^{+.004}_{-.004}$ & 0.114(3)$^{+.006}_{-.004}$ & 0.092(2)$^{+.002}_{-0}$ & 0.075(2)$^{+.005}_{-0}$ & 0.060(2)$^{+.006}_{-0}$ & 0.042(3)$^{+.013}_{-0}$ & 0.026(10)$^{+.014}_{-0}$ \\
\vspace{0.05in}
$A_{0}$ & 0.036(2)$^{+.002}_{-.002}$ & 0.024(2)$^{+.002}_{-.002}$ & 0.017(1)$^{+.003}_{-.003}$ & 0.013(1)$^{+.001}_{-.002}$ & 0.013(1)$^{+.001}_{-.002}$ & 0.007(1)$^{+.002}_{-.002}$ & 
  $-$ & $-$ & $-$ & $-$ & $-$ \\
\vspace{0.05in}
$\sigma_0$ & 0.625(26)$^{+.010}_{-.010}$ & 0.480(23)$^{+.018}_{-.018}$ & 0.398(31)$^{+.092}_{-.092}$ & 0.347(30)$^{+.053}_{-.057}$ & 0.233(20)$^{+.021}_{-.026}$ & 0.153(25)$^{+.022}_{-.029}$ & $-$ & $-$ & $-$ & $-$ & $-$ \\
\vspace{0.05in}
$A_3$ & -0.021(1)$^{+.002}_{-0}$ & -0.021(1)$^{+.002}_{-.001}$ & -0.021(1)$^{+.003}_{-.001}$ & -0.022(1)$^{+.003}_{-.001}$ & -0.023(2)$^{+.010}_{-.007}$ & -0.031(3)$^{+.016}_{-.010}$ & -0.059(6)$^{+.014}_{-.005}$ & -0.097(9)$^{+.020}_{-.008}$ & -0.110(9)$^{+.023}_{-.004}$ & -0.092(9)$^{+.029}_{-0}$ & -0.090(11)$^{+.036}_{-.002}$ \\

\hline\hline
         volume &  0.157 &  0.312 &  0.378 &  0.462 &  0.558 &  0.841 &  1.623 &  2.580 &  2.889 &  2.279 &  2.215 \\
$d\bar N_{\rm ch}/d\eta$ &   3.3  &   9.3 &  19.8 &  36.9   &  62.8  & 99.0  & 149  & 217  & 312 & 403 & 488  \\
        $\nu_{62}$ &   1.34 &   1.59 &   1.87 &   2.22 &   2.62 &   3.05 &   3.49 &   3.96 &   4.46 &   4.82 &   5.13 \\
     $N_{\rm part}$ &   4.0 &   9.3 &  18.7 &  33.8 &  55.7 &  84.8 & 122.0 & 171.6 & 238.4 & 297.6 & 344.6 \\
     $\chi^2$/DoF &  1.14  & 1.30    & 1.08   &  1.51  &  1.26  &  1.65  &  2.18  &  2.72  & 3.56   & 3.63   &  3.32  \\
\hline
  \end{tabular}
\end{sidewaystable*}





\begin{thebibliography}{99}

\bibitem{qgp1}
H. R. Schmidt and J. Schukraft, J. Phys. G {\bf 19}, 1705 (1993).

\bibitem{qgp2}  J.~W.~Harris and B.~Muller,
  Ann.\ Rev.\ Nucl.\ Part.\ Sci.\  {\bf 46}, 71 (1996).

\bibitem{hydro1}
D. Teaney, J. Lauret and E. Shuryak, Phys. Rev. Lett. {\bf 86}, 4783 (2001).

\bibitem{hydro2} P. F. Kolb, U. Heinz, P. Huovinen, K. J. Eskola and K. Tuominen,
Nucl. Phys. A {\bf 696}, 197 (2001).

\bibitem{hydro3} U. Heinz, J. Phys. G {\bf 31}, S717 (2005).

\bibitem{hydro4} P. Huovinen and P. V. Ruuskanen, Ann. Rev. Nucl. Part. Sci. {\bf 56}, 163 (2006).

\bibitem{perfect1} T. Hirano and M. Gyulassy, Nucl. Phys. A {\bf 769}, 71 (2006)

\bibitem{perfect2} L. P. Csernai, J. I. Kapusta and L. D. McLerran, Phys. Rev. Lett. {\bf 97}, 152303 (2006).

\bibitem{perfect3} M.~Gyulassy and L.~McLerran,
  Nucl.\ Phys.\  A {\bf 750}, 30 (2005).

\bibitem{perfect4} B.~M\"uller,
  Acta Phys.\ Polon.\  B {\bf 38}, 3705 (2007).

\bibitem{kovchegov}
Y. V. Kovchegov, Nucl. Phys. A {\bf 764}, 476 (2006).

\bibitem{powerlaw}
T. A. Trainor and D. J. Prindle, hep-ph/0411217.

\bibitem{starwp}
J. Adams {\em et al.,} (STAR Collaboration),
Nucl. Phys. A {\bf 757}, 102 (2005).

\bibitem{phenixwp}
K. Adcox {\em et al.,} (PHENIX Collaboration),
Nucl. Phys. A {\bf 757}, 184 (2005).

\bibitem{phoboswp}
B. B. Back {\em et al.,} (PHOBOS Collaboration),
Nucl. Phys. A {\bf 757}, 28 (2005).

\bibitem{brahmswp}
I. Arsene {\em et al.,} (BRAHMS Collaboration),
Nucl. Phys. A {\bf 757}, 1 (2005).

\bibitem{highpt}
D. d'Enterria, J. Phys. G {\bf 30}, S767 (2004).

\bibitem{starraa}  C.~Adler {\it et al.}  (STAR Collaboration),
  Phys.\ Rev.\ Lett.\  {\bf 89}, 202301 (2002).

\bibitem{staras} C.~Adler {\it et al.}  (STAR Collaboration),
  Phys.\ Rev.\ Lett.\  {\bf 90}, 082302 (2003).

\bibitem{minijet} X.-N. Wang,  Phys. Rev. D {\bf 46}, R1900 (1992);
X.-N. Wang and M. Gyulassy, Phys. Lett. B {\bf 282}, 466 (1992).

\bibitem{axialci}
J. Adams {\it et al.} (STAR Collaboration),
Phys. Rev. C {\bf 73}, 064907 (2006).

\bibitem{ptxptci}
J. Adams {\it et al.} (STAR Collaboration),
J. Phys. G {\bf 34}, 799 (2007).

\bibitem{bubbles}
J. Adams {\it et al.} (STAR Collaboration),
Phys. Rev. C {\bf 75}, 034901 (2007).

\bibitem{phobosci}
B. Alver {\em et al.,} (PHOBOS Collaboration),
Phys. Rev. C {\bf 75}, 054913 (2007).

\bibitem{fragevo}     T.~A.~Trainor,
  Phys.\ Rev.\  C {\bf 80}, 044901 (2009).

\bibitem{sarc} I.~Sarcevic, S.~D.~Ellis and P.~Carruthers,
  Phys.\ Rev.\  D {\bf 40}, 1446 (1989).

\bibitem{kll}  K.~Kajantie, P.~V.~Landshoff and J.~Lindfors,
  Phys.\ Rev.\ Lett.\  {\bf 59}, 2527 (1987).

\bibitem{brown} A. Einstein, Ann. Phys. {\bf 17}, 549 (1905).

\bibitem{jethydro}
L. Pang, Q. Wang, X.-N. Wang and R. Xu, Phys. Rev. C {\bf 81}, 031903 (2010).

\bibitem{mjmuel} A. H. Mueller, Nucl. Phys. B {\bf 572}, 227 (2000).

\bibitem{nayak} G.~C.~Nayak, A.~Dumitru, L.~D.~McLerran and W.~Greiner,
  Nucl.\ Phys.\  A {\bf 687}, 457 (2001).

\bibitem{mjshin} G. R. Shin and B. M\"{u}ller, J. Phys. G {\bf 29}, 2485 (2003).

\bibitem{hijing} 
X.-N. Wang, M. Gyulassy, Phys. Rev. D {\bf 44}, 3501 (1991); version 1.382.

\bibitem{auto} T. A. Trainor, R. J. Porter and D. J. Prindle, J. Phys. G {\bf 31}, 809 (2005).

\bibitem{axialcd}
J. Adams {\it et al.} (STAR Collaboration), Phys. Lett. B {\bf 634}, 347 (2006).

\bibitem{ptedep} J.~Adams {\it et al.}  (STAR Collaboration),
  J.\ Phys.\ G {\bf 34}, 451 (2007).

\bibitem{jeffpp1} R.\,J.~Porter and T.\,A.~Trainor (STAR Collaboration), J. Phys. Conf. Series {\bf 27}, 98 (2005).

\bibitem{aspect}  R.~J.~Porter and T.~A.~Trainor  (STAR Collaboration),
  PoS C {\bf FRNC2006}, 004 (2006).



\bibitem{phenix}  A.~Adare {\it et al.}  (PHENIX Collaboration),
  Phys.\ Rev.\ Lett.\  {\bf 104}, 252301 (2010).

\bibitem{trigger}   J.~Adams {\it et al.}  (STAR Collaboration),
  Phys.\ Rev.\ Lett.\  {\bf 95}, 152301 (2005).

\bibitem{phoboscorr}  B.~Alver {\it et al.}  (PHOBOS Collaboration),
  Phys.\ Rev.\  C {\bf 81}, 024904 (2010).

\bibitem{star} K. H. Ackermann {\em et al.}, Nucl. Instrum. Meth. A {\bf 499}, 624 (2003); see other STAR papers in volume A{\bf 499}.

\bibitem{daugherity}  M.~Daugherity  (STAR Collaboration),
  J.\ Phys.\ G {\bf 35}, 104090 (2008).

\bibitem{ampt}  Z.~W.~Lin, C.~M.~Ko, B.~A.~Li, B.~Hang and S.~Pal,
          Phys.\ Rev.\ C {\bf 72}, 064901 (2005).

\bibitem{nexsph} Y.~Hama, T.~Kodama and O.~Socolowski, 
Braz.\ J.\ Phys. {\bf 35}, 24 (2005).

\bibitem{hbtreview} 
U. A. Wiedemann and U. Heinz, Phys. Rep. {\bf 319}, 145 (1999).

\bibitem{phobos62} 
B. B. Back {\em et al.,} (PHOBOS Collaboration),
Phys. Rev. C {\bf 74}, 021901(R) (2006).

\bibitem{starspec200} J. Adams {\it et al.} (STAR Collaboration),
Phys. Rev. Lett. {\bf 92}, 112301 (2004).

\bibitem{molnarthesis} \mbox{L.\,Molnar, Ph.\,D. Thesis, Purdue U. (2006),} {B.\,I.~Abelev {\em et al.} (STAR Collaboration),} Phys. Rev. C {\bf 79}, 034909 (2009).   


\bibitem{ayathesis}
A. Ishihara, Ph.~D. Thesis, University of Texas at Austin, 2004.

\bibitem{ua5}
G. J. Alner {\em et al.} (UA5 Collaboration), Z. Phys. C {\bf 33}, 1 (1986);
Phys. Rep. {\bf 154}, 247 (1987).


\bibitem{raycentral} R.~L.~Ray and M.~Daugherity, J. Phys. G {\bf 35}, 125106 (2008). 

\bibitem{siginel200} G.~J.~Alner {\it et al.}  (UA5 Collaboration),
  Z.\ Phys.\  C {\bf 32}, 153 (1986).

\bibitem{kn} D.~Kharzeev and M.~Nardi, Phys.~Lett.~B {\bf 507}, 121 (2001).

\bibitem{starjets} B.\ I.\ Abelev {\em et al.} (STAR Collaboration), Phys.\ Rev.\ Lett.\ {\bf 97}, 252001 (2006).

\bibitem{tomv2method1}
T.~A.~Trainor and D.~T.~Kettler,
  Int.\ J.\ Mod.\ Phys.\  E {\bf 17}, 1219 (2008),
 arXiv:0704.1674.
  
\bibitem{v2cumul}
C. Adler {\em et al.,} (STAR Collaboration),
Phys. Rev. C {\bf 66}, 034904 (2002).

\bibitem{lund}
B. Andersson, G. Gustafson, G. Ingelman and T. Sj\"{o}strand, Phys. Rep.
{\bf 97}, 31 (1983).

\bibitem{tzyam}  T.~A.~Trainor,
  Phys.\ Rev.\  C {\bf 81}, 014905 (2010).

\bibitem{quadpaper} D.~Kettler  (STAR collaboration),
  Eur.\ Phys.\ J.\  C {\bf 62}, 175 (2009).

\bibitem{Whitmore}
J. Whitmore, Phys. Rep. {\bf 27} (1976) 187.

\bibitem{jetyield}  T.~A.~Trainor and D.~T.~Kettler,
  Phys.\ Rev.\  C {\bf 83}, 034903 (2011).

\bibitem{ppspectra} J.~Adams {\it et al.}  (STAR Collaboration),
Phys. Rev. D {\bf 74}, 032006 (2006).

\bibitem{TomAuAuspectra} T.~A.~Trainor,
  Int.\ J.\ Mod.\ Phys.\  E {\bf 17}, 1499 (2008), arXiv:0710.4504.
  
\bibitem{mevsim} R.~L.~Ray and R.~Longacre, nucl-ex/0008009.

\bibitem{lisazibi}
Z. Chajecki and M. Lisa, Phys. Rev. C {\bf 78}, 064903 (2008). 

\bibitem{v2eta}
B. I. Abelev {\it et al.}  (STAR Collaboration), Phys. Rev. C {\bf 77}, 054901 (2008);  
B. B. Back {\em et al.,} (PHOBOS Collaboration),
Phys. Rev. Lett. {\bf 94}, 122303 (2005); Phys. Rev. C {\bf 72}, 051901(R) (2005).

\bibitem{v1a}
J.~Adams {\it et al.}  (STAR Collaboration),
Phys. Rev. C {\bf 73}, 034903 (2006).

\bibitem{v1b} 
B. B. Back {\em et al.,} (PHOBOS Collaboration),
Phys. Rev. Lett. {\bf 97}, 012301 (2006). 

\bibitem{2004}  J.~Adams {\it et al.}  (STAR Collaboration),
  Phys.\ Rev.\  C {\bf 72}, 014904 (2005). 

\bibitem{joern}  J.~Putschke  (STAR Collaboration),
  Nucl.\ Phys.\  A {\bf 783}, 507 (2007).

\bibitem{davidhq}   D.~Kettler  (STAR Collaboration),
  J.\ Phys.\ Conf.\ Ser.\  {\bf 270}, 012058 (2011).

\bibitem{ppcms} T.~A.~Trainor and D.~T.~Kettler,
  Phys.\ Rev.\  C {\bf 84}, 024910 (2011).

\bibitem{ptscale}
J. Adams {\it et al.} (STAR Collaboration),
J. Phys. G {\bf 32}, L37 (2006).

\bibitem{ffprd}
T. A. Trainor and D. T. Kettler, Phys. Rev. D {\bf 74}, 034012 (2006).

\bibitem{nucleonkt}
A. H. Mueller, A. I. Shoshi and S. M. H. Wong, Eur. Phys. J. A {\bf 29}, 49 (2006).  

\bibitem{tomv2method2} T. A. Trainor, Mod. Phys. Lett. A, {\bf 23}, 569 (2008), arXiv:0708.0792.

\bibitem{jeffpp0}  R.~J.~Porter and T.~A.~Trainor  (STAR collaboration),
  arXiv:hep-ph/0406330.

\bibitem{multipoles}  T.~A.~Trainor,
  arXiv:1109.2540.

\bibitem{triflow} B.~Alver and G.~Roland, 
Phys. Rev. C {\bf 81}, 054905 (2010).

\bibitem{xuko} J.~Xu and C.~M.~Ko, 
Phys. Rev. C {\bf 84}, 044907 (2011).

\bibitem{mawang} G.-L.~Ma and X.-N.~Wang, 
Phys. Rev. Lett. {\bf 106}, 162301 (2011).

\bibitem{sharma} M.~Sharma, C.~Pruneau, S.~Gavin, J.~Takahashi, R.~D.~de Souza and T.~Kodama, 
Phys. Rev. C {\bf 84}, 054915 (2011).

\bibitem{werner} H.~J.~Drescher, M.~Hladik, S.~Ostapchenko, T.~Pierog and K.~Werner, 
Phys. Rep. {\bf 350}, 93 (2001).

\bibitem{ua1} C.~Albajar {\it et al.}  (UA1 Collaboration),
  Nucl.\ Phys.\  B {\bf 309}, 405 (1988).
  
\bibitem{durand} L.~Durand and H.~Pi,
  Phys.\ Rev.\  D {\bf 40}, 1436 (1989).

\bibitem{hijingpp} X.~N.~Wang and M.~Gyulassy,
 Phys.\ Rev.\  D {\bf 45}, 844 (1992).
  
\bibitem{cooper} F.~Cooper, E.~Mottola and G.~C.~Nayak,
  Phys.\ Lett.\  B {\bf 555}, 181 (2003).

\bibitem{aleph}  D.~Buskulic {\it et al.}  (ALEPH Collaboration),
  Z.\ Phys.\  C {\bf 55}, 209 (1992).
  
\bibitem{pythia} T. Sj\"ostrand and M. van Zijl, Phys. Rev. D {\bf 36}, 2019 (1987).





\end{thebibliography}
\end{document}